\documentclass[fleqn,usenatbib]{mnras}

\usepackage{newtxtext,newtxmath}

\usepackage{blindtext}

\usepackage[T1]{fontenc}
\usepackage{tikz}
\usepackage{subcaption}
\usepackage{tabularx}
\usepackage{amsmath}

\usetikzlibrary{matrix}
\DeclareRobustCommand{\VAN}[3]{#2}
\let\VANthebibliography\thebibliography
\def\thebibliography{\DeclareRobustCommand{\VAN}[3]{##3}\VANthebibliography}

\usepackage{gensymb}
\usepackage{float}

\usepackage{booktabs}
\usepackage{wrapfig}
\usepackage{subcaption}

\usepackage{hyperref}
\usepackage{fixltx2e}
\usepackage{graphicx}	
\usepackage{amsmath}	
\usepackage{tabularray}

\usepackage{subcaption}
\usepackage{stackengine}

\usepackage{cleveref}

\graphicspath{{Plots/}}

    \title[Quenching at cosmic noon]{Environmental vs. intrinsic quenching at cosmic noon: Predictions from cosmological hydrodynamical simulations for VLT-MOONRISE}

\author[P. H. Goubert et al.]{Paul H. Goubert,$^{1}$\thanks{E-mail: pgoubert@fiu.edu},
Asa F. L. Bluck ,$^{1}$\thanks{E-mail: abluck@fiu.edu},
Joanna M. Piotrowska$^{2}$,
Paul Torrey$^{3,4,5}$,
Roberto Maiolino$^{6,7,8}$,
\newauthor Thomas Pinto Franco$^{1}$,
Camilo Casimiro$^1$
and Nicolas Cea$^1$
\\
\\
$^{1}$Stocker AstroScience Center, Dept. of Physics, Florida International University,11200 SW 8th Street, Miami, 33199, Florida, USA\\
$^{2}$Department of Astronomy, California Institute of Technology, 1200 East California Boulevard, Pasadena, California 91125, USA\\
$^{3}$Department of Astronomy, University of Virginia, 530 McCormick Road, Charlottesville, VA 22904, USA\\
$^{4}$Virginia Institute for Theoretical Astronomy, University of Virginia, Charlottesville, VA 22904, USA\\
$^{5}$The NSF-Simons AI Institute for Cosmic Origins, USA\\
$^{6}$Kavli Institute for Cosmology, University of Cambridge, Madingley Road, Cambridge, CB3 0HA, UK\\
$^{7}$Cavendish Laboratory, University of Cambridge, 19 J. J. Thomson Avenue, Cambridge CB3 0HE, United Kingdom\\
$^{8}$Department of Physics and Astronomy, University College London, Gower Street, London WC1E 6BT, UK\\
}

\date{Accepted XXX. Received YYY; in original form ZZZ}

\pubyear{2025}

\begin{document}
\label{firstpage}
\pagerange{\pageref{firstpage}--\pageref{lastpage}}
\maketitle

\begin{abstract}
We present an investigation into the quenching of simulated galaxies across cosmic time, honing in on the role played by both intrinsic and environmental mechanisms at different epochs. In anticipation of VLT-MOONRISE, the first wide-field spectroscopic galaxy survey to target cosmic noon, this work provides clear predictions to compare to the future observations. We investigate the quenching of centrals, high-mass satellites, and low-mass satellites from two cosmological hydrodynamical simulations: IllustrisTNG and EAGLE. Satellites are split according to bespoke mass thresholds, designed to separate environmental and intrinsic quenching mechanisms. To determine the best parameter for predicting quiescence, we apply a Random Forest classification analysis for each galaxy class at each epoch. The Random Forest classification determines supermassive black hole mass as the best predictor of quiescence in centrals and high-mass satellites. Alternatively, the quenching of low-mass satellites is best predicted by group halo mass, at all epochs. Additionally, we investigate the evolution in the dependence of the quenched fraction with various parameters, revealing a more complex picture. There is strong evidence for the rejuvenation of star formation from z = 2 to z = 0 in EAGLE, but not in IllustrisTNG. The starkest discrepancy between simulations rests in the mass threshold analysis. While IllustrisTNG predicts the existence of environmentally quenched satellites visible within the survey limits of MOONRISE, EAGLE does not. Hence, MOONRISE will provide critical data that is needed to evaluate current models, and constrain future models, of quenching processes.
\end{abstract}




\section{Introduction}

Galaxies, both in the local Universe and at higher redshifts, demonstrate a strong bimodality in numerous parameters, including color, morphology, and star formation rate (SFR)~\citep[e.g.,][]{Strateva_2001, Brinchmann_2004, Williams_2009, Peng_2010, Peng_2012}. Indeed, when analyzing the distribution of galaxies along the star formation rate (SFR) - stellar mass ($M_*$) (\citealt{Brinchmann_2004}), color - magnitude~\citep{Strateva_2001, Williams_2009}, and color - density~\citep[][]{Woo_2013} planes, a clear segregation becomes apparent. 

Accordingly, galaxies may then be split into two populations. These are colloquially referred to as the `blue cloud', or main sequence, and the `red sequence', or quenched population. The former is comprised of galaxies with strong ongoing star formation, while the latter consists of galaxies whose star formation rates fall far below the main sequence at a fixed stellar mass and redshift~\citep[e.g.,][]{Peng_2010, Peng_2012}. The process a galaxy undergoes to transition from the former to the latter is known as quenching. Studying this critical period in galaxies' lifetimes plays a pivotal part in understanding galaxy formation and evolution in general.

Quenching has been shown to come in one of two main forms, intrinsic and environmental. Intrinsic, or `mass', quenching is connected to properties such as stellar mass, black hole mass, central velocity dispersion, and bulge mass~\citep{Peng_2010, Bluck_2014, Bluck_2016, Piotrowska_2022, Brownson_2022}. Environmental quenching links quiescence to a galaxy's surroundings, through parameters such as halo mass~\citep[][]{Woo_2013, Barsanti_2018, Goubert_2024} and local galaxy over-density~\citep{Dokkum_2007, Peng_2010, Peng_2012, Woo_2013, Contini_2020, Bluck_2020b}. It is important to note that certain intrinsic and environmental parameters are correlated with another, such as black hole mass with halo mass and local density. As such, for this study environmental quenching refers solely to effects due to the environment itself (e.g., as a consequence of ram pressure stripping, tidal stripping or galaxy-galaxy interactions).

Intrinsic quenching connects the quiescence of galaxies to mechanisms which are inherent to the system. Possible theoretical routes linked to intrinsic quenching include, supernova feedback~\citep[][]{Bower_2008,Schaye_2015}, virial shock heating~\citep[][]{Dekel_2006, Woo_2015, Gabor_2015}, and active galactic nuclei (AGN) feedback~\citep[][]{Croton_2006,Sijacki_2007,Fabian_2012,Maiolino_2012,Weinberger_2017}. However, it has previously been shown that supernova feedback does not provide the energy necessary to fully quench more massive systems~\citep{Croton_2006, Bower_2006, Bower_2008, Smith_2018, Henriques_2019}. On the other hand, AGN feedback has been shown to be the most likely culprit in quenching massive galaxies, both in simulations and observations, both in the local Universe~\citep[][]{Weinberger_2018,Zinger_2020,Bluck_2016,Bluck_2020a,Piotrowska_2022}, and at higher redshifts~\citep[][]{Bluck_2022, Bluck_2023, Bluck_2024}.

While intrinsic processes are efficient in quenching high-mass galaxies, the same can not be said for low-mass systems~\citep{Wang_2018,Contini_2020,Bluck_2020b,Goubert_2024}. The presence of quiescent low-mass galaxies in the local Universe therefore indicates the need for a separate path to achieve quiescence. Causal links between environmental parameters, such as local over-density and halo mass, and the star formation rate (SFR), or color, of galaxies are well documented for the local Universe~\citep[see, e.g.,][]{Peng_2010,Peng_2012, Woo_2013, Woo_2015, Woo_2017}. Furthermore, quenched galaxies tend to be preferentially found in denser environments, and more massive groups, when compared to their star forming counterparts. As such, the environment of galaxies must be at least partially responsible for the quenching process, and is likely crucial at lower stellar masses. However, the exact mechanisms at play in environmental quenching remain unknown. There are numerous candidates, including ram pressure stripping~\citep[e.g.,][]{Gunn_1972, Kapferer_2009,Cramer_2019, Lotz_2019, Rhee_2024}, galaxy-galaxy harassment~\citep[][]{Moore1996, Moore_1998,Marasco_2016}, tidal stripping~\citep{Moore_1999, Fang_2016}, and other forms of starvation~\citep[see][]{vandenBosch_2008,Cortese_2009,van_de_Voort_2017,Xie_2020, Wright_2022}.

The arrival of JWST~\citep{Finkelstein_2023,Kocevski_2023,Kartaltepe_2023, Eisenstein_2023,Eisenstein_2023b}, has enabled the study of intrinsic quenching at intermediate and high redshifts~\citep[e.g.,][]{Bluck_2024, Carnall_2023, Carnall_2024}. As in the local Universe, the quiescence of massive galaxies (primarily centrals) at higher redshifts has been linked to the mass of their supermassive black hole, or a well motivated proxy thereof~\citep{Bluck_2023, Bluck_2024}. Therefore, even in the early Universe, the probable role of intrinsic quenching operating via AGN feedback has been clearly demonstrated. 

On the other hand, while there exists strong evidence for environmentally driven quenching at low redshifts~\citep[see, e.g.,][]{Bundy_2006,Ji_2018,Foltz_2018,Goubert_2024,Hamadouche_2024}, establishing its importance beyond the local Universe has proven challenging. The difficulty lies in the lack of wide-field spectroscopic observational surveys at intermediate and high redshifts. The dearth of spectroscopic redshift values is particularly detrimental to investigations into environment, as a 2D+z observer-space is further reduced to a purely 2D phase space (with highly limited radial distance constraints from photometric redshifts). Wide spread group contamination is largely unavoidable, impacting central - satellite classification, as well as the estimation of group parameters, especially halo mass. 

Whilst some spectroscopic measurements do exist at these epochs (for up to $\sim$1000's of galaxies), these are highly limited in area~\citep[][]{zCOSMOS_2007,VIPERS_2014,Candelsz7, DEugenio_2025}. Consequently, measurements of environment at cosmic noon and above are highly incomplete, lacking constraints on the full dynamical range in the cosmic web. Fortunately, the upcoming arrival of the VLT-MOONRISE survey~\citep{Maiolino_2020, Cirasuolo_2020} will largely resolve this issue by probing environment across volumes approaching the local homogeneity scale. This represents the ideal scenario for investigating environmental quenching at, and around, cosmic noon.

VLT-MOONRISE is anticipated to gather data on half a million galaxies within a redshift range of 0.9 < z < 2.6. This will allow us to study the effects of environment on galaxies, including their star formation rates, star formation histories, quiescent fractions, morphologies, and many other properties. In particular, the MOONRISE survey is predicted to probe a density range spanning over four orders of magnitude, exploring from voids to dense clusters. These observations, alongside robust estimates for halo masses and central - satellite classification (see Bluck et al. 2025, in press), will prove indispensable for analyzing the environment of galaxies and its connection to galaxy evolution. Furthermore, the spectroscopic data obtained will aid in constraining SFR, stellar mass, and even black hole mass (via calibrations, see \citealt{Bluck_2023,Bluck_2024}), whilst also collecting valuable information on optical AGN. Consequently, VLT-MOONRISE is set to become a revolutionary survey for understanding the quenching of galaxies within the full environmental context at cosmic noon.

In the meantime, cosmological hydrodynamical simulations present a unique opportunity for developing testable predictions for what is expected to be found at cosmic noon in VLT-MOONRISE. A clear set of predictions will be invaluable for enabling observers to immediately test and constrain theoretical expectations. We choose to utilize two leading cosmological hydrodynamical simulations in this work (IllustrisTNG \& EAGLE), enabling comparison both to future observations and to each other. Yet, to be effective, one must first carefully account for the observational, and theoretical, limitations in the comparisons. 

Due to the resolution limits of hydrodynamical simulations, much of the physics of galaxy evolution is modeled subgrid. This includes, star formation, stellar evolution, stellar feedback, supermassive black hole formation \& growth, and AGN feedback (among many other processes). The environment of galaxies is allowed to evolve according to the prescribed laws of physics (particularly those of hydrodynamics and gravitation), with a more modest impact from subgrid physics than on galaxy scales. Indeed, processes such as the hierarchical assembly of groups and clusters, the clustering of dark matter, and mechanisms such as ram pressure stripping, tidal stripping, and galaxy-galaxy harassment, occur essentially organically within simulations, albeit with limitations imposed through the resolution in volume and mass~\citep[see,][]{Bahe_2017, DeLucia_2025_Cosmic}. Hence, cosmological hydrodynamical simulations offer powerful insights into the impact of both  environment and intrinsic properties on galaxy evolution.

The binary nature of quenching (i.e., that galaxies can be either star forming or quenched) leads to its analysis being ideally suited to classification. Classification aims to determine the fundamental discrepancies between objects of different groups, in order to best segregate them accurately. Therefore, we choose to employ a Random Forest classifier as our primary method for analysis. Random Forest (RF) classification is a supervised machine learning classification algorithm, focused on identifying the key differences between objects of separate classes, so as to optimally sort them. This is achieved by determining which parameter, and threshold, is best at separating galaxies into either class. As the RF classifier excels at parsing through nuisance parameters to establish causal links between features and classes~\citep[see,][]{Bluck_2022}, it is particularly well suited for analyzing quenching, due to its complex nature and the inter-correlation of parameters. Moreover, classification via deep learning~\citep[e.g.,][]{Dominguez_2018,Busca_2018,Pérez-Carrasco_2019, DeDiego_2020} is not suitable for our purposes as it offers limited interpretability (see \citealt{Bluck_2022} for a discussion).

The RF method has previously successfully established causal links between the quenching of central galaxies and AGN feedback in simulations and observations of local Universe with the SDSS~\citep[see][]{Bluck_2020b,Piotrowska_2022}, and at higher redshifts~\citep{Bluck_2022,Bluck_2023,Bluck_2024}, comparing simulations to the HST-CANDELS \& JWST-CEERS surveys. Furthermore, the work done in \citet{Piotrowska_2022} and in \citet{Bluck_2022} demonstrates the potential of Random Forest classification for identifying the true, causal connections between parameters and quenching, via reverse engineering contemporary cosmological simulations. Through RF classification analysis, the integrated AGN feedback (probed by supermassive black hole mass) is identified to be the dominant predictor for quenching of central galaxies, even when accounting for the instantaneous AGN output, stellar mass of galaxies, and halo mass of galaxy groups.

In this paper, we expand upon our previous quenching works ~\citep{Bluck_2022,Bluck_2023,Bluck_2024,Piotrowska_2022,Brownson_2022,Goubert_2024} by investigating the mechanisms at play in the quenching of simulated galaxies at z = 0, 1, 2, from the EAGLE~\citep{Schaye_2015, Crain_2015, MCALPINE_2016} and IllustrisTNG~\citep[][]{Marinacci_2018, Naiman_2018, Nelson_2018,Springel_2018, Pillepich_2018b} hydrodynamical simulations. Of particular relevance to the present work, RF classification has previously been used in \citet{Goubert_2024} to identify the features most predictive of quenching for central galaxies, high-mass satellites, and low-mass satellites at z = 0 for simulations, as well as in SDSS observations. Through this analysis we determine that, for both simulations and observations, the quenching process of centrals and high-mass satellites is remarkably similar at low redshift, directly related to intrinsic AGN feedback via black hole mass. On the other hand, low-mass satellites in the local universe are determined to quench solely via environmental mechanisms, particularly those connected to halo mass. In this work we expand our prior analysis to cosmic noon.

Furthermore, while previous work had determined a central - satellite classification to be sufficient in separating intrinsic and environmental quenching~\citep{Peng_2012}, more recent work has shown an additional mass cut to be necessary~\citep{Goubert_2024, Hamadouche_2024, Lim_2025}. Therefore, to best examine each quenching avenue we split galaxies into one of three groups, as motivated in \citet{Goubert_2024}. Explicitly, we separate galaxies into: (i) centrals (which only quench at relatively high masses in significant numbers), (ii) high-mass satellites (which present as either star forming or quenched in high numbers), and (iii) low-mass satellites (which also present as either star forming or quenched in high numbers). Applying RF classification, along with several other statistical tests, we aim to simultaneously study intrinsic and environmental quenching from the local Universe to cosmic noon. Furthermore, in anticipation of VLT-MOONRISE, we leverage the known physics of the simulations, and our findings from each, to form specific predictions regarding quenching for upcoming observations. Through this analysis we gain insight in understanding the change in the impact of environment on quenching galaxies across cosmic time, and how it differs between simulations.

This work is structured as follows. In Sec. \ref{sec:Data} we detail the simulations and observational data used in this work. In Sec. \ref{sec:Methods} we first outline the `MOONRISE-Like' data set, which aims to apply conditions similar to those expected for VLT-MOONRISE to the simulations, and describe how we compute the nearest neighbor densities. We then give an overview of our Random Forest classification methodology, and finally present our quenched - star forming segregation criteria. In Sec. \ref{sec:Results}, we first present the stellar mass functions of different galaxy populations at each redshift from both simulations, and then define our mass thresholds which are used to segregate high and low-mass satellites. We include the results for the Random Forest classification and quenched fraction analyses. Finally, in Sec. \ref{sec:Summary} we summarize our findings. In Appendix~\ref{appendix:B} we include additional information regarding the parameters for our Schechter fits from Sec. \ref{sec:HighLowThresh}, as well details of our RF classification tests, including hyper-parameter values, seen in Appendix~\ref{appendix:C}.

\section{Data}\label{sec:Data}

\subsection{Hydrodynamical Simulations}

In this paper we make use of two cosmological hydrodynamical simulations, IllustrisTNG and EAGLE. For each of these simulations we obtain data at three redshift snapshots, z = 0, 1, 2.

\subsubsection{IllustrisTNG}

From IllustrisTNG\footnote{IllustrisTNG Data Access: \url{www.tng-project.org/}} (hereafter TNG) we use the TNG100-1 simulation suite, for multiple epochs (z = 0, 1, 2), contained within a simulation volume of $\sim (100 \ \mathrm{cMpc})^3$~\citep[][]{Marinacci_2018, Naiman_2018, Nelson_2018,Springel_2018, Pillepich_2018b}. The simulation assumes the cosmological parameters from the Planck Collaboration~\citep[][]{Plank_2016} assuming a spatially flat $\Lambda$CDM universe. 

The TNG simulation builds upon the original Illustris simulation~\citep{Vogelsberger_2014a, Genel_2014, Sijacki_2015} and Illustris model~\citep{Vogelsberger_2014a, Torrey_2014}. TNG utilizes the moving mesh code \textsc{arepo}~\citep{Springel_2018} to model gravitation, hydrodynamics, and magnetic fields, and implements an updated subgrid galaxy formation model~\citep{Pillepich_2018a, Weinberger_2017}. This allows for the modeling of galaxies and their environment, as well their evolution across cosmic time~\citep[see][]{Pillepich_2018a}. Importantly, the interactions between galaxies and their group haloes, and centrals, are directly captured, although environmental effects on galaxies remain dependent on certain subgrid models, especially the feedback implementations. This is of particular interest as we aim to understand how galaxy environment impacts galaxy evolution at multiple epochs.

\par Due to the limited resolution of the simulation, some processes must be accounted for via subgrid models. These include the physics behind star formation as well certain feedback mechanisms, namely supernova and AGN feedback. In TNG, black holes are seeded with mass $M_{\mathrm{BH}} = 8 \times 10^{5} \, {\mathrm{M_{\odot}}}\, h^{-1}$, once a halo reaches the threshold mass of $M_{\mathrm{Halo}} = 5 \times 10^{10} \, {{\mathrm{M_{\odot}}}}\, h^{-1}$. Their accretion is regulated via Eddington-limited, Bondi-Hoyle-Lyttleton accretion~\citep[see,][]{Bondi_1944, Hoyle_Lyttleton}. The AGN feedback prescription allows for two main feedback modes, a high accretion rate (quasar) mode, and a low accretion rate (radio/ kinetic) mode~\citep{Weinberger_2017, Weinberger_2018}. The quasar mode models AGN driven outflows during periods of high accretion rate. The kinetic mode attempts to account for the effect of jets by injecting momentum kicks imparted in a random direction, into the interstellar medium (ISM). The kinetic `jets' disturb the ISM, slowing star formation, and heat the circum-galactic medium (CGM), preventing cooling and accretion of gas into the system, enabling long-term quiescence. The AGN feedback mode at any given time is determined by the ratio of the black hole's accretion rate and the Eddington accretion rate, $f_{\mathrm{Edd}}$. Specifically, to switch from `quasar' to `kinetic' mode, the Eddington ratio of an AGN must be below the black hole mass dependent threshold $\chi$, expressed as:

\begin{equation}
\chi = \mathrm{min}\left[ 0.002\frac{M_{\mathrm{BH}}}{10^8 \mathrm{M_{\odot}}} ,\ 0.1 \right].
\end{equation}

\noindent Therefore, the kinetic mode is usually only active for black holes with $M_{\mathrm{BH}} \geq 10^8 {\mathrm{M_{\odot}}}$~\citep{Weinberger_2017}. 

Additionally, the star formation rate of galaxies in IllustrisTNG is modeled following the \citet{Spingel_Hernquist} prescription~\citep{Pillepich_2018a}. For each individual gas cell within a subhalo, the gas becomes star forming once it reaches a threshold density of $n_{\mathrm{Halo}} = 0.1\,\mathrm{cm}^{-3}$. The gas cells are then attributed a star formation rate, dependent upon their density, so as to comply to the empirically determined Kennicutt-Schmidt relation~\citep{Schaye_Kennicutt}, assuming a Chabrier stellar initial mass function~\citep{Chabrier_2003,Pillepich_2018a}. Star particles are then stochastically created in accordance with the star formation rate of the gas cell. The total SFR of a galaxy is taken to be the sum of the SFRs for all gas cells within the subhalo. It is possible for certain galaxies to have no gas that meets the criteria for star formation, and thus are an assigned SFR value of zero. Since we work in logarithmic units (where this would yield -$\infty$), for all galaxies with SFR = 0, we set log(SFR) = -5, an arbitrary low value which is well below the star forming main sequence (and hence leads to no confusion with our star forming - quenched classification).

\subsubsection{EAGLE}

We also utilize EAGLE\footnote {Eagle Data Access: \url{http://icc.dur.ac.uk/Eagle/}} (Evolution and Assembly of GaLaxies and their Environment; \citealt{Schaye_2015, Crain_2015, MCALPINE_2016}), a hydrodynamical simulation run with a modified version of \textsc{GADGET-3} smoothed particle hydrodynamics solver~\citep{Springel_2005}. We use the EAGLE-RefL0100N1504 suite for three snapshots, at  z = 0, 1, 2. EAGLE utilizes the Plank Collaboration I~\citep[][]{Plank_2014} cosmological parameters and assumes a $\Lambda$CDM universe. Similar to IllustrisTNG, while the physics and interactions on large scales are simulated directly, the resolution limits necessitate the implementation of subgrid models at small scales, $\sim1$\,kpc, and for non-gravitational and non-hydrodynamical physics~\citep[see][for full details]{Schaye_2015}. 

\par In EAGLE, black holes are seeded with $M_{\mathrm{BH}} = 10^{5} \, {\mathrm{M_{\odot}}}\, h^{-1}$ once a halo reaches $M_{\mathrm{Halo}} =  10^{10} \, {\mathrm{M_{\odot}}}\, h^{-1}$, and then accrete matter according to Eddington-limited, Bondi-Hoyle-Lyttleton accretion~\citep[see][]{Bondi_1944, Hoyle_Lyttleton}. However, unlike TNG, the AGN feedback prescription for EAGLE comes in only one mode. It is most similar to the quasar, or high accretion rate, mode of TNG, operating by injecting stored energy into neighboring gas cells, with the injection rate given by $\dot{E}_\mathrm{feedback} =\epsilon_{f}\epsilon_{r} \, \dot{M}_{BH}c^{2}$, where $\epsilon_{f}$ is the fraction of energy coupled to the ISM and $\epsilon_{r}$ is the radiative efficiency of the accretion disk. 

The quasar mode leads to a strong ejective AGN feedback process, which works by blowing large amounts of gas out from the ISM, and into the CGM, of the galaxy. Each black hole has its own energy reservoir, which increases by $\dot{E}_\mathrm{feedback} \, \delta t$ for each time step, $\delta t$. The temperature of the surrounding gas will be increased once enough energy is stored to heat the said gas by some amount, $\Delta T_\mathrm{AGN}$, set to $\Delta T_\mathrm{AGN} = 10^{8.5}K$ for this specific simulation suite. Importantly, as opposed to TNG, EAGLE has no bespoke preventative AGN feedback mode. As such, EAGLE has no purpose-built mechanism in place to prevent galaxies from re-accruing previously expelled gas~\citep[for a detailed description of the black hole subgrid model we refer the reader to,][]{Booth_Schaye, Schaye_2015}. Such a preventative mode may not be necessary in the real Universe, and therefore the lack of its inclusion may well be correct. Nevertheless, this is a very interesting difference between the feedback - quenching connection in the two simulations, which in principle may lead to observable differences.

Similarly to the IllustrisTNG simulation, the star formation rate of the EAGLE simulation follows a subgrid model. Specifically, the star formation prescription follows the pressure law of \citet{Schaye_Kennicutt}, and makes use of a simplified Kennicutt-Schmidt star formation law~\citep{Kennicutt_1998}. Explicitly, in its pressure dependent form, this is shown to be:

\begin{equation}
\dot{m}_{\star } = m_{\mathrm{g}} \,\, A \,\, (1 \,\, \mathrm{M}_{{\odot }} \,\, \mathrm{pc}^{-2})^{-n} \, \left(\frac{\gamma }{G} \,\, f_{\mathrm{g}} \,\, P\right)^{(n-1)/2},
\end{equation}

\noindent where $m_{\mathrm{g}}$ is the mass of the gas particle, $\gamma$ is the ratio of specific heats, $f_{\mathrm{g}}$ is the mass fraction in gas, G is the gravitational constant, and P is the total pressure. The values of the free parameters A and $n$ are set by observational measurements, with $A = 1.515 \times 10^{-4} \dot{m}_{\star }\,\,\mathrm{yr}^{-1}\,\,\mathrm{kpc}^{-2}$ and $n$ = 1.4, and $\gamma = 5/3$~\citep[][]{Crain_2015}.

This allows for the subgrid model to be calibrated by parameters specified by observations. The EAGLE simulation suite employs a metallicity-dependent density threshold, $\overset{\star}{n_{H}}(Z)$, defined as:

\begin{equation}
    \overset{\star}{n_{H}}(Z) = 0.1 \mathrm{cm}^{-3}\left(\frac{Z}{0.002}\right)^{-0.64}
\end{equation}

\noindent where Z is the gas metallicity. This accounts for the increased cooling efficiency of metal-rich gas in the subgrid prescription.

\subsection{Sample Selection}

For all data, we make use of a stellar mass selection criteria of $M_{*}> 10^9 \, {\mathrm{M_{\odot}}}$ and a halo mass selection criteria of $M_{\mathrm{200}}> 10^{11} \, {\mathrm{M_{\odot}}}$. This is done to ensure galaxies are well above the resolution limits of the simulations. 

As we desire to probe both environmental and intrinsic quenching at numerous redshifts, we first subdivide galaxies as either centrals or satellites~\citep[see][]{Peng_2010,Peng_2012,Bluck_2016,Goubert_2024}. A central galaxy is defined as the most massive galaxy of its group, while satellites are simply all of the remaining galaxies within each group. As previously seen, satellites are subject to quenching via both intrinsic and environmental means, dependent upon their stellar mass~\citep{Goubert_2024, Lim_2025}. Therefore, we further categorize satellite galaxies as either low or high-mass, split according to the mass thresholds determined in Section \ref{sec:HighLowThresh} for the simulations. 

\begin{table}
    \centering
        \caption{The number of galaxies which make up each galaxy class at every redshift for both TNG and EAGLE. We use the acronyms `LMS' and `HMS' for low and high-mass satellites respectively, while `Q' refers to quenched galaxies.}
    \begin{tblr}{
             colspec = { c c c c c c },  
             rowsep=0.5ex, colsep =1ex
             }
      \hline
      Simulation & N\textsubscript{Cen} & N\textsubscript{LMS} & N\textsubscript{LMS;Q} & N\textsubscript{HMS} & N\textsubscript{HMS;Q}\\
      \hline
       TNG; z = 0 & 11726 & 8152 & 3801 & 1024 & 775\\
     
       TNG; z = 1 & 11415 & 5551 & 1264 & 651 & 330\\
      
       TNG; z = 2 & 8956 & 2792 & 353 & 277 & 102\\
  
       EAGLE; z = 0 & 7125 & 4676 & 2299 & 422 & 211\\
    
       EAGLE; z = 1 & 7796 & 3105 & 515 &  740 & 223\\
   
       EAGLE; z = 2 & 5894 & 1138 & 98 &  464 & 111\\
       \hline
    \end{tblr}
    \label{tab:GalaxyNumber}
\end{table}

Additionally, future analyses are performed for up to three separate data sets for each parent sample. These are: (i) a complete data set (for which the only restrictions are the stellar mass selection criteria of $M_{*}> 10^9 \, {\mathrm{M_{\odot}}}$ and the halo mass selection criteria of $M_{\mathrm{Halo}}> 10^{11} {\mathrm{M_{\odot}}}$); (ii) a 2D+z complete data set which follows the same mass criteria but operates in 3D `observer' space (2D coordinate space plus observation-like spectroscopic redshift), as opposed to the full 6D phase-space of the simulations; and (iii) a `MOONRISE-Like' data set (described in Section \ref{ssec:MOON-like}). We show the number of galaxies belonging to each class at each redshift in Tables~\ref{tab:GalaxyNumber} \& \ref{tab:GalaxyNumber_Moons}, for the complete and `MOONRISE-Like' data sets respectively.

\begin{table}
    \centering
        \caption{The number of galaxies which make up each galaxy class at every redshift for the `MOONRISE-Like' of both TNG and EAGLE.}
    \begin{tblr}{
             colspec = { c c c c c c },  
             rowsep=0.5ex, colsep =1ex
             }
      \hline
      Simulation & N\textsubscript{Cen} & N\textsubscript{LMS} & N\textsubscript{LMS;Q} & N\textsubscript{HMS} & N\textsubscript{HMS;Q}\\
      \hline
       TNG; z = 0 & 5365 & 2013 & 915 & 647 & 507\\
     
       TNG; z = 1 & 5074 & 1307 & 298 & 379 & 191\\
      
       TNG; z = 2 & 3588 & 530 & 51 & 183 & 58\\
  
       EAGLE; z = 0 & 3143 & 1269 & 578 & 256 & 132 \\
    
       EAGLE; z = 1 & 3158 & 406 & 45 & 431 & 134\\
   
       EAGLE; z = 2 & 2050 & 0 & 0 &  257 & 58\\
       \hline
    \end{tblr}
    \label{tab:GalaxyNumber_Moons}
\end{table}

\section{Methods}\label{sec:Methods}

In this section we first detail the construction of the `MOONRISE-Like' data set, which aims to form samples under similar conditions to what is expected to be observable in VLT-MOONRISE. We then describe how local galaxy over-density is measured, and give an overview of the Random Forest classification method. Finally in this section, we present our methodology for determining quiescence at each redshift, for both TNG and EAGLE.

\subsection{MOONRISE-Like Data}\label{ssec:MOON-like}

VLT-MOONRISE is anticipated to become the largest spectroscopic survey at intermediate-to-high redshifts, targeting up to half a million galaxies~\citep{Maiolino_2020,Cirasuolo_2020}. The survey will target the environment and behavior of galaxies at (and around) cosmic noon ($z \sim 1.5 - 2$). In this paper, we leverage the IllustrisTNG and EAGLE simulations to pinpoint the important processes, and parameters, with respect to quenching across cosmic time, specifically with the goal to test this with VLT-MOONRISE in the near future.

In order to make accurate predictions for observational data it is important to have comparable simulated data sets. Therefore, we choose to form `MOONRISE-Like' data sets for the simulations, applying restrictions similar to those expected for VLT-MOONRISE. These include a stellar mass selection criteria of $M_{*}> 10^{9.5} \, {\mathrm{M_{\odot}}}$ (above which MOONRISE will be approximately mass complete); operating within a 3D observer-space (2-coordinate space, X-Z in our case, \& observational-like redshift) rather than the full 6D phase space (3-coordinate and 3-velocity space) of simulations \footnote{While we choose the Y-axis as our line of sight (LoS) for all data sets, we find our results to be consistent independent of the axis upon which the LoS is projected.}; and randomly removing 30\% of the data to mimic the anticipated 70\% completeness expected for VLT-MOONRISE. 

The observational-like redshift is computed as the sum of a cosmological redshift, $z_{\mathrm{cos}}$, and a peculiar redshift, $z_{\mathrm{pec}}$, dependent on the peculiar velocity of each galaxy. As the simulation boxes are each at a specific redshift, the cosmological redshift is computed as:

\begin{equation}
    z_{\mathrm{cos}} = z(D_c(z_\mathrm{snapshot}) + Y_i)
\end{equation}

\noindent where $D_c(z_\mathrm{snapshot})$ is the comoving distance to the redshift of a specific snapshot (z = 0, 1, 2) and $Y_i$ is the comoving distance in the Y axis, the coordinate perpendicular to the `2D plane’ (X-Z plane here). The redshift at this total comoving distance can now be calculated from:

\begin{equation}
    \mathrm{D_c}(z) = \frac{c}{H_0}  \int_{0}^{z} \frac{dz'}{E(z')} \, 
\end{equation}

\noindent with E(z) is given by:

\begin{equation}
\mathrm{E(z)} = \sqrt{\Omega_{M,0}(1+z)^3 + \Omega_{k,0}(1+z)^2 + \Omega_{\Lambda,0}} \,\text{,}
\end{equation}

\noindent where $\Omega_{M}$ is the total matter density, $\Omega_{k}$ describes the curvature of the Universe, and $\Omega_{\Lambda}$ is the fraction of the critical density attributed to dark energy (or the cosmological constant $\Lambda$). Therefore, $z_{\mathrm{cos}}$ represents the cosmological redshift for each galaxy, accounting for both the snapshot redshift and the positioning of the galaxy within the simulation box.

A peculiar redshift is then evaluated for each galaxy, using their peculiar velocities in the Y axis, $V_{y}$. Specifically, $z_{\mathrm{pec}}$ is given by:

\begin{equation}
    z_{\mathrm{pec}} = \frac{V_{y}(1+z_\mathrm{snapshot})}{c} \,\, \mathrm{for} \,\, V_y \ll c \,.
\end{equation}

\noindent It is then possible to assign an observational-like redshift to every galaxy, computed as:

\begin{equation}
    z_{\mathrm{obs}} = z_\mathrm{cos} + z_\mathrm{pec}.
\end{equation}

\noindent As is done for observational data, we apply velocity restrictions prior to estimating distance measurements. This assures objects are likely to be physically connected. Specifically we require:

\begin{equation}
    \Delta V_\mathrm{los} = \frac{c|\Delta z|}{(1+\bar{z})} \leq 1500 \, \mathrm{km/s} \label{eq:Vlos}
\end{equation}

\noindent where $\Delta V_\mathrm{los}$ is the difference in line of sight velocities for two galaxies, $\Delta z$ the difference in observer like redshift, and $\bar{z}$ is the mean redshift of the two galaxies. 

For a more detailed description of the methodology used here, we refer the reader to Bluck et al. 2025 (in press). 

We present the amount of galaxies belonging to each class at each redshift in Table~\ref{tab:GalaxyNumber_Moons}, for the `MOONRISE-Like' data sets. From the information in Table~\ref{tab:GalaxyNumber_Moons}, the impact of the imposed criteria, particularly the mass completeness limit threshold of $M_{*}> 10^{9.5} \, {\mathrm{M_{\odot}}}$, on each population is made clear. Most notable is the depletion of low-mass satellites for EAGLE beyond $z\geq1$. We will return to this important difference between EAGLE and TNG throughout Section 4.

\subsection{Nearest Neighbor Density}\label{ssec:local density}

In this work we define a galaxy's local over-density as:

\begin{equation}\label{eq:6D}
\delta_{n} = \log({\rho_n}) - {\langle \log( \rho_n})\rangle; \,\, \mathrm{where,} \,\, \rho_n = \left(\frac{3n}{4\pi R^{3}}\right) \,\, \text{for 6D,} 
\end{equation}
\noindent and
\begin{equation}\label{eq:2D+z}
    \delta_{n} = \log({\Sigma_n}) - \langle \log(\Sigma_n) \rangle; \,\, \mathrm{where,} \,\, \Sigma_n = \left(\frac{n}{\pi R^{2}}\right) \,\, \text{for 2D+z.}  
\end{equation}

\noindent Here $n$ denotes the nearest neighbor (third, fifth, tenth, etc...) and R is the distance to the n\textsuperscript{th} nearest neighbor. For all data sets, we determine the local density to the third, fifth, and tenth nearest neighbors, yielding $\delta_{3}$, $\delta_{5}$, and $\delta_{10}$. We present the nearest neighbor density as an over-density, normalized by the geometric mean in log space. For the complete, 6D phase space, sample sets we utilize equation~\eqref{eq:6D}, while for the `2D complete' and `MOONRISE-Like' data sets we employ equation~\eqref{eq:2D+z}. 

To properly account for the periodic boundary conditions of the simulation boxes, we create duplicates of the original box, and arrange one to each of its sides and corners~\citep[see][for more detail]{Goubert_2024}. For the relatively small scales of clusters (compared to the simulation box) this poses no concern of duplications via multiple box crossing on the scales of interest in this work.

\subsection{Random Forest Classification}\label{ssec:RF_meth}

In this work we utilize a Random Forest classifier, constructed from the {\small SCIKIT-LEARN PYTHON} package~\citep{scikit-learn}, to determine the parameters most important for predicting quiescence. Random Forest classification is a supervised machine learning classifier that works by sorting data into types given a set of input features. A Random Forest is composed of multiple decision trees, each of which is given a subset of randomly bootstrapped data samples. In this way, Random Forest classification is highly efficient at determining whether parameters are causally linked, or simply inter-correlated nuisance parameters~\citep[see,][]{Bluck_2022}. 

For all trees within the Random Forest, the input data serves as a `root' node, and is repeatedly split into `parent' and `daughter' nodes. The `parent' nodes are split so as to generate increasingly more homogeneous `daughter' nodes (better segregated according to class). There are multiple techniques that may be employed to yield an optimal split. For our purposes, we choose to focus on the Gini impurity. The Gini impurity measures the probability of selecting a misclassified object in a random node, given by:

\begin{equation}
    G(n) = 1 - \sum_{i}^{c} p_{i}(n^{2}),
\end{equation}

\noindent where the summation is performed over all classes, $c$, and where, $p(n)$ indicates the probability of randomly selecting class, $i$, at node, $n$. At each node of each tree, the input feature and threshold which best minimizes the Gini impurity is chosen to split the remaining data. 

As opposed to some machine learning techniques, such as artificial neural networks, Random Forest classification is not a black box ML algorithm. Explicitly, this enables the extraction of the performance of each parameter in classifying the data individually (rather than in aggregate in more complex methodologies). The `performance' of a parameter is then interpreted as the relative importance of a given input feature when classifying. Computing the relative importance for each input feature is done via the following equation:

\begin{equation}
        I_{R}(k) = \frac{1}{N_\mathrm{trees}} \sum_\mathrm{trees}\Biggl\{ \frac{\sum_{nk}N(n_{k})\Delta G(n_{k})}{\sum_{n}N(n)\Delta G(n)}\Biggl\},
\end{equation}

\noindent where $I_R(k)$ is the relative importance of input feature $k$~\citep{scikit-learn}. The numerator is a sum over all nodes that split with feature $k$, and the denominator is summed over all of the nodes. Here, $\Delta G$ corresponds to the improvement of the Gini impurity from parent to daughter node, weighed by the number of input items to be classified which reach the parent node. The mean relative importance of each given feature, $k$, is computed by averaging $I_{R}(k)$ over the totality of the Random Forest. In our case, this returns the importance to quenching of each input feature, controlling for all other available features simultaneously. For deeper insight on Random Forest classification, we refer the reader to \citet{Bluck_2022}, where a number of detailed tests on mock and simulation data are performed in their Appendix B.

\begin{figure*}
    \centering
    \begin{subfigure}{\textwidth}
    \caption{TNG}
        \includegraphics[width=\textwidth]{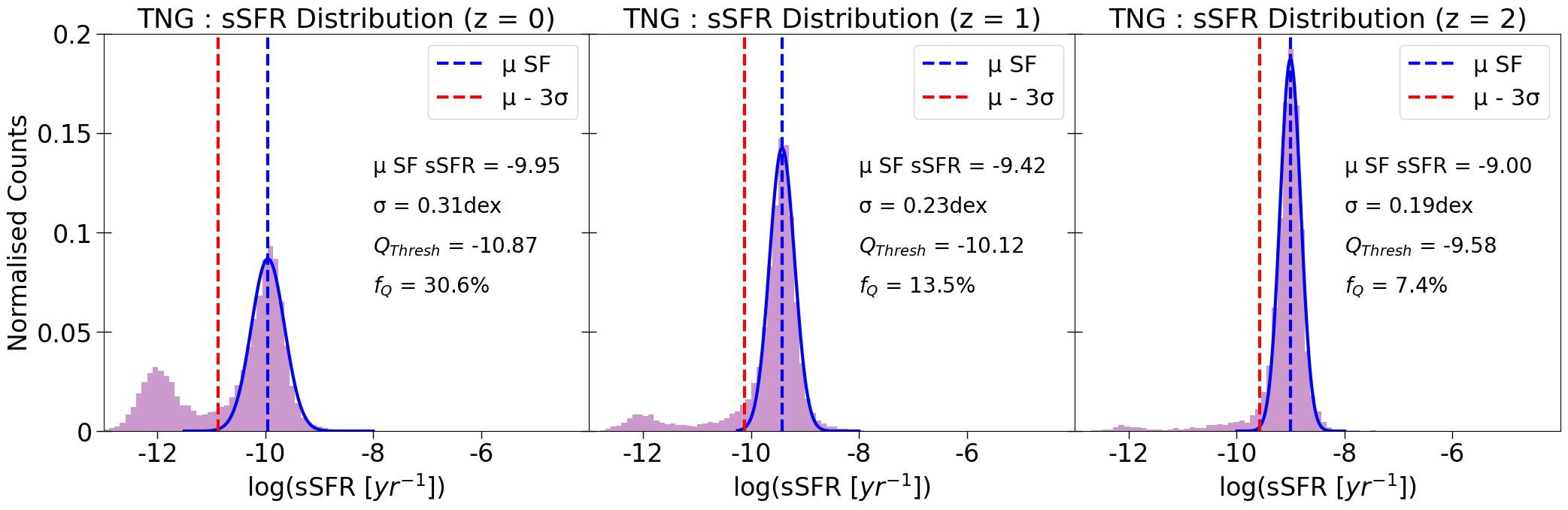}  
        \label{fig:QuenchThresh_TNG}
    \end{subfigure}
    \hfill
    \begin{subfigure}{\textwidth}
    \caption{EAGLE}
        \includegraphics[width= \textwidth]{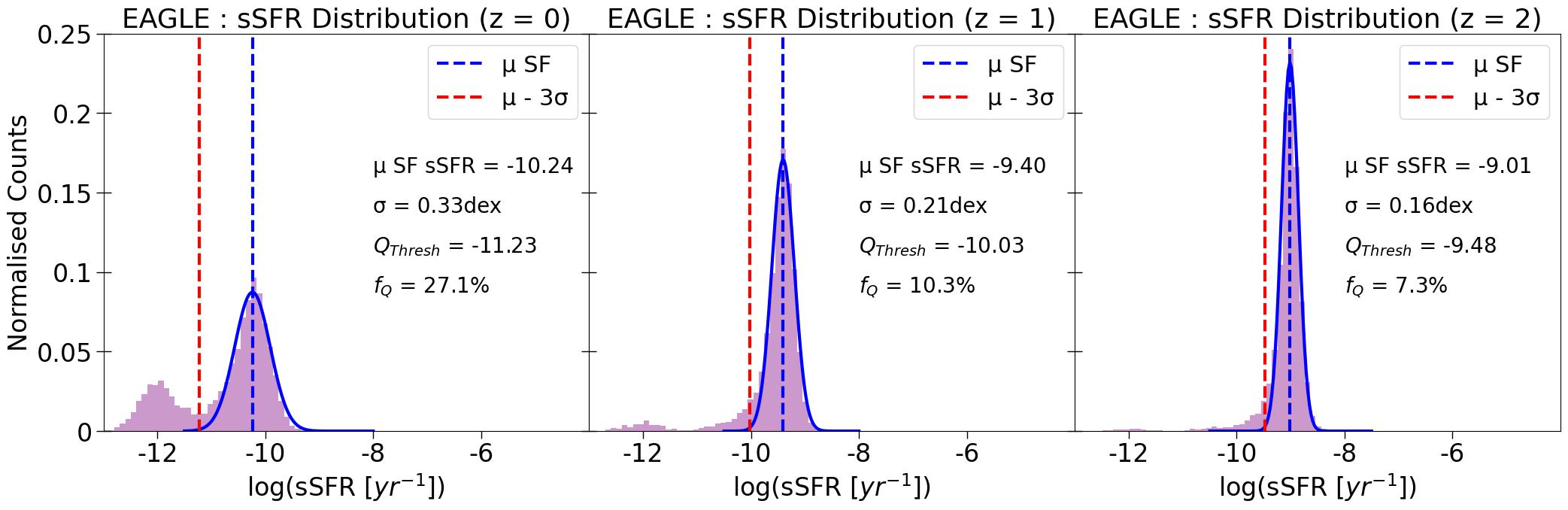} 
        \label{fig:QuenchThresh_EAGLE}
    \end{subfigure}

\vspace*{-3mm}
\caption{Illustration of our method to determine the quenched threshold for galaxies from TNG (top panels) and EAGLE (bottom panels) at z = 0, 1, 2 (along each row). The panels show the sSFR distribution at each redshift for each simulation. Additionally, a Gaussian fit to the star forming population is overlaid (removing obviously quiescent systems at $\log(\mathrm{sSFR}/{\mathrm{yr}}^{-1}) < -11.5$). On each panel we present the mean sSFR of star forming systems (indicated by a blue dashed line) and the determined quenching threshold (shown as a dashed red line), see equation~\eqref{eq:SSFR_THRESH}. Note that we redistribute very low sSFR values to a normal distribution centered around $\log(\mathrm{sSFR}/\mathrm{yr}^{-1}) = -12$, in order to represent all galaxies (which may extend to $\log(\mathrm{sSFR}/\mathrm{yr}^{-1}) = -\infty$) on each panel. The mean sSFR increases with redshift and the width of the distribution decreases with redshift in a similar manner in both simulations. The fraction of quenched galaxies expected from this methodology at each epoch is displayed on the panels. In both simulations, there is a clear reduction in the quenched fraction as redshift increases, from $f_{\mathrm{Q}}\sim 30\%$ at z = 0, to $f_{\mathrm{Q}}\sim 7\%$ at z = 2. Clearly, the majority of quenching takes place between z = 2 and z = 0.} \label{fig:QuenchThresh}
\end{figure*}

For our purposes, galaxies are taken to be in one of two classes, either quenched or star forming. The input features for each RF test are some combination of the following parameters: black hole mass ($M_{\mathrm{BH}}$), halo mass ($M_{\mathrm{Halo}}$), stellar mass
($M_{\mathrm{*}}$), nearest neighbor density ($\delta$), the distance to the central galaxy ($D_{\mathrm{cen}}$), and a random number in the range 0 - 1 (Rand.). The relative performance of $\delta$ is computed as the sum of the relative importances of $\delta_{3}$, $\delta_{5}$, and $\delta_{10}$. For all future Random Forest classification results, the relative importance presented represents the mean relative importances of a given parameter averaged over 100 Random Forest classification runs. The uncertainty in relative importance is given as the standard deviation across the 100 training and testing runs~\citep[as in][]{Goubert_2024}.

\subsection{Determining Quiescence}\label{ssec:Q_thresh}

\par Since the SFR - $M_{*}$ relation evolves with redshift, so must the threshold at which a galaxy is to be deemed quenched~\citep[e.g.,][]{Madau_2014,Ilbert_2015,Girelli_2019,Bluck_2024}. As such, it is important to establish a method to determine this threshold in a consistent manner throughout varying epochs and simulations. In this work, we accomplish this by leveraging the characteristics of the sSFR ($\equiv \mathrm{SFR}/M_*$) distributions at each relevant redshift of the TNG and EAGLE simulations. We present the sSFR distributions in Figs. \ref{fig:QuenchThresh_TNG} \& \ref{fig:QuenchThresh_EAGLE}, for TNG and EAGLE, respectively. 

A Gaussian curve is fit around the peak of the star forming region of each distribution, from which the mean and standard deviation are extracted. We exclude obviously quenched galaxies prior to fitting (explicitly, those at $\log(\mathrm{sSFR}/\mathrm{yr}^{-1}) < -11.5$, which is significantly offset from the star forming peak at all epochs in both simulations). 

Given that the large majority of a Gaussian distribution is contained within 3$\sigma$ of the mean, few star forming galaxies are expected to lie beyond this threshold. Therefore, we choose a robust quenching threshold of: 

\begin{equation}\label{eq:SSFR_THRESH}
    {\mathrm{sSFR}}_{\mathrm{QT}}\,(z) \equiv \mu_\mathrm{SF}\,(z) - 3\sigma_\mathrm{SF}\,(z),
\end{equation}

\noindent where $\mu_{\mathrm{SF}}\,(z)$ is the mean of the star forming Gaussian distribution, and $\sigma_{\mathrm{SF}}\,(z)$ its standard deviation, at a given redshift $z$. This results in a bespoke threshold at each redshift for each simulation. Explicitly, galaxies are defined as quenched if:

\begin{equation}
\mathrm{sSFR}_i\,(z_i) \leq \mathrm{sSFR}_{\mathrm{QT}}\,(z_i).
\end{equation}

\noindent If the above criterion is not met, we define galaxies to be star forming.

We illustrate the process of determining the quenched threshold in Fig. \ref{fig:QuenchThresh}. On each panel we display the mean sSFR of the star forming population, the standard deviation of the Gaussian fit, and the associated quenching thresholds at z = 0 - 2 (across each row) for TNG (upper panels) and EAGLE (lower panels). In both simulations, the mean sSFR (and hence quenching threshold) increases with redshift, while the star forming Gaussian distributions become progressively tighter in log-space. The location of the quenched threshold is displayed as a red dashed line on each panel, with the location of the mean of the star forming distribution shown as a dashed blue line on each panel. Galaxies with sSFR below the quenched threshold form our quenched population at each epoch, with galaxies above this threshold forming our star forming population at each epoch.

It is important to note that the estimated quenching thresholds are likely to lead to some contamination of the quenched sample with `Green Valley' (GV) galaxies (those residing between the star forming and quenched peaks). However, these galaxies must still exhibit properties of the quiescent sample as they transition from the Main Sequence. Therefore, the `GV contamination' of the quenched sample may lend additional insight to the early stages of quenching, particularly at higher redshift, where there is little-to-no established quiescent population.

From Fig. \ref{fig:QuenchThresh}, it is clear that the fraction of quenched galaxies, indicated within each panel as $f_{\mathrm{Q}}$, decreases as redshift increases, in both simulations. Indeed, at z = 0 we find $f_{\mathrm{Q}} \sim 30\%$, while at z = 2 this is reduced to $<$10\%, for both TNG and EAGLE. As such, a large portion of quenching occurs within this redshift range (z = 2 - 0).

\section{Results}\label{sec:Results}

\subsection{Identifying Quenching Mass Thresholds}{\label{sec:HighLowThresh}}

\begin{figure*}
    \centering
    \begin{subfigure}{\textwidth}
        \caption{}
        \includegraphics[width = \textwidth]{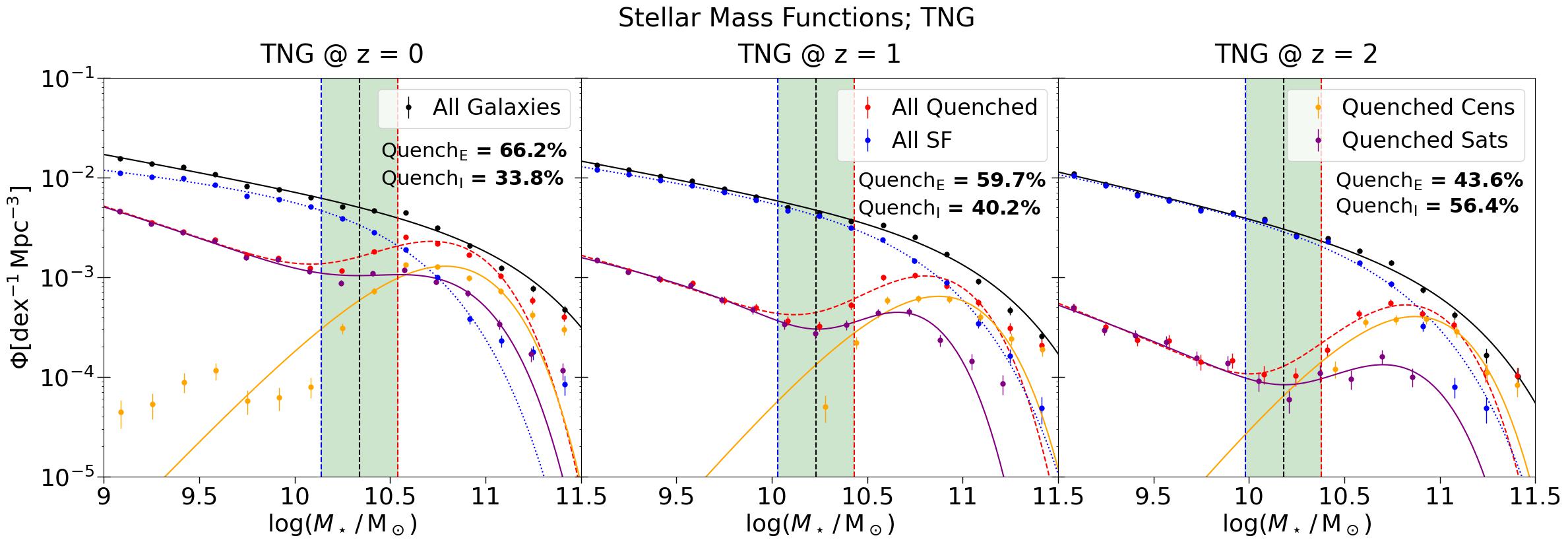} \label{fig:TNG_SF_Q}
    \end{subfigure}
    \hfill
    \begin{subfigure}{\textwidth}
        \caption{}
        \includegraphics[width= \textwidth]{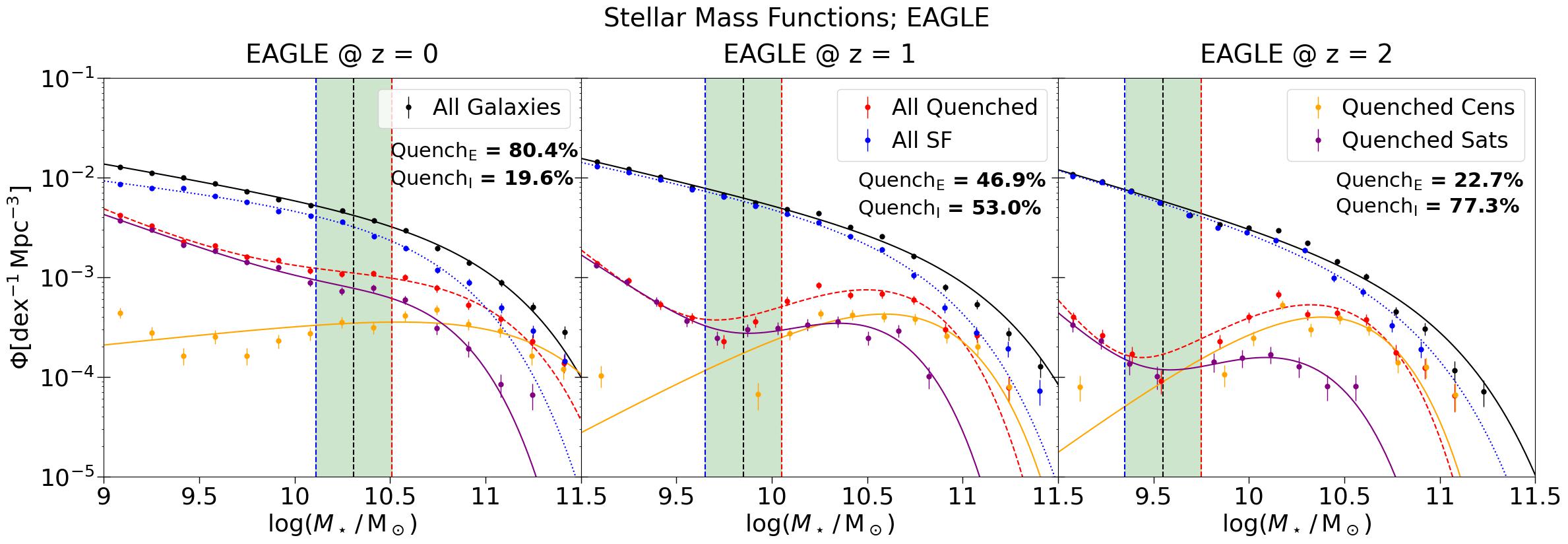} \label{fig:EAGLE_SF_Q}
    \end{subfigure}
    
\vspace*{-3mm}
\caption{The stellar mass functions of various galaxy populations for the TNG (Fig. \ref{fig:TNG_SF_Q}) and EAGLE (Fig. \ref{fig:EAGLE_SF_Q}) simulations. Explicitly, we show stellar mass functions for: (i) all galaxies (in black); (ii) quenched galaxies (in red); (iii) star forming galaxies (in blue); (iv) quenched satellites (in purple); and (v) quenched centrals (in orange). The rows correspond to a given simulation (TNG in the top row and EAGLE in the bottom row), while the columns display redshift snapshots (z = 0, 1, 2, from left to right). In most cases (all redshifts for both simulations except for EAGLE at z = 0), the stellar mass functions of the full quenched and quenched satellite populations are clearly bimodal. We leverage this bimodality, specifically of the quenched satellite population, for each simulation to determine a `high-mass/low-mass' separation in the quenched population of satellites. This is achieved by locating the local minimum of the gradient of its Schechter fit (displayed as a vertical dotted black line on each panel). We introduce a buffer zone of $\pm$0.2\,dex between the low and high-mass thresholds (displayed as a green shaded region in each panel). This enables a separation of the quenching population on the basis of stellar mass. Note that while the quenched satellite population is bimodal, similarly to the full quenched population, the quenched central population is largely uni-modal. This indicates that centrals quench predominantly at high-masses, yet satellites may quench throughout the full stellar mass range probed. The corresponding stellar mass thresholds for each simulation and redshift are provided in Table~\ref{tab:Mass_Thresh}.}
\label{fig:Q_SMF_Schechter}
\end{figure*}

In Fig. \ref{fig:Q_SMF_Schechter}, we present the stellar mass functions for different galaxy populations, for both simulations. Explicitly, we show: all galaxies (in black), quenched galaxies (in red), star forming galaxies (in blue), quenched satellites (in purple), and quenched centrals (in orange). The top and bottom rows correspond to TNG and EAGLE simulations (respectively), while the panels are organized by redshift, from z~=~0~-~2 (along each row). A Schechter~\citep[][]{Schechter_1976}, or double Schechter, functional fit is applied to each stellar mass function, with the following functional forms:\\

\noindent Single:

\begin{equation}
    \phi(M)dM = \mathrm{ln}(10)\phi^{*}\bigg(10^{(M-M^{*})(\alpha + 1)}\bigg) \, \mathrm{exp}\bigg\{-10^{M-M^{*}}\bigg\}dM
\end{equation} \\

\noindent Double:

\begin{equation}
\begin{aligned}
  &\phi(M)dM = \mathrm{ln}(10)\bigg[\phi_{1}^{*}\bigg(10^{(M-M^{*})(\alpha_{1} + 1)}\bigg) + \phi_{2}^{*}\bigg(10^{(M-M^{*})(\alpha_{2} + 1)}\bigg)\bigg] 
  \\
  &\times \mathrm{exp}\bigg\{-10^{M-M^{*}}\bigg\}dM
\end{aligned}
\end{equation}

\noindent where, M $\equiv$ log($M_{*} / {\mathrm{M}_\odot}$), $\phi^{*}$ is the normalization constant, $\alpha$ is the slope of the mass function, and $M^{*}$ is the characteristic mass, associated with the `knee' of the stellar mass function. The values of each parameter for each fit are presented in Appendix~\ref{appendix:B}, in Tables~\ref{tab:Schech_TNG_Param} \& \ref{tab:Schech_EAGLE_Param}, for TNG and EAGLE respectively. Additionally, included within each panel is the percentage of the population quenched via environmental means, $\mathrm{Quench_{E}}$, and via intrinsic means, $\mathrm{Quench_{I}}$, demonstrating their respective importance to quiescence at each redshift. The values of $\mathrm{Quench_{E}}$ and $\mathrm{Quench_{I}}$ are determined by evaluating the integral of the quenched Schechter function for the low and high-mass regimes (respectively), normalized by the total area under the quenched stellar mass function. 

Quenching in the low-mass regime is expected to be dominated by environmental processes~\citep[e.g.,][]{Contini_2020,Mao_2022,Goubert_2024, Hamadouche_2024}, while quenching in the high-mass regime is linked to intrinsic mechanisms~\citep[e.g.,][]{Bluck_2022,Bluck_2023, Bluck_2024,Piotrowska_2022,Goubert_2024}. However, the stellar mass at which this turnover is expected, and its evolution with redshift, remains unclear. 

Considering the stellar mass functions of quenched galaxies for both simulations reveals a bimodality at all redshifts, though this is quite weak for EAGLE at z = 0. Furthermore, while the quenched satellite stellar mass function retains this bimodality, it is largely absent for quenched centrals, whereby only higher mass centrals are able to quench in significant numbers. This strongly implies that there must be a quenching mechanism available only to low-mass satellites, unrelated to intrinsic processes. High-mass satellites, on the other hand, are likely subject to quenching similar to centrals, due to the similarity of their mass functions~\citep[see also][]{Bluck_2019,Bluck_2020b,Goubert_2024,Hamadouche_2024,Lim_2025}. Therefore, there is clear evidence for the presence of separate quenching avenues at low and high masses, which cannot be isolated by a simple central - satellite split.

Therefore, to best segregate these two quenching routes we develop a method to determine where to bisect the low and high mass galaxy regimes. This is done by leveraging the stellar mass functions of the quenched satellite population at each epoch (although similar thresholds can be obtained analyzing the full quenched population stellar mass functions). First, we ascertain the position of lowest gradient along the Schechter fit of the quenched satellite stellar mass function, which identifies the local minimum between the bimodal peaks. We present its location as a dotted black line traversing each panel in Figs. \ref{fig:TNG_SF_Q} \& \ref{fig:EAGLE_SF_Q}. While this offers a possible stellar mass at which to separate the populations, one would expect environmental and intrinsic mechanisms to share a role in quenching galaxies near this limit. To remedy this potential issue, we choose to introduce a buffer zone of $\pm$ 0.2\,dex about this point, indicated by the area shaded in green on each panel, to reduce contamination of environmental quenching in the intrinsic sample, and vice versa. We are then left with bespoke mass thresholds for TNG and EAGLE at all redshifts studied in this work, which are presented in Table~\ref{tab:Mass_Thresh}.  

However, as shown in \citet{Kukstas_2022}, there is evidence of over-quenching of low-mass satellites in simulations when compared to observational data, particularly in clusters. Importantly, this over-quenching could result in a non-physical (seen solely in simulations) upturn of the quenched (total and satellite) stellar mass function at low masses in simulations. It follows then that the mass threshold methodology implemented in this study to separate quenching avenues would not necessarily map onto observational data, as the quenched stellar mass functions may not present with such strong bimodality. 

Nonetheless, the low mass upturn in the quenched stellar mass function is real in simulations, and its presence, or absence, in VLT-MOONRISE will provide insight on the validity (or otherwise) of their quenching prescriptions at cosmic noon. Furthermore, a similar method leveraging the bimodality of the quenched stellar mass function has been used in \citet{Hamadouche_2024} and was shown to be efficient in isolating environmental and intrinsic quenching in observations, up to z~$\sim$~2. It should be noted that in \citet{Hamadouche_2024} galaxies are split at $M_{*} = 10^{10} \mathrm{M_{\odot}}$ at all redshifts. Therefore, whether the upturn of the quiescent stellar mass function at low stellar masses is present in both simulations and observations, or solely in simulations due a low-mass satellite over-quenching, is yet to be fully determined. In fact, as VLT-MOONRISE will be the first wide field spectroscopic survey at intermediate and high redshifts, it will provide the ideal testing data to resolve this ambiguity.

When comparing the evolution of the TNG and EAGLE mass thresholds with redshift, shown in Table~\ref{tab:Mass_Thresh}, a distinct difference is apparent. The TNG thresholds evolve very little from z = 0 - 2 (decreasing by only 0.1\,dex in total). On the other hand, the mass thresholds determined for the EAGLE simulation decrease significantly with redshift, first by 0.45\,dex from z = 0 - 1, then by a further 0.3\,dex from z = 1 - 2. This may indicate a weakening impact on quenching carried by environment, as well as the presence of an intrinsic quenching mechanism more effective at higher redshift (see later for further discussion). Crucially, these predicted differences in the stellar mass functions of quiescent objects can be tested directly in VLT-MOONRISE. This will offer a powerful test on which, if either, of the simulations is more effective at describing quenching in nature.

Furthermore, the sharp decrease in the low-mass threshold in EAGLE directly impacts the `MOONRISE-Like' data sets for the EAGLE simulation. In fact, the low-mass thresholds at z = 1 \& 2 (i.e., $\log(M_{*}/{\mathrm{M_{\odot}}}) \leq 9.65$ and $\log(M_{*}/{\mathrm{M_{\odot}}}) \leq 9.35$, respectively) lie near, or even below, the MOONRISE survey mass completeness limit. Therefore, the `MOONRISE-Like' sample sets at z $\geq$ 1 for EAGLE are essentially devoid of `low-mass' satellites, i.e., satellites which are expected to quench solely via environmental means. Conversely, since the mass thresholds for TNG shift only slightly with redshift, they do not have the same impact on the `MOONRISE-Like' data sets. This leads to important testable differences between the two simulations: TNG predicts that MOONRISE will observe low-mass satellite quenching via environmental processes, but EAGLE does not.

\begin{table}
    \centering
        \caption{The high and low-mass thresholds determined via the stellar mass functions of the quenched satellite population for TNG and EAGLE.}
    \begin{tblr}{
             colspec = { c c },  
             row{1}  = {font=\bfseries, c},
             column{1}  = {font=\bfseries, c},
             rowsep=0.5ex, colsep =1ex
             }
      \hline
      \textbf{Simulation} & \textbf{bisecting-mass; log($M_{*}/{\mathrm{M_{\odot}}}$)}\\
      \hline
       TNG; z = 0 & 10.30 $\pm$ 0.2\\
     
       TNG; z = 1 & 10.25 $\pm$ 0.2\\
      
       TNG; z = 2 & 10.15 $\pm$ 0.2\\
  
       EAGLE; z = 0 & 10.30 $\pm$ 0.2\\
    
       EAGLE; z = 1 & 9.85 $\pm$ 0.2\\
   
       EAGLE; z = 2 & 9.55 $\pm$ 0.2\\
       \hline
    \end{tblr}
    \label{tab:Mass_Thresh}
\end{table}

\subsection{Evolution of Environment}

\begin{figure*}
    \centering
    \begin{subfigure}{\textwidth}
        \caption{TNG - 3D $\delta_{10}$ Distribution}
        \includegraphics[width = \textwidth]{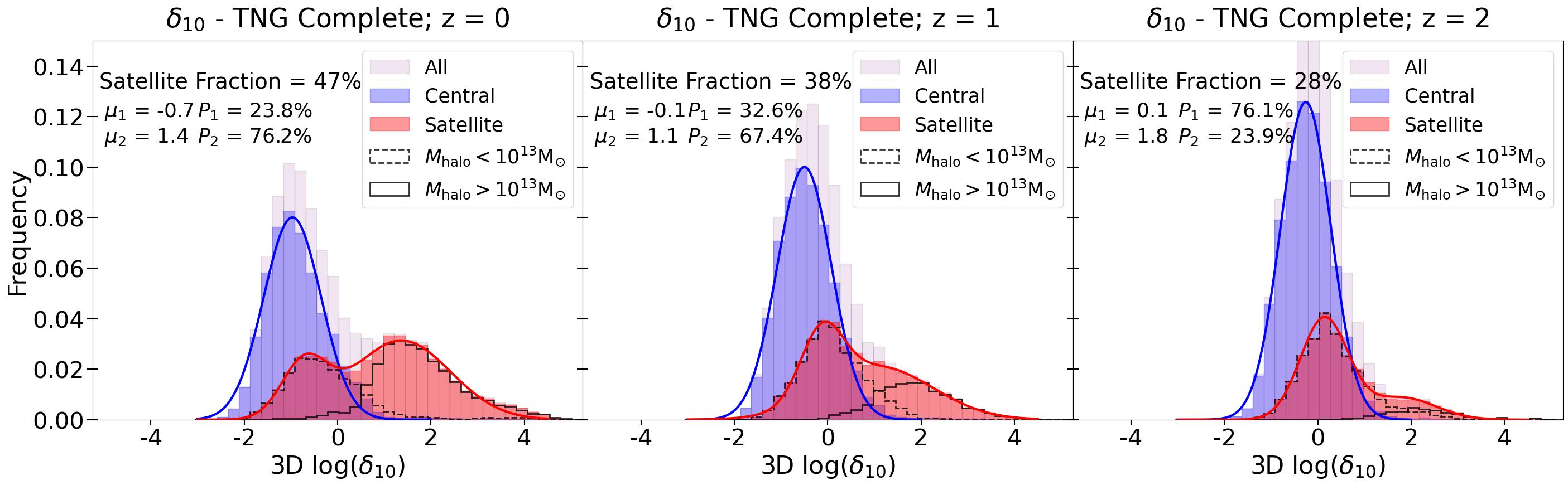}  
    \end{subfigure}
    \hfill
    \begin{subfigure}{\textwidth}
        \caption{TNG - MOONRISE-Like $\delta_{10}$ Distribution}
        \includegraphics[width= \textwidth]{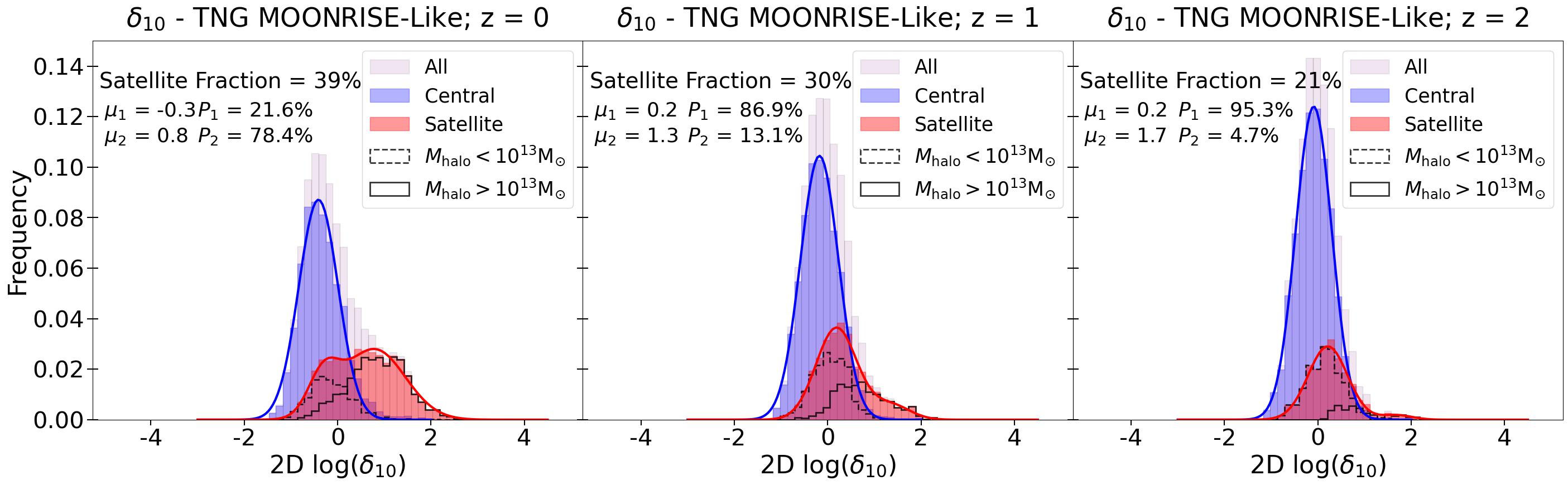}\label{fig:TNG_Den_Moons} 
        
    \end{subfigure}
    
\vspace*{-3mm}
\caption{The distributions of $\log(\delta_{10})$ for the entire sample (shown in light grey), centrals (shown in blue), and satellites (shown in red) for the complete TNG data set (top panels) and the `MOONRISE-Like' data set (bottom panels). We fit a single Gaussian to the uni-modal central distribution, and a double Gaussian to the bimodal satellite distribution. The mean of each Gaussian ($\mu_{1}$ \&  $\mu_{2}$) of the satellites distributions are displayed on each panel. Additionally, we present the fraction of satellites belonging to both the low and high density regimes of the bimodal distribution ($P_{1}$ \& $P_{2}$, respectively). The distribution in $\delta_{10}$ of satellites in high and low-mass groups ($M_{\mathrm{Halo}} \geq 10^{13} {\mathrm{M_{\odot}}}$ and $M_{\mathrm{Halo}} \leq 10^{13} {\mathrm{M_{\odot}}}$) are included as solid and dashed open histograms respectively. There is a rising fraction of satellites found in the high density regime, and corresponding decreasing fraction found in the low density peak, as a function of cosmic time. This, in combination with a growing satellite fraction (indicated within each panel), indicates a clear growth in environment as large groups and galaxy clusters form. Furthermore, it is clear that splitting satellites into high and low-mass groups is effective in recovering the bimodality in density present for satellites. As such, local over-density is a strong tracer of halo mass, or at least efficient in segregating satellites into high and low-mass groups. Whilst the results are qualitatively similar in the `MOONRISE-Like' data sets, the range of $\delta_{10}$ is much more curtailed and shows much less prominent bimodality, than for the complete data set. The group mass split is also much less effective at reproducing the density bimodality. This highlights a significant loss of information between 6D and 2D+z density measurements.} \label{fig:TNG_Den}
\end{figure*}

\begin{figure*}
    \centering
    \begin{subfigure}{\textwidth}
        \caption{EAGLE - 3D $\delta_{10}$ Distribution}
        \includegraphics[width = \textwidth]{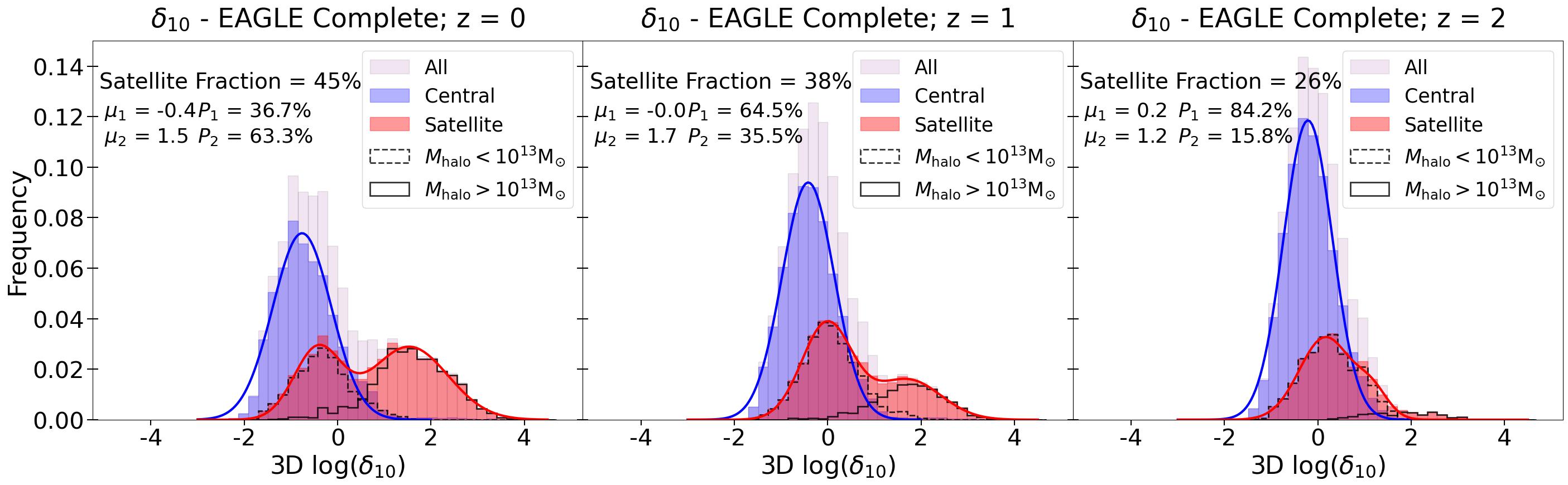}  
    \end{subfigure}
    \hfill
    \begin{subfigure}{\textwidth}
        \caption{EAGLE - MOONRISE-Like $\delta_{10}$ Distribution}
        \includegraphics[width= \textwidth]{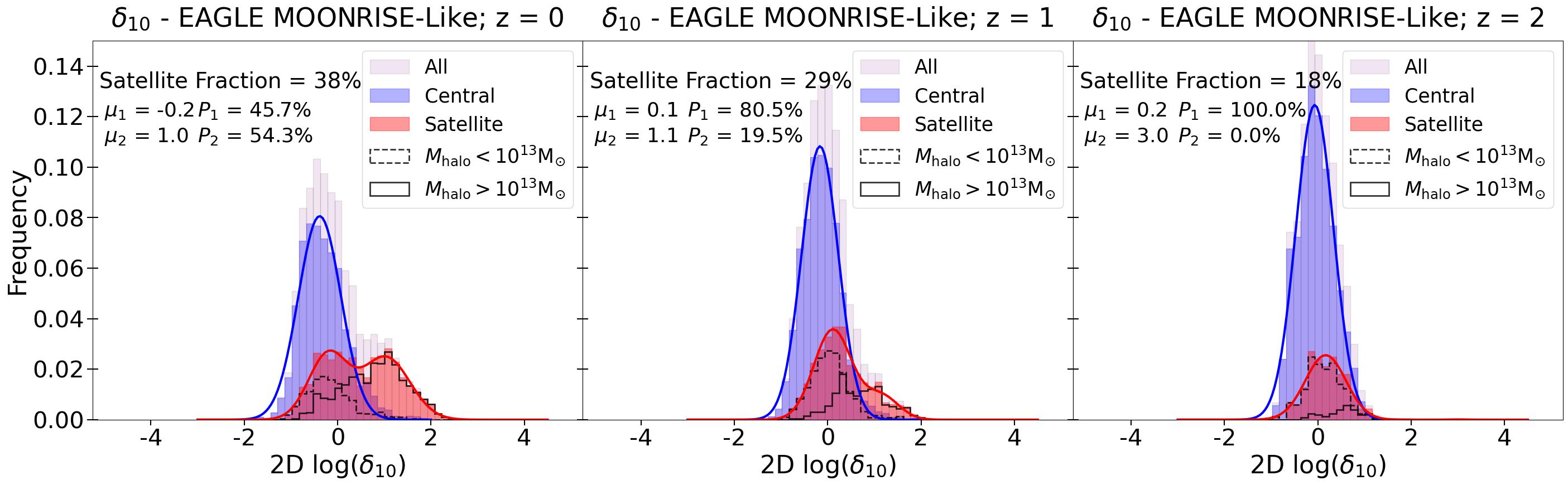}\label{fig:EAGLE_Den_Moons}  
        
    \end{subfigure}
    
\vspace*{-3mm}
\caption{Identical in structure and method to Fig. \ref{fig:TNG_Den}, but here presenting the evolution of the distribution in galaxy over-density for the EAGLE simulation. The high density regime satellite population, and total satellite fraction, increase with cosmic time, indicating the growth of environment (as seen before for TNG). Furthermore, the distribution becomes increasingly bimodal, becoming a better tracer to identify differences in the environment. This is further highlighted by the distributions of satellites in high and low-mass groups, which become increasingly separate as redshift decreases. As seen for TNG in Fig. \ref{fig:TNG_Den}, the distributions of $\delta_{10}$ for the `MOONRISE-Like' data sets are less bimodal and more curtailed in their density range than for the complete data set. This again indicates information loss going from 6D phase space to an observation like 2D+z phase space. Nonetheless, it is still possible to see a growth in the high density population as redshift decreases.} \label{fig:EAGLE_Den}
\end{figure*}

To explore the evolution of environment with cosmic time, we present the distributions of the tenth nearest neighbor over-density ($\log(\delta_{10})$) at z = 0, 1, 2, for TNG (Fig. \ref{fig:TNG_Den}) and EAGLE (Fig. \ref{fig:EAGLE_Den}). We choose to show $\delta_{10}$, as it presents with the strongest bimodality of all nearest neighbor densities, although similar trends are observed for all, as can be seen in Appendix~\ref{appendix:A}. We show these results separately for the complete and `MOONRISE-Like' data sets (top and bottom rows in each figure). This serves to highlight the differences in the distributions expected to occur as data moves from a simulated 6D phase space to an observational-like 2D+z space. Each panel (z = 0 - 2, from left to right) contains the distributions for the total sample (shown in faint gray), centrals (shown in blue), and satellites (shown in red). The distributions of satellites in high and low-mass groups ($M_{\mathrm{Halo}} \geq 10^{13}\, {\mathrm{M_{\odot}}}$ and $M_{\mathrm{Halo}} \leq 10^{13}\, {\mathrm{M_{\odot}}}$, respectively) are included as solid and dashed open histograms, respectively.

The central galaxy population is clearly uni-model and Gaussian-like in both simulations and at all epochs probed. Conversely, the satellite population is bimodal, most pronouncedly at z = 0. As such, we fit a single Gaussian to the central population within each panel, and a double Gaussian to the satellite populations. This enables us to determine the characteristics of the low and high density populations in satellites, and their evolution with redshift. The mean of each Gaussian of the bimodal fits are included within each panel. To quantify the change in the environment of satellites with redshift, we display the total fraction of satellites, as well as the fraction of satellites which belong to each Gaussian of the bimodal fit, expressed as $P_1$ (under-dense region) and $P_2$ (over-dense region) in the panels.

As redshift increases, the population fraction of the over-dense peak decreases, and the distributions begin to concentrate toward the under-dense peak, independent of simulation. At z = 2, there is a nearly uni-modal distribution, with very few satellites left in high density regions. Furthermore, there is a clear anti-correlation between the total satellite fraction and redshift, which grows from 28\% to 47\% from z = 2 to z = 0 in TNG, and grows from 26\% to 45\% over the same time period for EAGLE. 

The evolution of the over-dense region of the bimodal distribution with redshift, in combination with a growing satellite fraction, points towards an increase in the abundance of high-mass groups and clusters, as groups merge, over cosmic time. This can also be seen by the evolution in the distribution of satellites in low and high-mass groups. There is a clear growth in the fraction of satellites in high-mass groups as redshift decreases. Additionally, the low and high-mass group split performs well in recovering the bimodality in $\delta_{10}$ present for all satellites. This is true at all redshifts, but is increasingly apparent as redshift decreases. This points towards density as an effective tracer of halo mass. Furthermore, this hierarchical assembly will convert past centrals into current satellite galaxies. As such, it is interesting to probe the effect of an ever changing environment on the quenching of satellites across redshift.  

The general trends in the evolution of local galaxy density are stable to the `MOONRISE-Like' sample selection (lower panel sets in Figs. \ref{fig:TNG_Den} \& \ref{fig:EAGLE_Den}). However, there is an evident reduction the density range, and, to a lesser degree, in the bimodality of the distributions. It can also be seen that splitting satellites by halo mass is less effective at separating the low and high density satellite populations when compared to the complete data set. This indicates a significant loss of information when moving from a 6D phase space to 2D+z observer space. As such, we will evaluate all quenching trends for both the base simulation (revealing the inner workings of the model) and for the `MOONRISE-Like' sample (revealing what ought to be seen in MOONRISE, if the simulation is an accurate representation of nature).

\subsection{Random Forest Classification}\label{sec:RF}

\begin{figure*}
    \includegraphics[width=\textwidth]{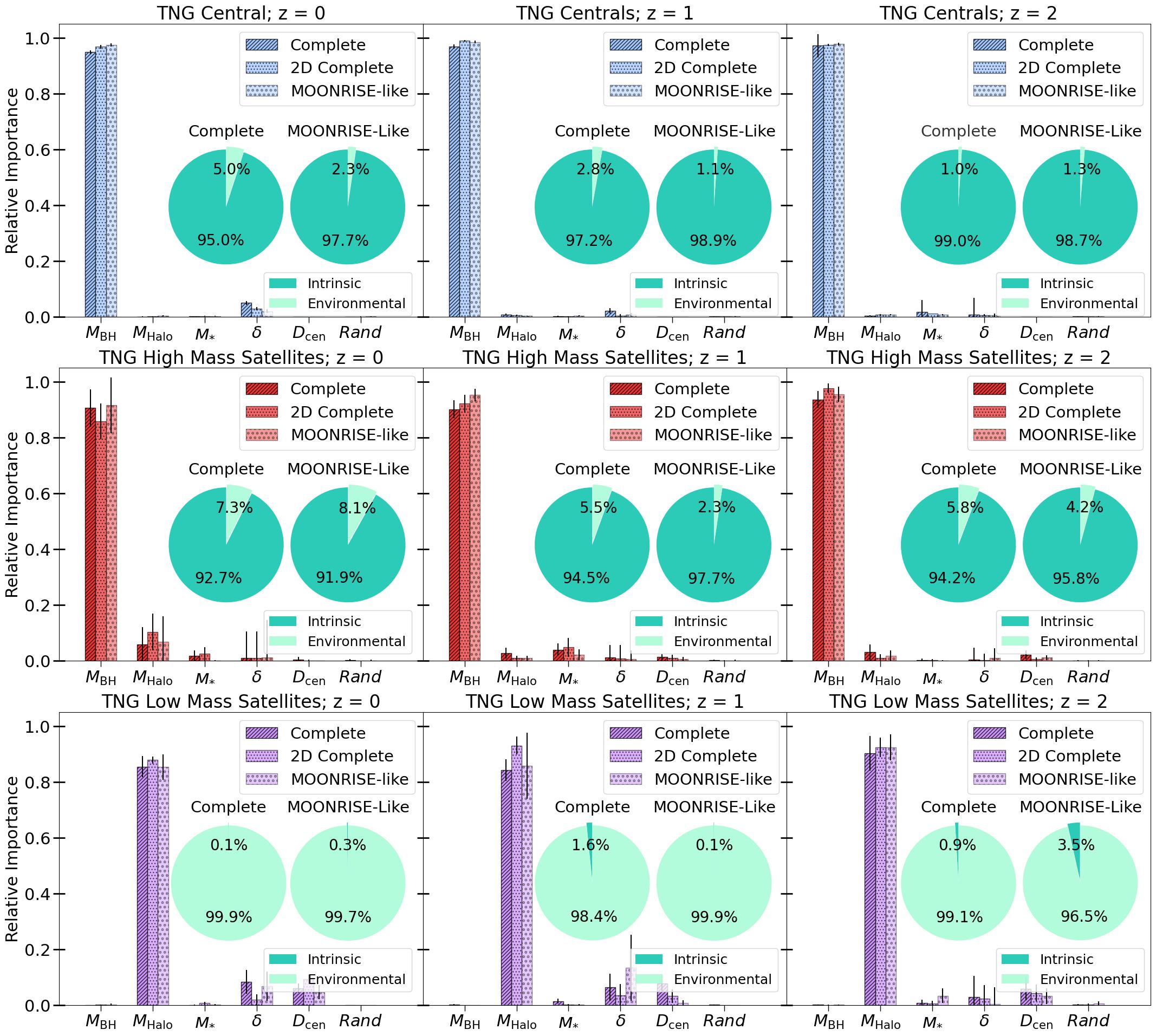}
    \vspace*{-3mm}
    \caption{Random Forest quenching classification analyses for TNG. Each row corresponds to a galaxy class (from top to bottom: centrals, high-mass satellites, and low-mass satellites), and each column to a given redshift (z = 0 - 2, from left to right). The input features tested are listed along the $x$-axis, with their relative quenching importances indicated by the $y$-axis bar heights. Uncertainties are given by the variance of 100 independent classification runs. Separate bars are shown for the complete, the 2D complete, and `MOONRISE-Like' datasets, indicated by different hatches on the bars. A pair of pie charts are displayed on each panel, indicating the total importance to quenching of intrinsic and environmental parameters for the complete (left) and `MOONRISE-Like' (right) data sets. Across all redshifts and data samples, we find that centrals and high-mass satellites quench predominantly by intrinsic processes, with $M_{\mathrm{BH}}$ being by far the best predictor of quiescence. Conversely, low-mass satellites are determined to quench through environmental processes, with $M_{\mathrm{Halo}}$ found to be the dominant parameter. This indicates that in the TNG simulation both centrals and high-mass satellites quench predominantly via AGN feedback, yet low-mass satellites quench via environmental mechanisms at all epochs from z = 0 - 2. This provides a clear, testable prediction for the VLT-MOONRISE survey.}\label{fig:TNG_RF}
\end{figure*}

\begin{figure*}        
    \includegraphics[width=\textwidth]{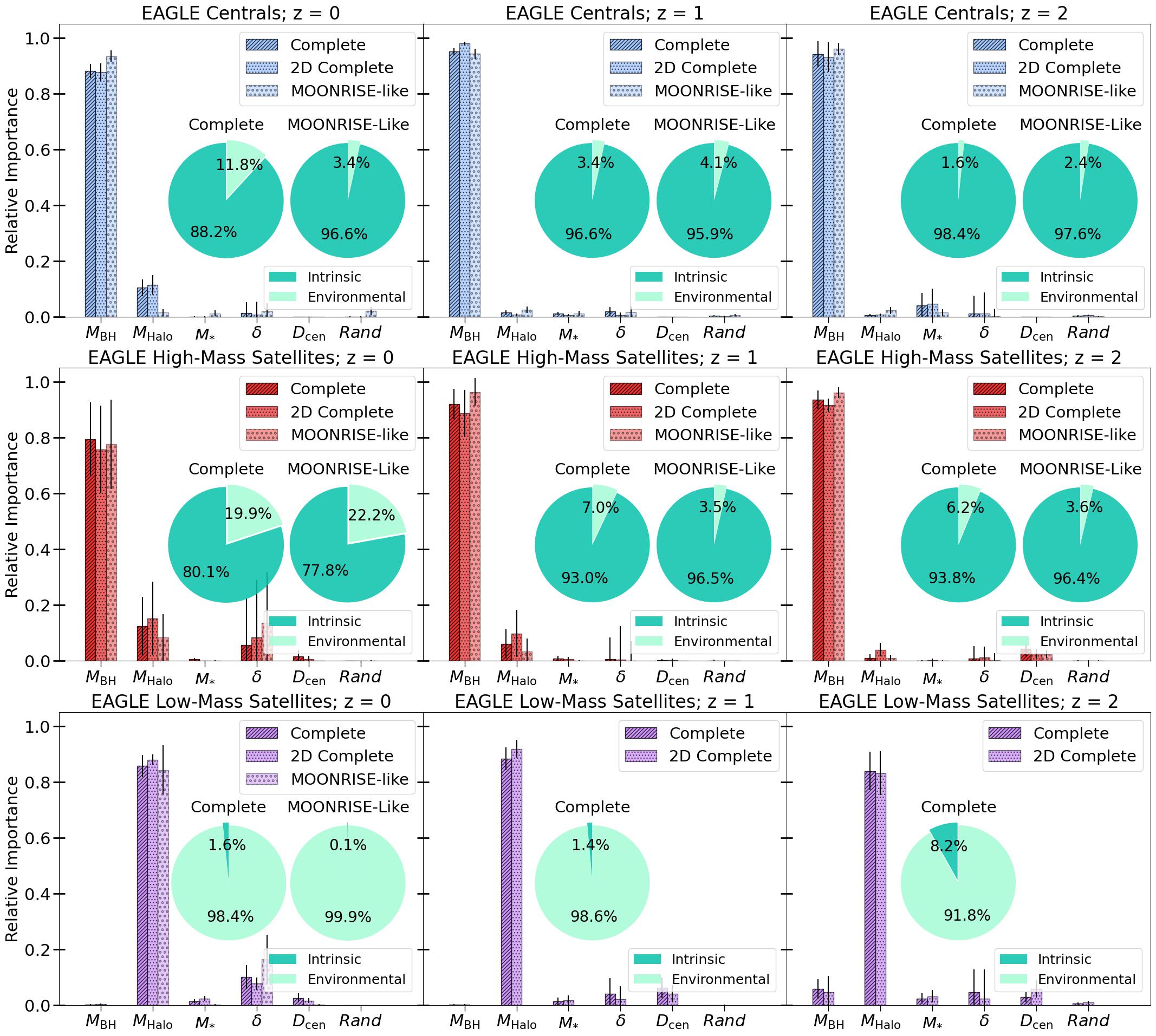}
    \vspace*{-3mm}
    \caption{Random Forest quenching classification analyses for EAGLE. This figure is structured exactly like Fig. \ref{fig:TNG_RF}. As seen for TNG, at all redshifts and independent of data sample, the quiescence of centrals and high-mass satellites is best predicted by intrinsic parameters, with $M_{\mathrm{BH}}$ being the dominant feature. For the complete and 2D complete data sets, low-mass satellites are determined to quench through environmental processes at all redshifts, with $M_{\mathrm{Halo}}$ found to be the dominant parameter. However, as the mass completeness threshold of the `MOONRISE-Like' data set is near, or greater than, the low-mass thresholds at z = 1 and z = 2, virtually no low-mass satellites are present in these samples. This implies that environmental quenching would not be detectable in MOONRISE, if the EAGLE model is correct. The results for the low-mass satellites of the `MOONRISE-Like' data set at z = 0 are in agreement with the complete and 2D complete results, with environment dominating. Taken together, this figure provides numerous clear predictions for quenching across cosmic time.}\label{fig:EAGLE_RF}
\end{figure*} 

\par We present the results from a series of Random Forest quenching classification analyses of the TNG and EAGLE simulations in Figs. \ref{fig:TNG_RF} \& \ref{fig:EAGLE_RF}, respectively. Each bar represents the mean of the relative importances of a given parameter from a series of 100 independent Random Forest classification runs. The errors are given by the standard deviation across the 100 runs. Every row corresponds to a galaxy class (from top to bottom: centrals, high-mass satellites, low-mass satellites), and each column to a redshift (z = 0, 1, 2; from left to right). Within each panel, results are shown separately for three data sets: `Complete', `2D Complete', and `MOONRISE-Like', as indicated by different hatches on the corresponding bars (see legends). A pair of pie charts are displayed on each panel, indicating the total importance to quenching of intrinsic and environmental parameters for the complete (left) and `MOONRISE-Like' (right) data sets. We present the hyperparameters used for each RF classification test presented in this section in Appendix \ref{appendix:B}.

The input features tested are: $M_{\mathrm{BH}}$, $M_{\mathrm{*}}$, $M_{\mathrm{Halo}}$, $\delta$, $D_{\mathrm{cen}}$, where $\delta$ is the sum of the relative importances of the third, fifth, and tenth nearest neighbor densities. Additionally, a number drawn from a random distribution between 0 and 1 is added to gauge the impact of random uncertainty in this method. The panels are accompanied by a pair of pie charts, showing the total relative importance of intrinsic ($M_{\mathrm{BH}}$ and $M_{\mathrm{*}}$) and environmental ($M_{\mathrm{Halo}}$, $\delta$, $D_{\mathrm{cen}}$) parameters for the complete (left pie plot) and `MOONRISE-Like' (right pie plot) data sets. This is to facilitate the comparison in importance to quenching between intrinsic and environmental processes as a function of galaxy type, as well as the evolution with redshift. 

\par Focusing first on central galaxies, we find that $M_{\mathrm{BH}}$ dominates quenching at all redshifts, independent of the sample set, for both TNG and EAGLE. This implies an intrinsic quenching mechanism coupled to black hole growth, explicitly AGN feedback driven quenching~\citep[as seen in][]{Piotrowska_2022, Bluck_2023, Bluck_2024}. Furthermore, the pie plots clearly identify quenching as being intrinsic in origin, with environment attributed $\lesssim 15$\% of importance across redshifts.

The second row of plots presents the classification results for high-mass satellites. As with the central galaxies, for both simulations we find black hole mass to be the best predictor of quiescence at all redshifts. This points to an AGN feedback driven quenching process for high-mass satellites, in both simulations. However, a secondary dependence on environment (specifically halo mass) is clearly present. The `quenching importance' pie plots confirm and highlight this intrinsic quenching dominance in both the complete and `MOONRISE-Like' data sets. Environmental parameters make up less than 10\% and 20\% of the pie at all redshifts for the complete data sets of TNG and EAGLE, respectively. This is particularly interesting because historically many authors have considered all satellites to be environmentally quenched (see, e.g., \citealt{Peng_2012}).

\begin{figure*}

    \includegraphics[width=\textwidth]{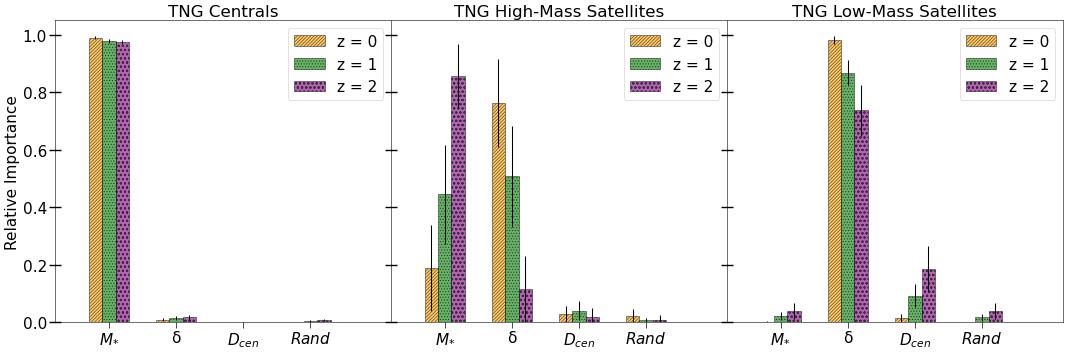}
    \vspace*{-3mm}
    \caption{Random Forest quenching classification analyses for the TNG `MOONRISE-Like' data set, absent $M_{\mathrm{BH}}$ and $M_{\mathrm{Halo}}$. Each panel corresponds to a galaxy class (centrals, high-mass satellites, low-mass satellites), with the results at each redshift presented with different colored bars on each panel. The parameters used for training the Random Forest are listed along the $x$-axis, with their relative quenching importances indicated by the $y$-axis bar heights. Uncertainties are given by the variance of 100 independent classification runs. We find the quiescence of centrals to be unanimously predicted by stellar mass at all redshifts. In contrast, the quenching of high-mass satellites is best predicted by $\delta$ at z = 0, though a secondary dependence on $M_{*}$ is present. As redshift increases, so does the relative importance of $M_{*}$. At z = 1, stellar mass and $\delta$ share similar importance (50-40\%), whereas $M_*$ dominates at z = 2. Finally, we find that $\delta$ dominates at all redshifts for low-mass satellites, although a small contribution from $D_{\mathrm{cen}}$ is present at z = 1 \& 2.}
    \label{fig:TNG_RF_Moons}
\end{figure*}

\begin{figure*}

    \includegraphics[width=\textwidth]{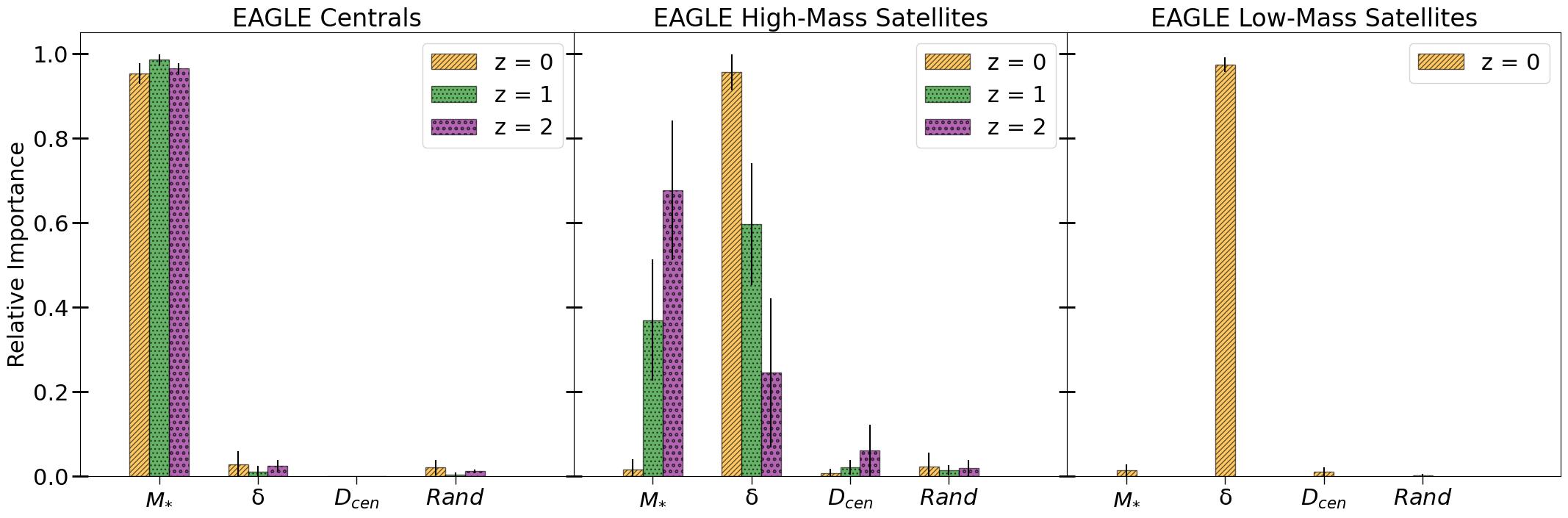}
    \vspace*{-3mm}
    \caption{Random Forest quenching classification analyses for the EAGLE `MOONRISE-Like' data set, absent $M_{\mathrm{BH}}$ and $M_{\mathrm{Halo}}$. This figure is structured identically to Fig. \ref{fig:TNG_RF_Moons}. We find that central galaxies are predicted to quench intrinsically, with $M_{*}$ dominating at all redshifts, as for TNG. High-mass satellites present with $\delta$ dominating at z = 0, and it remains the best predictor at z = 1. At z = 2 stellar mass is the best predictor of quiescence. Finally, at z = 0, the quenching of low-mass satellites is best predicted by $\delta$. The results for low-mass satellites are absent at z = 1 and z = 2, due to the mass completeness limit of the `MOONRISE-Like' data sets and the relatively low mass threshold for environmental quenching in EAGLE.}
    \label{fig:EAGLE_RF_Moons}
\end{figure*}

Overall, the results of the Random Forest classification analyses for the TNG and EAGLE simulations are in strong agreement for centrals and high-mass satellites. We find their quiescence to be best predicted by black hole mass, which is known to be intricately linked to AGN feedback in these simulations. This is consistent across redshifts and data sets. However, it seems the AGN feedback mechanism may be more effective at higher redshifts, as the relative importance in $M_{\mathrm{BH}}$ increases with redshift. This is most apparent for the EAGLE simulation, particularly regarding high-mass satellites, which lacks the preventative AGN feedback mode present in TNG. We further explore this later in the paper, specifically in Sec. \ref{sec:QuenchedFrac}. 

The results of the RF classification analyses for low-mass satellites are shown in the bottom rows of Figs. \ref{fig:TNG_RF} \& \ref{fig:EAGLE_RF}. In both TNG and EAGLE, it is clear that this population is quenched via a purely environmental mechanism at all redshifts, and for all data sets. Specifically, we find $M_{\mathrm{Halo}}$ to be the dominant parameter, with weak secondary dependencies on $\delta$ and $D_{\mathrm{cen}}$. These lesser importances indicate preferential quenching in the highest density environments, denoted by the importance of $\delta$, and towards the center of haloes, suggested by the importance of $D_{\mathrm{cen}}$ (see Sec. \ref{sec:QF_Loc}).  

However, a glaring difference between the simulations appears when focusing on the `MOONRISE-Like', low-mass satellite populations. While TNG shows consistent results across redshifts, the EAGLE simulation suffers from a dearth of low-mass satellites at z = 1 and z = 2, due to increasingly low mass thresholds, as seen in Table~\ref{tab:Mass_Thresh}. Therefore, as opposed to TNG, EAGLE predicts that environmental quenching will be undetectable in VLT-MOONRISE. This is a fundamental difference in the predictions formed from the simulations, and will prove useful in determining which model is more accurate.

\subsubsection{MOONRISE-Like with limited observational parameters}

Stellar mass and local over-density, rather than black hole mass and halo mass, are often used as parameters to investigate intrinsic and environmental quenching when analyzing observational data~\citep{Peng_2010, Peng_2012}. Therefore, to obtain results comparable to other studies, we conduct the Random Forest classification tests for the `MOONRISE-Like' data sets, absent both black hole mass and halo mass, so as to highlight the performances of stellar mass and local over-density. The input features tested for here are therefore: $M_{*}$, $\delta$, $D_{\mathrm{cen}}$, and a randomly generated number between 0 and 1. We present the hyperparameters used for each RF classification test in this section in Appendix \ref{appendix:B}.

The results for these classification tests are shown in Figs. \ref{fig:TNG_RF_Moons} \& \ref{fig:EAGLE_RF_Moons}, for TNG and EAGLE (respectively). Importantly, as VLT-MOONRISE will be a wide field spectroscopic survey it will be possible to conduct accurate central - satellite classification. As such, the panels are organized by galaxy class, i.e., centrals, high-mass satellites, and low-mass satellites (from left to right). Within each panel, we present the results at z = 0, 1, and 2, indicated by the different colored bars (see legends).

The results of the Random Forest classification tests for centrals and low-mass satellites of the TNG `MOONRISE-Like' and EAGLE `MOONRISE-Like' data sets, absent $M_{\mathrm{BH}}$ and $M_{\mathrm{Halo}}$, are broadly consistent with the results presented in Figs. \ref{fig:TNG_RF} \& \ref{fig:EAGLE_RF}. The quiescence of central galaxies is best predicted by intrinsic parameters, here stellar mass. However, the underlying mechanism of AGN feedback could be missed if the black hole mass is absent as a parameter. This result establishes that mass quenching~\citep[e.g., as in][]{Peng_2010,Peng_2012} is ultimately a result of black hole mass quenching (at least in these simulations), and hence ultimately AGN feedback~\citep[see also][for further evidence of this in both simulations and observation]{Terrazas_2016,Terrazas_2017,Piotrowska_2022,Bluck_2016,Bluck_2020a,Bluck_2020b,Bluck_2022,Bluck_2023,Bluck_2024}. On the other hand, quenching in low-mass satellites is determined by the local over-density parameter, indicating environmental quenching, at all redshifts (absent EAGLE at z = 1 \& 2, as nearly no quenched low-mass satellites are present). However, crucially, halo mass is predicted to be the superior environmental parameter regulating quenching in both simulations (and at all epochs), when available.

Comparing the results of the classification tests for high-mass satellites, absent black hole and halo mass (Figs. \ref{fig:TNG_RF_Moons} \& \ref{fig:EAGLE_RF_Moons}) to those seen in Figs. \ref{fig:TNG_RF} \& \ref{fig:EAGLE_RF}, a notable difference in the performance of intrinsic parameters is apparent. Absent $M_{\mathrm{BH}}$, $\delta$ is found to best predict quiescence at z = 0 for both simulations. For TNG, the dependence on local over-density wains in strength as redshift increases, as $M_{*}$ carries a similar importance as $\delta$ at z = 1, and dominates at z = 2. However, for EAGLE, $\delta$ remains the dominant parameter at z = 1. At z = 2, $M_{*}$ becomes the best predictor of quiescence, although a strong secondary dependence on environmental parameters remains. We find these results to be consistent with \citet{Peng_2010, Peng_2012}, wherein the quenching of satellites in the local Universe is determined to operate primarily environmentally (regardless of mass).

However, ultimately the quiescence of high-mass satellites in the local Universe is directly linked to their supermassive black hole (as demonstrated in Section \ref{sec:RF}). This disagreement can be resolved by linking a galaxy's local density to it's black hole mass. According to the morphology-density relation~\citep[e.g.,][]{Postman_1984, Goto_2003, Teklu_2017}, galaxies become progressively more spheroidal in higher density environments. Additionally, supermassive black holes tend to be more massive in spheroidal galaxies than disk galaxies, at a fixed stellar mass~\citep[see, e.g.,][]{Saglia_2016,Piotrowska_2022}, establishing a link between a galaxy's local over-density and the mass of its black hole.

Therefore, in the case of high-mass satellite quenching, $\delta$ is likely correlated with $M_{\mathrm{BH}}$ (at a fixed stellar mass). Hence, when black hole mass is absent, and the range of stellar masses is reduced (as it is for this sub-sample), the local density may trace black hole mass, or at least its link to quiescence, better than any other available parameter, especially for z < 2. Consequently, it is crucial to obtain as accurate as possible spectroscopic measurements to best estimate values of black hole mass in observational surveys (e.g., VLT-MOONRISE). Failure to do so may result in erroneously attributing intrinsic satellite quenching to environment.

\subsection{Quenched Fraction Evolution}\label{sec:QuenchedFrac}

\begin{figure*}
    \centering
        \includegraphics[width= \textwidth]{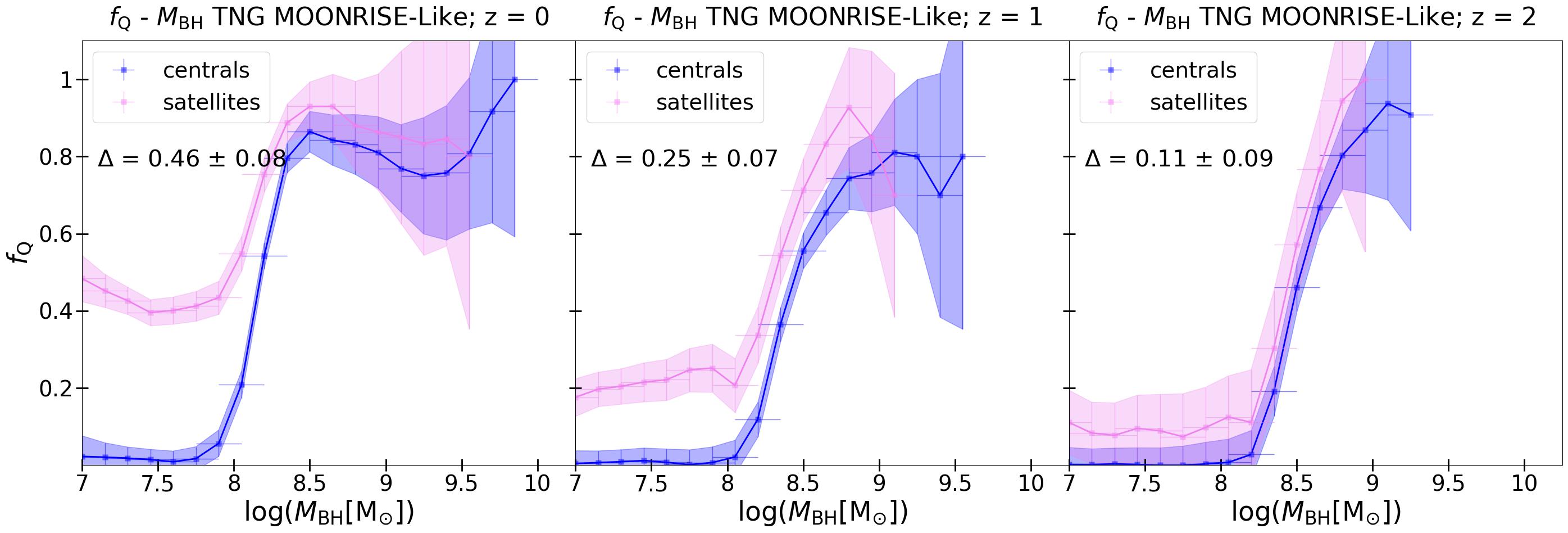} 
\vspace*{-3mm}
\caption{The quenched fraction of TNG central (shown in blue) and satellite (shown in violet) galaxies as a function of black hole mass for the `MOONRISE-Like' data set. The results are shown for different redshifts from left to right along each row (as labeled by the panel titles). At z = 0, there is a clear separation of the two populations in the low-mass regime, wherein satellites are more likely to be quenched than centrals. This is a strong indicator of environment quenching, operating at fixed black hole mass. The magnitude of the offset, indicated by $\Delta$, decreases strongly with redshift. Hence, we conclude that environmental quenching is more effective at late cosmic times. On the other hand, at high masses, and all redshifts, the satellite quenched fraction resembles the central quenched fraction. As such, one would expect quenching at high mass to be operated by the same mechanism (AGN feedback), regardless of central - satellite classification.}\label{fig:TNG_FQ_Z}
\end{figure*}

\begin{figure*}
    \centering
        \includegraphics[width= \textwidth]{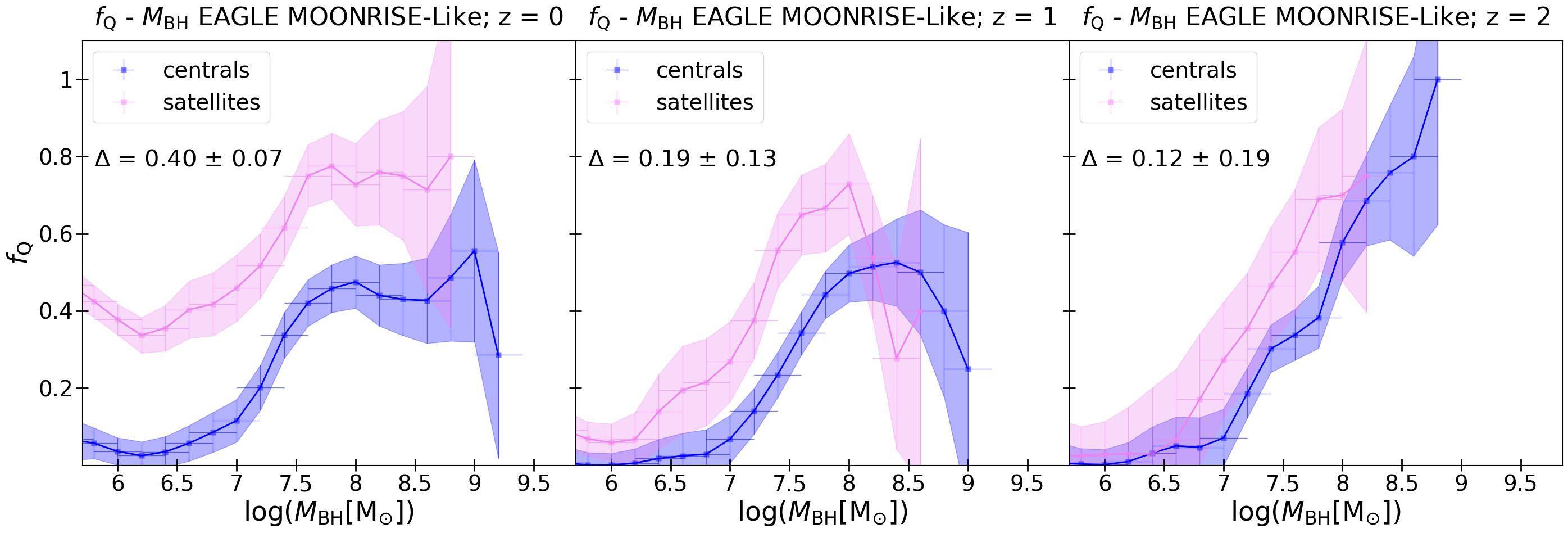} 

\vspace*{-3mm}
\caption{This figure is identical in structure to Fig. \ref{fig:TNG_FQ_Z}, but here showing results from the EAGLE simulation. In comparison to TNG, most trends are similar, with one important exception. The quenched fraction of central galaxies plateaus at a much lower value at z = 0 \& 1. This can also be seen for satellite galaxies at z = 1, though it is less apparent at z = 0. This is indicative of the EAGLE quenching mechanism failing to keep galaxies quenched over long cosmic times, leading to star formation rejuvenation. There remains a noticeable gap between the quenched fractions of satellites and centrals at low black hole mass, especially at z = 0. This gap points towards an environmentally dominated quenching process for low-mass satellites, which decreases as redshift increases. Additionally, while the satellite and central quenched fractions are similar in trends at high masses, the satellite quenched fraction is elevated in comparison to the central quenched fraction. Therefore, while this does imply a similar causal origin to the quenching of high mass satellites to that of centrals, it is clear other factors are aiding in the quenching of satellites, even at higher masses.}\label{fig:EAGLE_FQ_Z}
\end{figure*}

In this sub-section we explore the evolution with redshift of the quenched fraction of central and satellite galaxies, for the `MOONRISE-Like' data sets of both TNG and EAGLE. In Figs.\ref{fig:TNG_FQ_Z} (TNG)  \&  \ref{fig:EAGLE_FQ_Z} (EAGLE) we present the quenched fractions of centrals and satellites as a function of $M_{\mathrm{BH}}$. Black hole mass represents the most predictive quenching parameter for central galaxies, enabling one to assess how other quenching routes impact galaxies once this channel is accounted for. The quenched fraction of central galaxies is shown in blue, while the quenched fraction for satellite galaxies is presented in violet. We compute the uncertainty as the Poisson statistics uncertainty in each bin.

We first focus on the results for central galaxies from the TNG and EAGLE simulations. Across all redshifts, the quenched fraction as a function of $M_{\mathrm{BH}}$ sits at nearly zero for low black hole mass for both simulations ($\log(M_{\mathrm{BH}}/{\mathrm{M}_\odot}) \leq 8$ for TNG, and $\log(M_{\mathrm{BH}}/{\mathrm{M}_\odot}) \leq 7$ for EAGLE). Concentrating on TNG centrals, the quenched fraction behaves similarly at all redshifts, rapidly increasing once beyond the low mass regime, then plateauing around $f_{\mathrm{Q}} = 0.9$. On the other hand, the quenched fraction of centrals from the EAGLE simulation differs greatly from redshift to redshift. At z = 2, $f_{\mathrm{Q}}$ steadily rises, resulting in a near fully quenched population at the highest black hole masses. At z = 1, however, the quenched fraction first rises then quickly plateaus, reaching a maximal value of $f_{\mathrm{Q}} \simeq 0.6$. There is even a downward trend once $\log(M_{\mathrm{BH}}/{\mathrm{M}_\odot}) \geq 8$. At z = 0, $f_{\mathrm{Q}}$ reaches a maximum at the relatively low value of $f_{\mathrm{Q}} = 0.5$, and there is further evidence of a downwards trend at high black hole masses, as seen at z = 1. Indeed, EAGLE explicitly predicts that the higher $M_{\rm{BH}}$ central galaxies will be star forming at least $\sim$50\% of the time, which is in stark contrast to observations (see, e.g., \citealt{Peng_2010, Bluck_2016, Terrazas_2016, Terrazas_2017, Piotrowska_2022, Bluck_2022}).

Whilst the results for centrals from TNG and EAGLE agree that low black hole mass centrals are unable to quench, they exhibit a very different evolution with redshift for higher black hole masses. The evolution of the quenched fraction of centrals for the EAGLE simulation demonstrates clear signs of rejuvenation - i.e., formally quenched galaxies restart star formation at late cosmic times. This is likely due to the absence of an explicitly preventative AGN feedback mode in the EAGLE simulation~\citep{Booth_Schaye, Schaye_2015, Crain_2015}. 

The EAGLE simulations makes use of an effective thermal feedback mode which drives outflows and enables the rapid quenching of galaxies by expelling gas at high redshifts (where the most powerful AGN reside,~\citep[][]{Aird_2010,Fabian_2014, Aird_2018}. However, the feedback prescription employed in EAGLE weakens as low Eddington ratios dominate. Therefore, as $f_{\mathrm{Edd}}$ decreases with redshift, so does the potential impact incurred by AGN on their galaxies and surrounding haloes. Furthermore, there is no explicit mechanism to prevent the galaxy from accreting gas from the hot halo via cooling at late cosmic times. These effects in combination may lead to the rejuvenation of star formation. This situation is in stark contrast with TNG, which shows no sign of late-time rejuvenation. This is almost certainly a result of TNG employing a `kinetic mode' of AGN feedback, which acts as a preventative feedback mode, keeping the halo gas hot. Crucially, the kinetic mode is most efficient at low Eddington ratios, which are more common at late cosmic times. As the gas of the CGM is unable to cool (due to continued heating), collapse into the galaxy at late cosmic times is suppressed, enabling long-term quiescence~\citep[see,][]{Weinberger_2017,Nelson_2018, Pillepich_2018a, Zinger_2020}. 

We turn our attention now to satellite galaxies. For TNG, at higher black hole masses the quenched fractions of satellites and centrals are remarkably similar at all redshifts. However, for EAGLE, while the satellite quenched fraction is similar in trend to the central quenched fraction, it is markedly more elevated than the latter at high mass, especially at z = 0. On the other hand, while rejuvenation is best seen in the quenched fraction of centrals, there is also clear evidence of it for high black hole mass satellites as well, particularly at z = 1. This demonstrates the similarity in the quenching mechanisms of high-mass satellites and central galaxies, driven by their supermassive black hole (as confirmed by the RF analyses in the previous subsection). However, given the elevated quenched fraction of high black hole mass satellites at z = 0, there appears to be some processes, likely environmental, which aids quenching in, or prevents the rejuvenation of, high-mass satellites. 

At low black hole mass, the satellite and central quenched fractions differ to varying degrees at different redshifts. To quantify the offset between the central and satellite quenched fractions in the low $M_{\mathrm{BH}}$ (or low $
M_{\mathrm{*}}$) regime, we define $\Delta$, the maximum difference between the satellite and central quenched fractions in said region, indicated within each panel. The error in $\Delta$ is determined by the sum in quadrature of the Poisson uncertainties for the associated bins of the satellite and central quenched fractions. As such, $\Delta$ serves as an indicator for the impact of environment on quenching at each redshift. For both TNG and EAGLE at z = 2, the satellite quenched fraction is only slightly raised when compared to that of centrals. However, as redshift decreases, $\Delta$ increases. For TNG, the offset increases from $\Delta = 0.11$ at z = 2, to $\Delta = 0.25$ at z = 1, and $\Delta \simeq 0.45$ at z = 0. Focusing on the EAGLE simulation, there is a similar trend as $\Delta$ grows from about $\Delta \simeq 0.12$ at z = 2 to $\Delta \simeq 0.4$ at z = 0. The growth in separation of the quenched fractions of each population indicates an increase in the impact of environmental quenching for low-mass satellites as redshift decreases. This is likely due to the formation and growth of high-mass cluster environments, and therefore high density regions, through hierarchical assembly over cosmic time (as is also evident in Figs. \ref{fig:TNG_Den} \& \ref{fig:EAGLE_Den}).

It is important to note that a similar analysis, showing the quenched fraction as a function of stellar mass, is performed for EAGLE in \citet{Schaye_2015} at z~=~ 0.1, albeit with a different definition for the quenching threshold. Comparing our findings for EAGLE at z~=~0 to those of the aforementioned study, there is good agreement at low mass but a stark difference at higher stellar, and therefore black hole, masses ($M_{*}\sim 10^{11}{\mathrm{M_{\odot}}}$). Indeed, while we find the quenched fraction, particularly of central galaxies, to plateau and even show a downturn at high black hole mass, this is not present in \citet{Schaye_2015}. However, due to the sparsity of data at these stellar masses in EAGLE, in \citet{Schaye_2015} the quenched fraction beyond $M_{*}\sim 10^{11}{\mathrm{M_{\odot}}}$ is said to be `unreliable', and this region is presented as an extrapolation (shown as a dotted line). The results shown in Fig. \ref{fig:EAGLE_FQ_Z} are supported by several previous works~\citep[][]{Bahe_2017,Donnari_2021, Piotrowska_2022, Bluck_2023}. Therefore, while there may appear to be disagreement between the results presented and previous analyses, this is due to a difference in methodology. The actual quenched fraction at high masses in EAGLE is lower than presented in \citet{Schaye_2015}. Whilst it remains possible that this is partly due to a lack of high mass galaxies simulated, the fact that this trend is absent in TNG (which simulates a similar volume) leads us to conclude that this is more likely a result of the EAGLE AGN feedback mechanism not being effective at maintaining long-term quiescence in high-mass galaxies and haloes.

\subsection{Satellite quenching as a function of environment}

In the following two sub-sections we present the quenched fraction of satellites as function of several environmental parameters, for the `MOONRISE-Like' data sets of TNG and EAGLE, at z = 0, 1, 2. To ensure solely environmental effects are probed, we compute and analyze: $\Delta f_{\mathrm{Q}}$. This statistic is defined as the difference in the quenched state of a given satellite to a control sample of centrals with similar black hole masses ($|\log(M_{\mathrm{BH}_{i}}) - \log(M_{\mathrm{BH}_{cen}})| \leq 0.2$). Given intrinsic quenching (specifically AGN feedback) is the sole avenue to quench centrals in both simulations, $\Delta f_{\mathrm{Q}}$ probes the effect of additional quenching channels (e.g., through environment) on the satellite population in different scenarios.  

In Sec. \ref{sec:QF_ClusterGroup} we compute and compare $\Delta f_{\mathrm{Q}}$ as a function of stellar mass for two satellite populations, those in high-mass groups, defined as $M_{\mathrm{Halo}} \geq 10^{13} {\mathrm{M_{\odot}}}$, and low-mass groups, $M_{\mathrm{Halo}} \leq 10^{13} {\mathrm{M_{\odot}}}$. We then explore how a galaxy's location within its group or cluster affects quenching, presented in Sec. \ref{sec:QF_Loc}. To do so we evaluate $\Delta f_{\mathrm{Q}}$ as a function of $D_{\mathrm{cen}}$, the distance of a satellite to its central, normalized by the virial radius of the halo.

\subsubsection{Quenched Fraction of satellites in clusters and groups as a function of stellar mass}\label{sec:QF_ClusterGroup}

\begin{figure*}
    \centering

    \includegraphics[width=\textwidth]{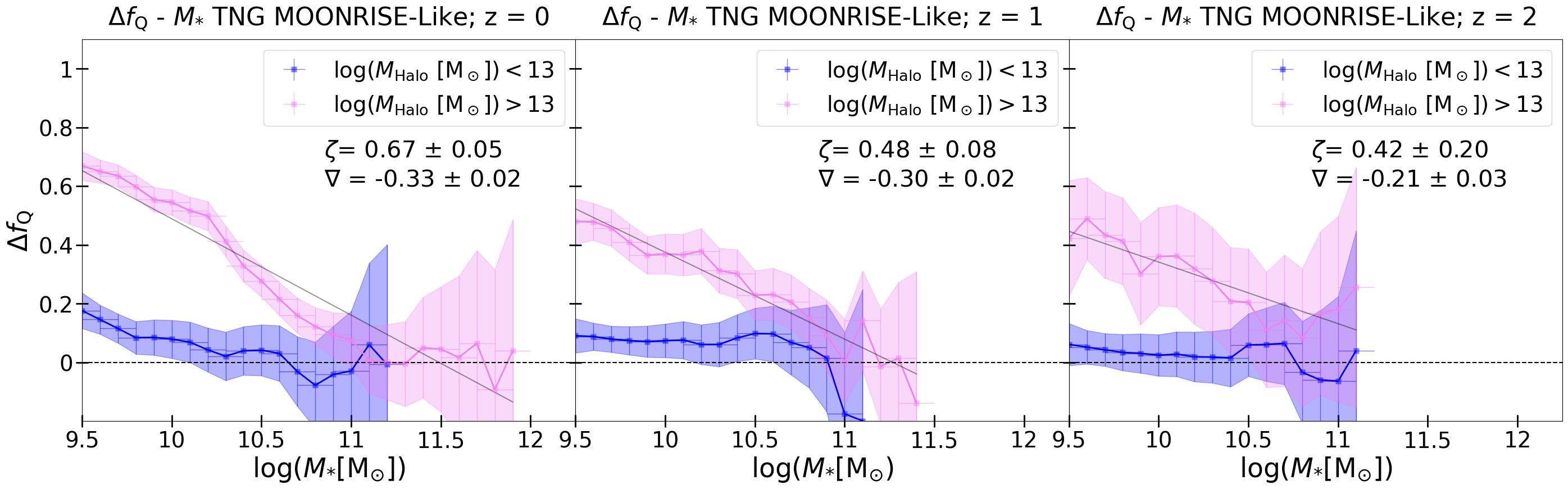}
    \vspace*{-3mm}
    
\vspace*{-3mm}
\caption{The quenching of satellites in TNG as a function stellar mass, separated by the mass of their parent halo (see legends). We present the delta quenched fraction ($\Delta f_{\mathrm{Q}}$) as a function of stellar mass for the `MOONRISE-Like' data sets. This statistic quantifies the enhancement in quenching of satellites over centrals, evaluated by an offset in quenched fraction from a central control sample with similar central black hole masses. The panels correspond to a given redshift, from z = 0 to z = 2 (left to right). Within each panel we include the values of the $\zeta$ (maximum offset) and $\nabla$ (gradient) parameters, for the high mass group sample. Low mass groups experience weak environmental enhancement of quenching throughout the full range in stellar mass. At all redshifts, and independent of data set, low-mass satellites in high-mass groups/clusters undergo stronger quenching than centrals with a similar black hole mass. Meanwhile, $\Delta f_{\mathrm{Q}}$ has a clear anti-correlation with satellite stellar mass, demonstrated by the negative values of $\nabla$ steadily declining as $M_{*}$ increases, which is a strong indication that intrinsic quenching grows in relative importance with stellar mass. Low-mass satellites quench preferentially in high-mass groups, whereas the highest mass satellites are no more frequently quenched than centrals in either halo mass population at any epoch.} \label{fig:TNG_Group_Clust}   
\end{figure*}

\begin{figure*}
    \centering

    \includegraphics[width=\textwidth]{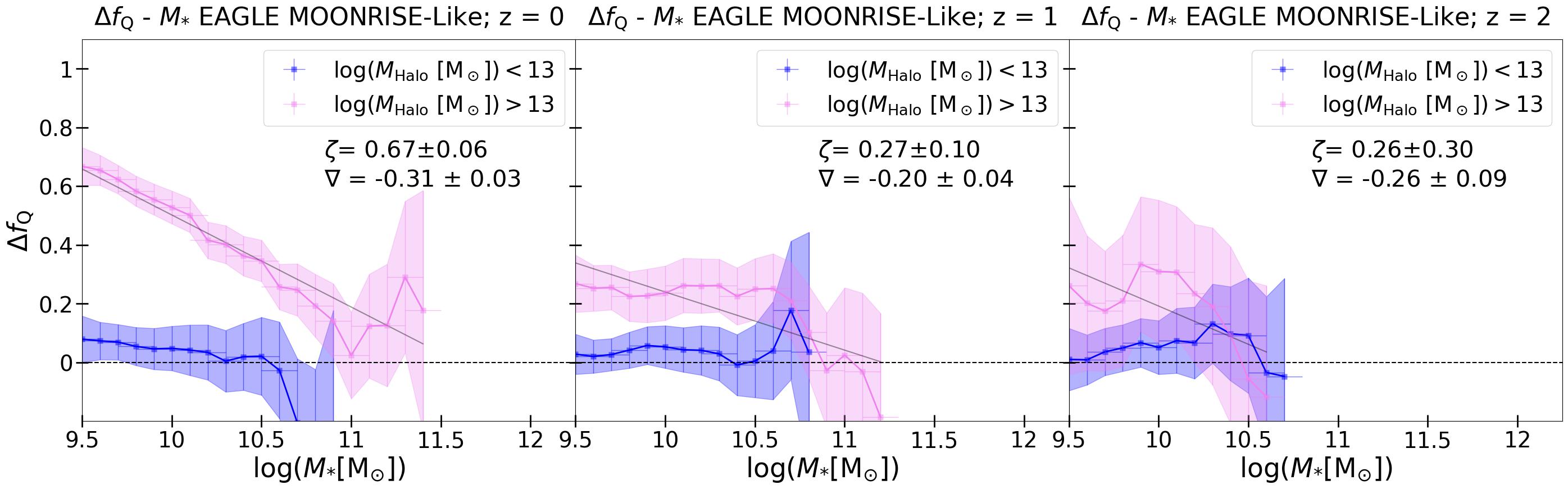}
    \vspace*{-3mm}
    
\vspace*{-3mm}
\caption{Identical in structure to Fig. \ref{fig:TNG_Group_Clust}, but here presenting results for EAGLE. The effect of environment on quenching is strongest at z = 0 and weakens significantly as redshift increases. This is demonstrated by the decrease of $\Delta f_{\mathrm{Q}}$ for the least massive satellites in high-mass groups as redshift increases, as quantified by $\zeta$ (displayed on panels). This is a strong indication that the impact of environment is weak at higher redshifts for the EAGLE simulation. On the other hand, the anti-correlation of $\Delta f_{\mathrm{Q}}$ with stellar mass, indicated by $\nabla$, demonstrates a growing similarity in quenching between satellites and centrals as black hole and stellar masses increase, as is also seen in TNG.} \label{fig:EAGLE_Group_Clust}   
\end{figure*}

In Figs. \ref{fig:TNG_Group_Clust} (TNG) \& \ref{fig:EAGLE_Group_Clust} (EAGLE) we present $\Delta f_{\mathrm{Q}}$ for satellites in high-mass groups (presented in violet), and low-mass groups (presented in blue), as a function of stellar mass for the `MOONRISE-Like' data sets. This enables direct comparison to the upcoming VLT-MOONRISE galaxy survey. The panels are ordered with increasing redshift from left to right. To quantify the maximal impact environment may carry on quenching, we define $\zeta$, which is computed by the $\Delta f_{\mathrm{Q}}$ value at the lowest stellar mass probed (where environmental effects dominate). Furthermore, we perform an OLS fit to $\Delta f_{\mathrm{Q}}$ for satellites residing in high-mass groups, where environmental effects are significant. From these fits we extract the gradient, $\nabla$, quantifying the (anti) correlation between $\Delta f_{\mathrm{Q}}$ and stellar mass. We include the value of each of these parameters within each panel. In essence, this analysis reveals the stabilizing effect of increasing stellar mass on the impact of environment on star formation within satellite galaxies.

We start by analyzing the results for the TNG simulation, seen in Fig. \ref{fig:TNG_Group_Clust}. It is clear that satellites preferentially quench in high-mass groups at low stellar masses, while even low-mass satellites are unlikely to quench in low-mass groups. Furthermore, these features remains constant across redshifts. Notably, the quenching of satellites in halos is most efficient at z = 0, likely due to an increased abundance of the highest-mass groups. Most interesting is an anti-correlation of $\Delta f_{\mathrm{Q}}$ with stellar mass at all redshifts, as demonstrated by $\nabla$. 

Honing in on z = 0 for the TNG simulation, $\Delta f_{\mathrm{Q}} \sim 70\%$ at the lowest stellar mass, and then steadily decreases as stellar mass increases, reaching $\Delta f_{\mathrm{Q}} = 0$ when $\log(M_{*}/{\mathrm{M_{\odot}}}) \simeq 11$. While the specific values differ slightly across redshift ranges, the trend of an anti-correlation with stellar mass, as indicated by the negative values of $\nabla$, is constant across cosmic time.

The strong anti-correlation with stellar mass points towards an increased dependence on intrinsic mechanisms for satellite quenching as stellar mass increases. Therefore, it is not the central - satellite class that truly matters when comparing quenching mechanisms, but rather a low/ high-mass class. Low-mass satellite galaxies are subject to environmental quenching, but all high-mass galaxies are primarily subject to intrinsic (i.e., AGN feedback) quenching. This provides further evidence for the similar conclusions reached in Section 4.3. via the Random Forest analyses.

When comparing the results for the TNG and EAGLE simulations, a clear difference appears in the impact environment carries on quenching as redshift increases. It is evident in EAGLE that as redshift increases, environment plays a lesser role in quenching satellites. This is demonstrated by both a decreasing value of $\zeta$, from 0.67 at z = 0 to 0.26 at z = 2, and a flattening of the decline in gradient of $\Delta f_{\mathrm{Q}}$ with increasing stellar mass (see panels).

This decline in the impact environment has on quenching is in line with the evolution of the $\mathrm{Quench_{E}}$ and $\mathrm{Quench_{I}}$ parameters shown in Fig.~\ref{fig:Q_SMF_Schechter}. While it is clear both TNG and EAGLE show an increasing intrinsic quenching fraction with redshift, the effect is more drastic in EAGLE. In the TNG simulation, $\mathrm{Quench_{I}}$ shifts from 40\% at z = 0 to 60\% at z = 2, while for EAGLE $\mathrm{Quench_{I}}$ moves from 20\% to 80\% over the same redshift interval. This points to less effective environmental quenching at higher redshifts, especially for EAGLE, presumably due to a lack of high-mass halos and clusters at these very early cosmic time. Hence, the early Universe is predicted to be progressively more dominated by intrinsic quenching~\citep[as also seen in][]{Kawinwanichakij_2017,Xie_2024}.

Therefore, the simulations predict significantly different environmental impact on quenching at higher redshifts. The VLT-MOONRISE survey will provide ideal observational data to test these differing predictions in the coming years.

\subsubsection{Satellite quenching as a function of location within the halo}\label{sec:QF_Loc}

\begin{figure*}
    \centering
    \includegraphics[width=\textwidth]{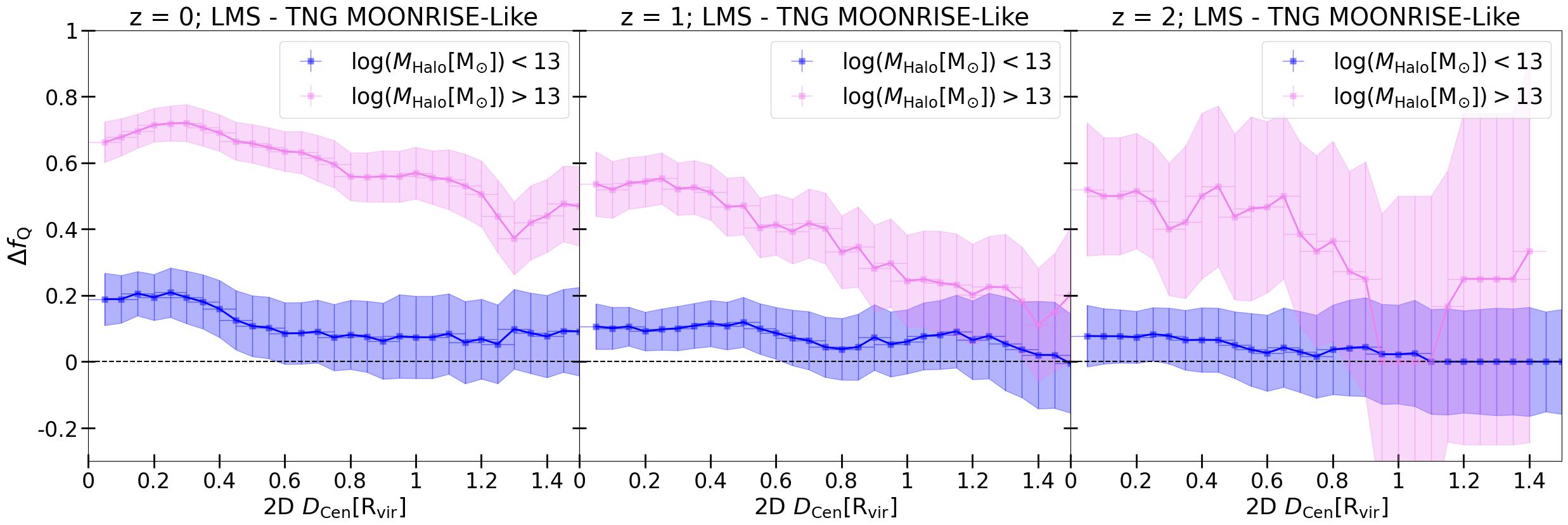}
\caption{The quenching of low-mass satellites (`LMS' in titles) as a function of location within their groups or clusters for TNG. We present the delta quenched fraction ($\Delta f_{\mathrm{Q}}$) as a function of the distance to the nearest central galaxy ($D_{\mathrm{cen}}$) separated into bins of halo mass (as indicated on the legends). The bins are formed from objects that are within a defined $D_{\mathrm{cen}}$ range of $\pm$0.15$R_\mathrm{vir}$, while the shaded regions are the margins of error for each bin, computed via Poisson statistics. We present these plots for the `MOONRISE-Like' data sets. As previously, $\Delta f_{\mathrm{Q}}$ measures the offset in quenched fraction for each satellite population relative to a control sample of central galaxies with similar black hole masses. Low-mass satellites are strongly offset to higher quenched fractions at close distances to their centrals, especially within high-mass groups/clusters, at all redshifts. Furthermore, there is a clear anti-correlation between $\Delta f_{\mathrm{Q}}$ and $D_{\mathrm{cen}}$, as $\Delta f_{\mathrm{Q}}$ decreases as $D_{\mathrm{cen}}$ increases. In low-mass groups however, even low-mass satellites exhibit difficulties in achieving quiescence.} \label{fig:TNG_Dc}   
\end{figure*}

In Fig. \ref{fig:TNG_Dc}, we present $\Delta f_{\mathrm{Q}}$ as a function of satellite location within a group, represented by the distance to the central galaxy (in units of virial radii), for two halo mass ranges (see legends) for the TNG `MOONRISE-Like' data sets. We perform this analysis solely for low-mass satellites, as the quenching of high-mass satellites is largely independent of environment. As there are few quenched low-mass satellites present in the EAGLE `MOONRISE-Like' data sets at z = 1, and none at z = 2, we present the results for TNG only. However, the results for EAGLE at z = 0 are consistent with those obtained for TNG.

At all redshifts, low-mass satellites struggle to quench when in a low-mass halo ($\log(M_{\mathrm{Halo}}/{\mathrm{M_{\odot}}} < 10^{13}$). However, $\Delta f_{\mathrm{Q}}$ is slightly elevated for satellites near the center of their group, and gradually decreases as $D_{\mathrm{cen}}$ increases. 

Analyzing $\Delta f_{\mathrm{Q}}$ in the most massive groups and clusters, the quenching of low-mass satellites occurs most efficiently near the center of these systems. Similar to what is seen for lower mass halos, $\Delta f_{\mathrm{Q}}$ is clearly inversely correlated with $D_{\mathrm{cen}}$, indicating an increasing quenching efficiency as satellites approach the center of their groups. Furthermore, at a fixed $D_{\mathrm{cen}}$, $\Delta f_{\mathrm{Q}}$ is much greater for satellites of the most massive groups than those with $\log(M_{\mathrm{Halo}}/{\mathrm{M_{\odot}}}) < 13$. Therefore, we conclude that while the mass of a satellites' group is most important to its quenching process (as revealed by the random forest analysis of Section 4.3), its location in the group additionally factors in reducing star formation. We find the results for low-mass satellites for EAGLE at z = 0 to be consistent with those of TNG.

In summary, the location of satellites within their group or cluster haloes results in a significant secondary quenching trend for low-mass systems within high mass haloes. This operates such that quenching is enhanced as satellites approach closer to their centrals, and hence the center of their groups or clusters. However, for low mass satellites in low mass haloes, the location within the halo is of little impact to quenching. These results are seen primarily from TNG, because EAGLE lacks the low-mass quenching channel accessible to MOONRISE. Hence, there are clear predictions provided here to be tested by future observations at cosmic noon with VLT-MOONRISE.

\section{Summary}\label{sec:Summary}

Over the course of this paper we analyze the quenching of different galaxy classes across cosmic time in two cosmological hydrodynamical simulations: IllustrisTNG and EAGLE. Furthermore, we employ these simulations to extract specific predictions for quenching at cosmic noon in anticipation of VLT-MOONRISE, the first wide-field spectroscopic galaxy survey at intermediate-to-high redshifts~\citep[see][for full details]{Cirasuolo_2020, Maiolino_2020}. 

We make use of sSFR distributions to determine bespoke quenching thresholds at each redshift for each simulation. Furthermore, with the aim of isolating intrinsic and environmental quenching avenues, we determine the optimal mass thresholds for splitting low and high-mass satellites in terms of quenching by leveraging the bimodality of the stellar mass function of quenched satellites. This approach is vindicated by the results from the Random Forest analyses, whereby we show that this method is highly effective at separating intrinsically from environmentally quenched galaxies. The final mass thresholds for each simulation and redshift are presented in Table~\ref{tab:Mass_Thresh}.

To investigate the evolution of different quenching avenues between z = 0 and z = 2, we split galaxies into three classes: (i) central galaxies, (ii) high-mass satellites, and (iii) low-mass satellites. We leverage the power of machine learning, specifically the Random Forest classification approach, to determine the dominant parameter in predicting quenching for each galaxy class at z = 0 - 2, for both simulations. This technique enables us to carefully control for nuisance variables before establishing the underlying causal relationships at play~\citep[for further discussion see,][]{Bluck_2022}. Additionally, we perform more traditional experiments, including probing the quenched fraction of centrals and satellites in different situations as a function of various parameters. \\

Our primary results, and predictions for VLT-MOONRISE, are as follows:

\begin{enumerate}

    \item TNG predicts the existence of environmentally quenched low-mass satellites at cosmic noon, which will be visible to the survey limits of VLT-MOONRISE ($\log(M_{*}/{\mathrm{M_{\odot}}}) \geq 9.5$), as seen in Fig. \ref{fig:Q_SMF_Schechter}. Conversely, EAGLE predicts that low-mass environmental quenching will be nearly absent at this mass completeness limit. Hence, the identification and abundance of environmentally quenched objects in MOONRISE (or lack thereof) will provide an excellent comparative test of these two galaxy evolution models. \\

    \item As demonstrated in Figs. \ref{fig:TNG_Den} \& \ref{fig:EAGLE_Den}, satellites exhibit a strong bimodality in over-density, for the complete, 6D phase space data sets of both simulations, especially at z = 0. While this bimodality weakens as redshift increases it can still be detected. Moreover, this bimodality in density clearly maps onto a separation in parent halo mass, whereby the high density peak corresponds to high-mass haloes and the low density peak corresponds to low-mass haloes. This highlights $\delta$ as an effective tracer for halo mass, at least in simulation space.\\
    
    \item However, as seen in the lower row of plots in Figs. \ref{fig:TNG_Den} \& \ref{fig:EAGLE_Den}, this bimodality is much weaker for the observation-like (2D+z observer space) `MOONRISE-Like' data sets. In fact, for both simulations, the distribution appears nearly unimodal for $z\geq1$. Furthermore, the range of over-density is greatly diminished when compared to that obtained for the complete data sets. This loss of information moving from the simulation space to an observation-like space is highly likely to impact the performance of $\delta$ as a tracer for environment in observational surveys, and serves to emphasize the need for accurate measurements of halo mass (e.g., Bluck et al. 2025, in press).\\

    \item Both TNG and EAGLE predict that central galaxies quench via intrinsic mechanisms at all redshifts, specifically via a clear dependence on supermassive black hole mass. This is clearly indicated by the dominant predictive power of black hole mass in the Random Forest classification analyses for this galaxy class, presented in Figs. \ref{fig:TNG_RF} \& \ref{fig:EAGLE_RF}. Hence, there is complete agreement between TNG and EAGLE as to the underlying mechanism of central galaxy quenching at cosmic noon~\citep[see also][for similar results]{Bluck_2022,Bluck_2023,Bluck_2024}. This is a clear testable prediction for VLT-MOONRISE, in which black hole mass may be estimated via the $M_{BH} - \sigma$ relation. \\

    \item However, an analysis of the evolution of the quenched fraction across cosmic time reveals EAGLE to have a much more effective AGN feedback prescription at higher redshifts than at lower redshifts, whereby high-mass galaxies are more frequently quenched at early cosmic times than at later epochs, see Fig. \ref{fig:EAGLE_FQ_Z}. This is a clear sign of rejuvenation, likely a consequence of the absence of a bespoke AGN `maintenance' mode in EAGLE. Conversely, TNG is equally effective at quenching high-mass systems across all cosmic times, presented in Fig. \ref{fig:TNG_FQ_Z}. This is expected due to its 'kinetic/ radio' mode feedback, which is most effective at low Eddington ratios, and hence within high mass black holes at late cosmic times. Ultimately, these discrepancies between TNG and EAGLE yield a key method to test which simulation better approximates the real Universe. If there is significant rejuvenation, EAGLE performs better; yet, if there is no (or little) rejuvenation, TNG performs better.\\

    \item Both TNG and EAGLE predict high-mass satellites to quench via intrinsic mechanisms across all redshifts, specifically connected with black hole mass, with a weak secondary dependence on environmental parameters (which is completely absent for centrals), seen in Figs. \ref{fig:TNG_RF} \& \ref{fig:EAGLE_RF}. In both simulations, this secondary dependence weakens with redshift. This clearly identifies AGN feedback as the driving quenching process for high-mass satellites in both simulations, aided by environmental effects such as starvation in some cases. Again this is a clear testable prediction for MOONRISE at intermediate-to-high redshifts.\\

    \item Both TNG and EAGLE predict low-mass satellites to quench due to environment, with halo mass found to be the best parameter to determine quiescence at all redshifts, as presented in Figs. \ref{fig:TNG_RF} \& \ref{fig:EAGLE_RF}. Of note, secondary dependencies on $D_{\mathrm{cen}}$ and $\delta$ are found when performing the Random Forest analysis. \\

    \item While several studies have supported a central - satellite decomposition as sufficient for separating intrinsic and environmental quenching, the Random Forest classification results (seen in Figs. \ref{fig:TNG_RF} \& \ref{fig:EAGLE_RF}) alongside the numerous quenched fraction analyses, point to stellar mass as a better tool to isolate quenching avenues. This can also be seen from the total quenched population stellar mass functions, which present with a strong bimodality for both simulations at all epochs, shown in Figs. \ref{fig:TNG_SF_Q} \& \ref{fig:EAGLE_SF_Q}. This bimodality is present for the quenched satellite population but absent from the quenched central population. As such, a low-mass quenching mechanism must be accessible to satellites, but not centrals. Crucially, both avenues are open to high-mass satellites, but intrinsic quenching dominates. \\

    \par For both simulations, the low-mass satellite quenched fraction as a function of location within a halo, seen in Fig. \ref{fig:TNG_Dc}, and as a function of stellar mass for low and high-mass groups, presented in \ref{fig:TNG_Group_Clust} \& \ref{fig:EAGLE_Group_Clust}, reinforce the secondary dependencies on $D_{\mathrm{cen}}$ and $\delta$ found from the Random Forest analyses. Taken together, this points towards low-mass satellite quenching being most effective towards the center of high mass groups and clusters. This indicates that mechanisms such as ram pressure stripping, galaxy-galaxy interaction, and starvation are the likely candidates responsible for this low-mass quenching. However, while this trend remains essentially constant across redshift for TNG, the impact of environment weakens significantly with increasing redshift in EAGLE.

    \par Conversely, high-mass satellites are not more frequently quiescent as one moves towards the center of even the most massive groups and clusters. This finding underscores the intrinsic nature of the quenching of high mass satellites. As the first wide-field spectroscopic survey at intermediate and high redshift, VLT-MOONRISE will be able to test the impact of environment on quenching from z $\simeq$ 1 to 2.5, discovering whether these simulations accurately predict the quenching process, or not.\\

    \item In the absence of $M_{\mathrm{BH}}$ and $M_{\mathrm{Halo}}$ as parameters (which are often lacking in observational analyses), the RF analysis of the `MOONRISE-Like' data point to $\delta$ as the best predictor of quiescence for high-mass satellites at z = 0 (and z = 1 for EAGLE), see Figs. \ref{fig:TNG_RF_Moons} \& \ref{fig:EAGLE_RF_Moons}. This results in a misidentification of environmental quenching in these cases, since the true parameter regulating quenching is actually supermassive black hole mass in these galaxies. This may be explained by the morphology - density relation, whereby at a fixed stellar mass galaxies in higher density environments tend to be more bulge-dominated and hence are likely to host more massive central black holes (e.g., \citealt{Saglia_2016, Bluck_2022}). Ultimately, this highlights the need for supermassive black hole mass estimates, particularly at z < 2, when investigating galaxy quenching in both central and satellite galaxies. Fortunately, this will be possible with VLT-MOONRISE, enabling rigorous tests to these predictions from simulations.

\end{enumerate}

In summary, we find EAGLE and TNG to agree on the quenching mechanisms of all galaxy classes from cosmic noon to the present epoch, which points towards a theoretical consensus. Centrals and high-mass satellites are predicted to quench via AGN feedback, although environment plays a small secondary role in the latter class. Conversely, low-mass satellites are predicted to quench via purely environmental means, with ram pressure stripping, galaxy-galaxy interactions, and starvation working in tandem as the leading candidates for the mechanism. These are all testable predictions which can be verified via VLT-MOONRISE.

In addition to the important similarities between EAGLE and TNG outlined above, there are also important differences between the two simulations with respect to quenching. Most notably, the mass thresholds determined for environmental quenching of satellites are very different between the two simulations. In fact, EAGLE predicts that the mass completeness limit of MOONRISE is too high to detect environmentally quenched satellites. Alternatively, TNG predicts that this quenching avenue should be clearly visible in this survey. As such, the abundance (or even the presence or absence) of low-mass environmentally quenched satellites in the MOONRISE survey will provide strong evidence for or against the TNG and EAGLE models.

In conclusion, while TNG and EAGLE agree in large part on the fundamental quenching mechanisms responsible in different galaxy classes, they disagree on the details. In combination with the VLT-MOONRISE survey, this will enable the astrophysics community to rigorously test these galaxy evolution models in an unprecedented parameter space, clearly identifying the strengths and weaknesses of each approach. From this, we will be in an excellent position to improve on the current status quo of the theory of galaxy evolution.

\section{Acknowledgments}

We are very grateful to the anonymous reviewer for many insightful and helpful comments on this work, which have helped to significantly improve the presentation of our results. AFLB gratefully acknowledges support from an NSF research grant: NSF-AST 2408009, in addition to the ORAU Ralph E. Powe Junior Faculty Enhancement Award in Physical Sciences, and research start-up funds from FIU. PG also acknowledges support from NSF-AST 2408009. RM acknowledges support from the Science and Technology Facilities Council (STFC), ERC Advanced Grant 695671 “QUENCH", and by the UKRI Frontier Research grant RISEandFALL. RM also acknowledges funding from a research professorship from the Royal Society. PT acknowledges support from NSF-AST 2346977 and the NSF-Simons AI Institute for Cosmic Origins which is supported by the National Science Foundation under Cooperative Agreement 2421782 and the Simons Foundation award MPS-AI-00010515.

\section{Data Availability}

 All data used in this study have been previously published and are
available at the following online locations:
\begin{itemize}
    \item EAGLE: \url{http://icc.dur.ac.uk/Eagle/}
    \item TNG: \url{https://www.tng-project.org/}   
\end{itemize}

\bibliographystyle{mnras}
\bibliography{Citations}

\appendix\label{appendix:Appendix}

\section{Evolution of Environment}\label{appendix:A}

\begin{figure*}
    \centering
    \begin{subfigure}{\textwidth}
        \caption{TNG - 3D $\delta_{5}$ Distribution}
        \includegraphics[width = \textwidth]{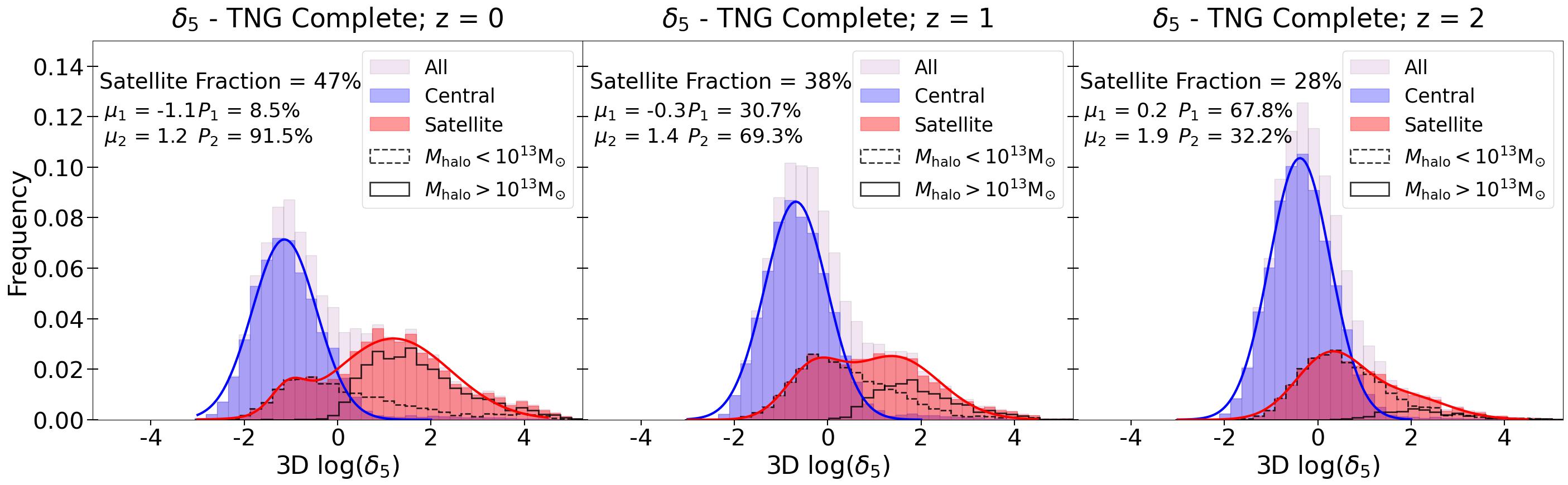}  
    \end{subfigure}
    \hfill
    \begin{subfigure}{\textwidth}
        \caption{TNG - MOONRISE-Like $\delta_{5}$ Distribution}
        \includegraphics[width= \textwidth]{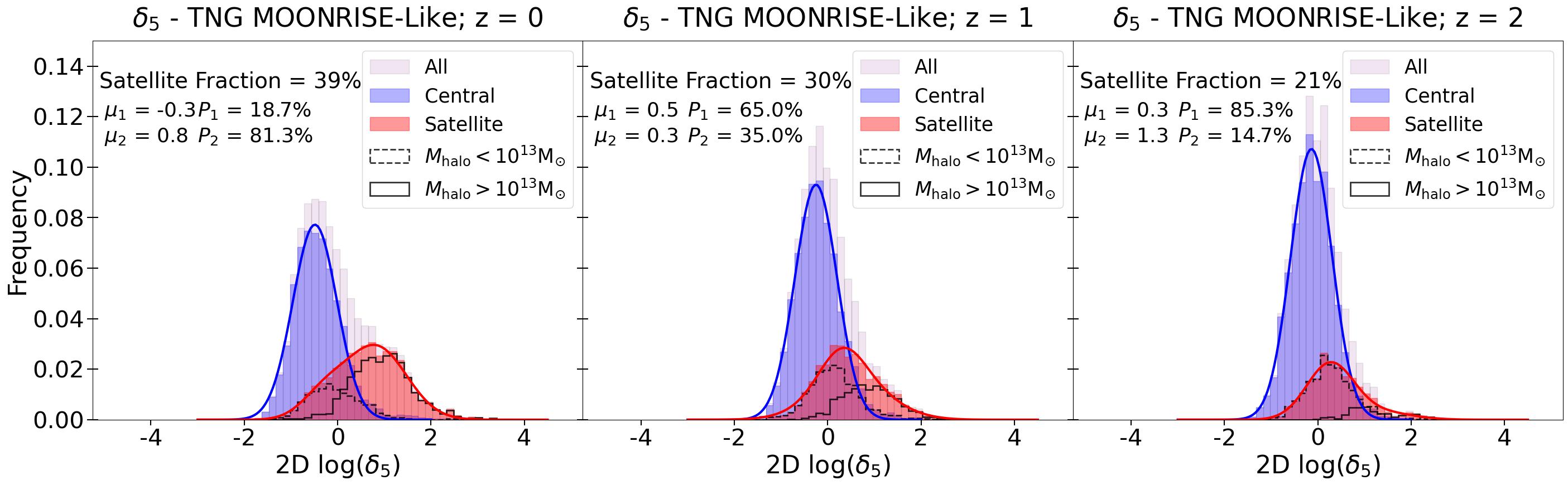} 
        
    \end{subfigure}
    
\vspace*{-3mm}
\caption{Identical in structure and method to Fig. \ref{fig:TNG_Den}, but here presenting the evolution of the distribution in galaxy over-density of the fifth nearest neighbor density, $\delta_5$ for the TNG simulation.} \label{fig:TNG_Den5}
\end{figure*}

\begin{figure*}
    \centering
    \begin{subfigure}{\textwidth}
        \caption{TNG - 3D $\delta_{3}$ Distribution}
        \includegraphics[width = \textwidth]{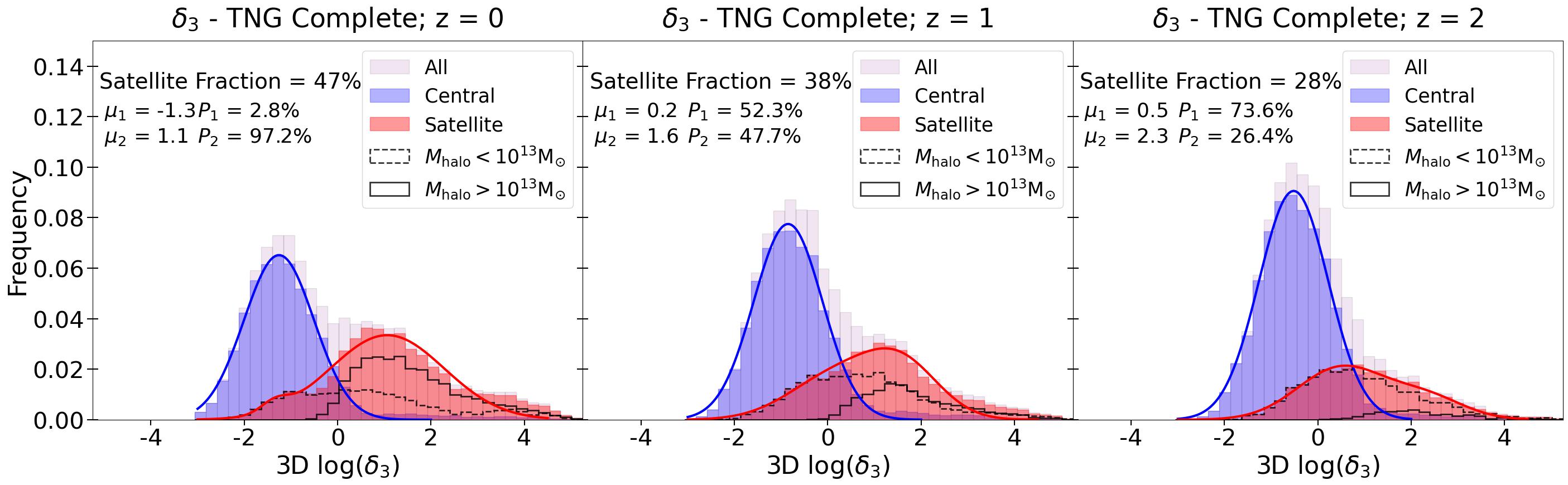}  
    \end{subfigure}
    \hfill
    \begin{subfigure}{\textwidth}
        \caption{TNG - MOONRISE-Like $\delta_{3}$ Distribution}
        \includegraphics[width= \textwidth]{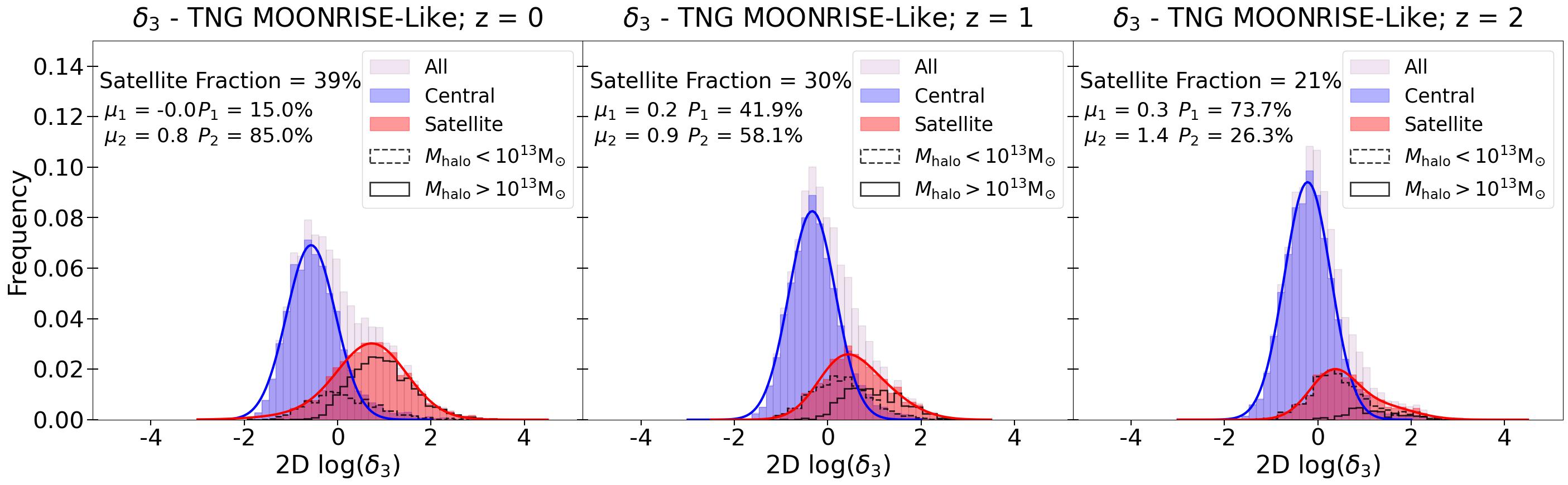} 
        \label{fig:TNG_Den3_Moons}
    \end{subfigure}
    
\vspace*{-3mm}
\caption{Identical in structure and method to Fig. \ref{fig:TNG_Den}, but here presenting the evolution of the distribution in galaxy over-density for the third nearest neighbor density, $\delta_3$, for the TNG simulation. } \label{fig:TNG_Den3}
\end{figure*}

\begin{figure*}
    \centering
    \begin{subfigure}{\textwidth}
        \caption{EAGLE - 3D $\delta_{5}$ Distribution}
        \includegraphics[width = \textwidth]{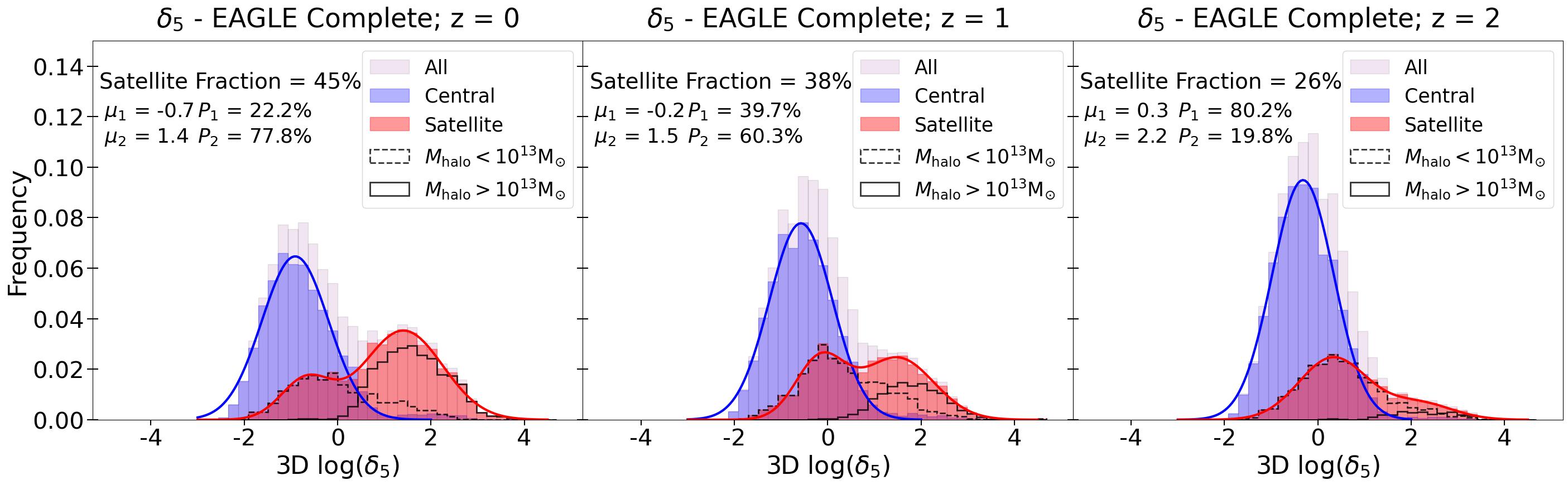}  
    \end{subfigure}
    \hfill
    \begin{subfigure}{\textwidth}
        \caption{EAGLE - MOONRISE-Like $\delta_{5}$ Distribution}
        \includegraphics[width= \textwidth]{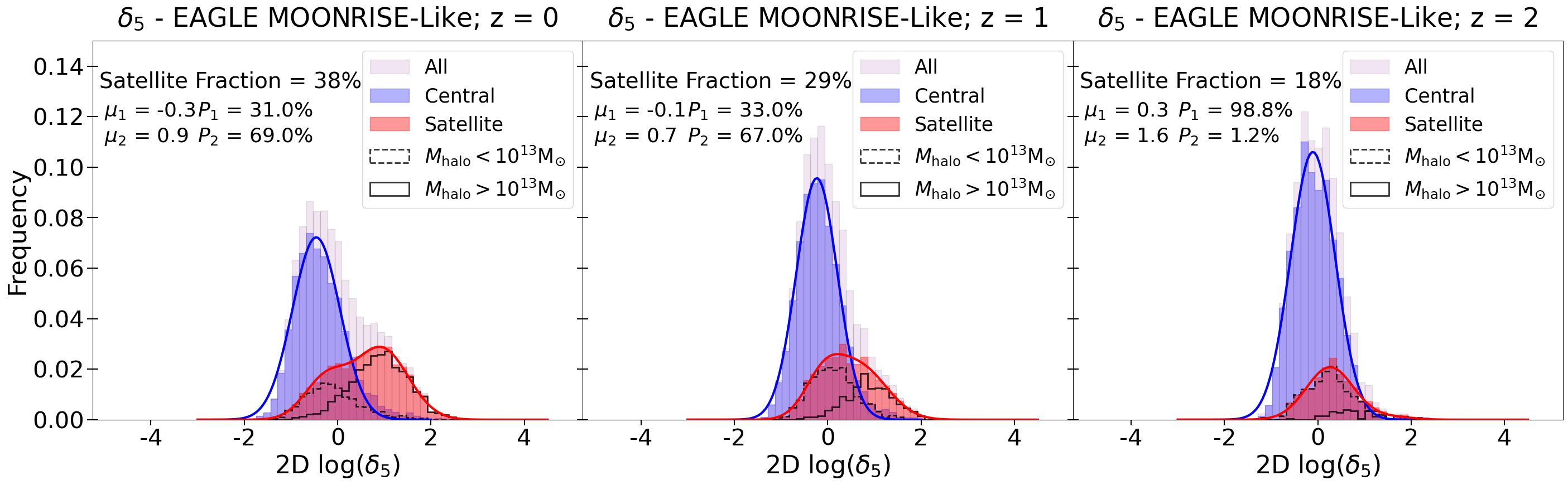} 
        
    \end{subfigure}
    
\vspace*{-3mm}
\caption{Identical in structure and method to Fig. \ref{fig:EAGLE_Den}, but here presenting the evolution of the distribution in galaxy over-density of the fifth nearest neighbor density, $\delta_5$ for the EAGLE simulation.} \label{fig:EAGLE_Den5}
\end{figure*}

\begin{figure*}
    \centering
    \begin{subfigure}{\textwidth}
        \caption{EAGLE - 3D $\delta_{3}$ Distribution}
        \includegraphics[width = \textwidth]{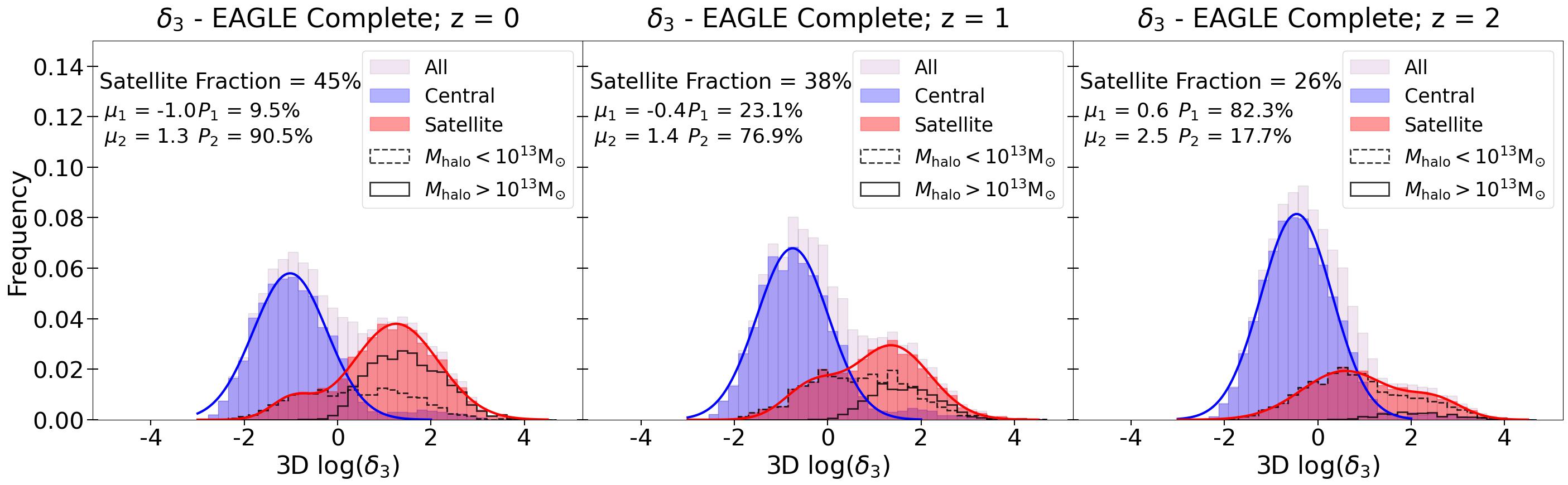}  
    \end{subfigure}
    \hfill
    \begin{subfigure}{\textwidth}
        \caption{EAGLE - MOONRISE-Like $\delta_{3}$ Distribution}
        \includegraphics[width= \textwidth]{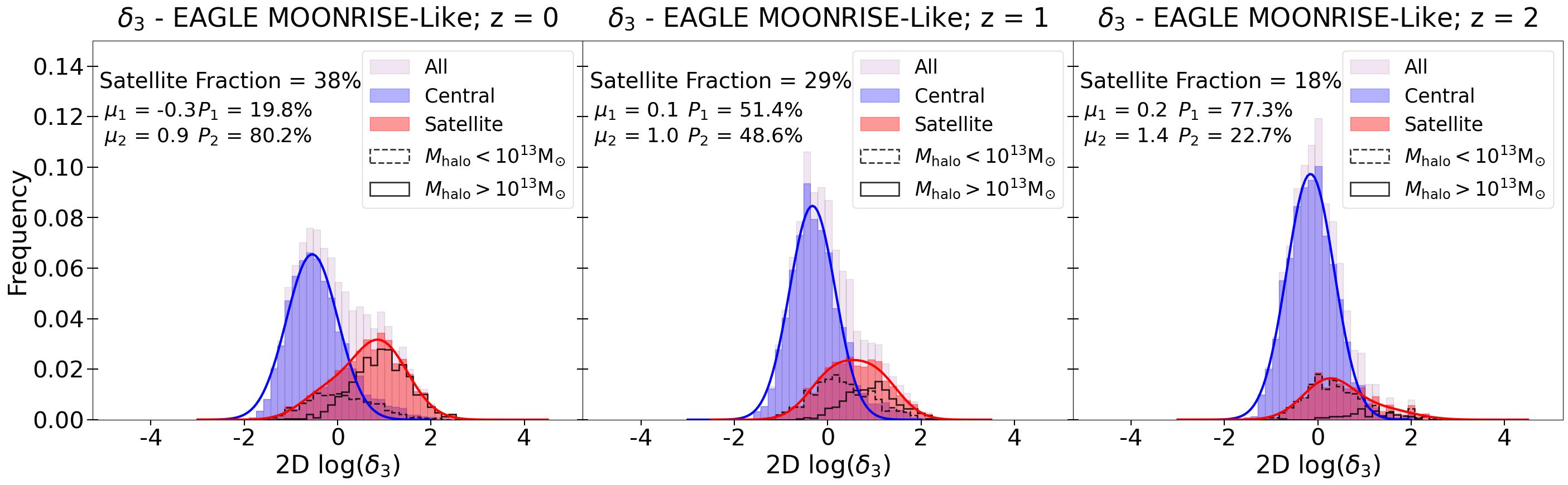} 
        \label{fig:EAGLE_Den3_Moons}
        
    \end{subfigure}
    
\vspace*{-3mm}
\caption{Identical in structure and method to Fig. \ref{fig:EAGLE_Den}, but here presenting the evolution of the distribution in galaxy over-density for the third nearest neighbor density, $\delta_3$, for the EAGLE simulation.} \label{fig:EAGLE_Den3}
\end{figure*}

As previously seen for $\delta_{10}$ in Figs. \ref{fig:TNG_Den} \& \ref{fig:EAGLE_Den}, here we additionally present the distributions of $\delta_3$ and $\delta_5$ for the total (in faint gray), central (in blue), and satellite (in red) populations of TNG, Figs. \ref{fig:TNG_Den3} \& \ref{fig:TNG_Den5}, and EAGLE, Figs. \ref{fig:EAGLE_Den3} \& \ref{fig:EAGLE_Den5}. We also include the distributions in $\delta_3$ and $\delta_5$ of satellites in high-mass groups, as solid line, and low-mass groups, as dashed line. The top row of each figure we show the distributions for the complete, 6D phase space data set and in the bottom rows we present the results for the `MOONRISE-Like' data sets, which operate in an observer like 2D+z phase space.

When analyzing the results for the complete data set, we find that for both simulations there is a clear bimodality in the distribution of both $\delta_5$ and $\delta_3$ for satellites at z = 0. However, as seen for $\delta_{10}$, this bimodality weakens significantly as redshift increases, with the low density regime growing in abundance as the high density population decreases. This is likely due to a decreasing amount of high-mass groups/ clusters as redshift increases, and points towards environmental growth with cosmic time within the simulations. 

Importantly, the bimodality observed for satellites in the complete data set is difficult to ascertain when moving to the 2D+z phase space, `MOONRISE-Like' data sets. This is true even at z = 0, where one would expect satellites to demonstrate the strongest bimodality in their density due to an increasing amount of high-mass haloes. In fact, as seen in Figs. \ref{fig:TNG_Den3_Moons} \& \ref{fig:EAGLE_Den3_Moons}, the distribution of $\delta_3$ for satellites appears to be nearly unimodal in the observer-like 2D+z phase space, at all redshifts. On the other hand, the high/low-mass group split remains effective in separating the peaks in $\delta_3$ of the high and low density populations. Therefore, while information in the distribution of $\delta$ for satellites, such as its bimodal nature, could be lost moving from a 6D phase space to 2D+z phase space, halo mass remains an effective parameter to separate two populations of satellites.  

\section{Schechter Fit Coefficients}\label{appendix:B}

We include in this appendix (see Tables~\ref{tab:Schech_TNG_Param} \& \ref{tab:Schech_EAGLE_Param}) the values of the coefficients, along with their associated uncertainties, for the best fit Schechter functions for all populations and redshifts of both simulations, as seen in Fig. \ref{fig:Q_SMF_Schechter}. The only uncertainties present for the simulations are from Poisson counting statistics, since stellar masses are known precisely. The tables' columns present the median of the following parameters: log($M^*$); $\phi$; $\alpha$; $\phi_2$; and $\alpha_2$. The uncertainties are presented as the total 1$\sigma$ errors.

Here, log($M^*$) corresponds to the `knee' or characteristic mass of the Schechter function. This is to say the mass at which the stellar mass function turns over from a power law to an exponential function. The normalization, $\phi$, corresponds to the number density at the characteristic mass, and $\alpha$ is defined as the slope of the power law which describes the stellar mass function prior to the characteristic mass. For double Schechter functions, the sum of two Schechter functions with the same characteristic mass, there are two additional parameters: $\phi_2$, the number density at the characteristic mass of the second Schechter function; and $\alpha_2$, the power law slope corresponding to the second Schechter function.

\begin{table*}
    \centering
        \caption{The best-fit Schechter function parameters in TNG. Each row corresponds to a sample subset for the TNG simulation at each redshift, while the columns list the parameters along with their associated uncertainties.}
        
    \begin{tblr}{
             colspec = { c c c c c c },  
             row{1}  = {font=\bfseries, c},
             rowsep=1ex, colsep =1ex
             }
      \hline
      \textbf{Sample Subset} & $\log(M^*)/{\mathrm{M_{\odot}}}$ & $\phi [10^{-3}\mathrm{Mpc}^{-3}]$ & $\alpha$ & $\phi_{2}[10^{-3}\mathrm{Mpc}^{-3}]$ & $\alpha_{2}$\\
      \hline
    
       \textbf{TNG - All Galaxies; z = 0} & $11.22\tiny{\Vectorstack{+0.02 -0.01}}$ & $1.911 \tiny{\Vectorstack{+0.035 -0.049}}$ & $-1.359 \tiny{\Vectorstack{+0.005 -0.006}}$ & $\cdots$ & $\cdots$ \\
    
       \textbf{TNG - All Star Forming; z = 0} & $10.55 \tiny{\Vectorstack{+0.01 -0.02}}$ & $2.428 \tiny{\Vectorstack{+0.111 -0.094}}$ & $-1.221 \tiny{\Vectorstack{+0.014 -0.009}}$ & $\cdots$ & $\cdots$ \\
    
       \textbf{TNG - All Quenched; z = 0} & $10.48 \tiny{\Vectorstack{+0.03 -0.03}}$ & $1.931 \tiny{\Vectorstack{+0.103 -0.103}}$ & $0.834 \tiny{\Vectorstack{+0.119 -0.109}}$ & $0.301 \tiny{\Vectorstack{+0.039 -0.030}}$ & $-1.597 \tiny{\Vectorstack{+0.028 -0.026}}$\\
    
       \textbf{TNG - Quenched Satellites; z = 0} & $10.45 \tiny{\Vectorstack{+0.05 -0.05}}$ & $0.905 \tiny{\Vectorstack{+0.056 -0.080}}$ & $0.669 \tiny{\Vectorstack{+0.222 -0.203}}$ & $0.333 \tiny{\Vectorstack{+0.061 -0.052}}$ & $-1.574 \tiny{\Vectorstack{+0.037 -0.041}}$\\
    
       \textbf{TNG - Quenched Centrals; z = 0} & $10.51 \tiny{\Vectorstack{+0.02 -0.03}}$ & $1.067 \tiny{\Vectorstack{+0.041 -0.064}}$ & $0.942 \tiny{\Vectorstack{+0.125 -0.072}}$ & $\cdots$ & $\cdots$\\
    
       \textbf{TNG - All Galaxies; z = 1} & $11.12 \tiny{\Vectorstack{+0.02 -0.02}}$ & $1.129 \tiny{\Vectorstack{+0.058 -0.041}}$ & $-1.354 \tiny{\Vectorstack{+0.008 -0.007}}$ & $\cdots$ & $\cdots$ \\
    
       \textbf{TNG - All Star Forming; z = 1} & $10.76 \tiny{\Vectorstack{+0.02 -0.02}}$ & $1.679 \tiny{\Vectorstack{+0.068 -0.080}}$ & $-1.297 \tiny{\Vectorstack{+0.013 -0.014}}$ & $\cdots$ & $\cdots$ \\
    
       \textbf{TNG - All Quenched; z = 1} & $10.46 \tiny{\Vectorstack{+0.03 -0.03}}$ & $0.624 \tiny{\Vectorstack{+0.058 -0.060}}$ & $1.316 \tiny{\Vectorstack{+0.117 -0.131}}$ & $0.129 \tiny{\Vectorstack{+0.0185 -0.0137}}$ & $-1.517 \tiny{\Vectorstack{+0.034 -0.034}}$\\
    
       \textbf{TNG - Quenched Satellites; z = 1} & $10.27 \tiny{\Vectorstack{+0.05 -0.03}}$ & $0.153 \tiny{\Vectorstack{+0.044 -0.023}}$ & $1.825 \tiny{\Vectorstack{+0.175 -0.311}}$ & $0.209 \tiny{\Vectorstack{+0.020 -0.034}}$ & $-1.424 \tiny{\Vectorstack{+0.036 -0.050}}$\\
    
       \textbf{TNG - Quenched Centrals; z = 1} & $10.58 \tiny{\Vectorstack{+0.07 -0.07}}$ & $0.497 \tiny{\Vectorstack{+0.079 -0.133}}$ & $0.945 \tiny{\Vectorstack{+0.402 -0.238}}$ & $\cdots$ & $\cdots$\\
    
       \textbf{TNG - All Galaxies; z = 2} & $11.14 \tiny{\Vectorstack{+0.03 -0.03}}$ & $0.546 \tiny{\Vectorstack{+0.032 -0.034}}$ & $-1.449 \tiny{\Vectorstack{+0.009 -0.008}}$ & $\cdots$ & $\cdots$ \\
    
       \textbf{TNG - All Star Forming; z = 2} & $10.85 \tiny{\Vectorstack{+0.02 -0.02}}$ & $0.813 \tiny{\Vectorstack{+0.048 -0.050}}$ & $-1.414 \tiny{\Vectorstack{+0.010 -0.012}}$ & $\cdots$ & $\cdots$ \\
    
       \textbf{TNG - All Quenched; z = 2} & $10.43 \tiny{\Vectorstack{+0.04 -0.04}}$ & $0.268 \tiny{\Vectorstack{+0.044 -0.043}}$ & $1.54 \tiny{\Vectorstack{+0.203 -0.212}}$ & $0.023 \tiny{\Vectorstack{+0.007 -0.005}}$ & $-1.706 \tiny{\Vectorstack{+0.077 -0.0725}}$\\
    
       \textbf{TNG - Quenched Satellites; z = 2} & $10.31 \tiny{\Vectorstack{+0.07 -0.08}}$ & $0.077 \tiny{\Vectorstack{+0.017 -0.031}}$ & $1.337 \tiny{\Vectorstack{+0.637 -0.337}}$ & $0.032 \tiny{\Vectorstack{+0.015 -0.008}}$ & $-1.653 \tiny{\Vectorstack{+0.083 -0.095}}$\\
    
       \textbf{TNG - Quenched Centrals; z = 2} &$ 10.51 \tiny{\Vectorstack{+0.04 -0.06}}$ & $0.235 \tiny{\Vectorstack{+0.038 -0.038}}$ & $1.367 \tiny{\Vectorstack{+0.256 -0.202}}$ & $\cdots$ & $\cdots$\\
      \hline
    
    \end{tblr}
    \label{tab:Schech_TNG_Param}
\end{table*}

\begin{table*}
    \centering
        \caption{The best-fit Schechter function parameters in EAGLE. Each row corresponds to a sample subset for the EAGLE simulation at each redshift, while the columns list the parameters along with their associated uncertainties.}
        
    \begin{tblr}{
             colspec = { c c c c c c },  
             row{1}  = {font=\bfseries, c},
             rowsep=1ex, colsep =1ex
             }
    
       \hline
   
      \textbf{Sample Subset} & $\log(M^*)/{\mathrm{M}_\odot}$ & $\phi [10^{-3}\mathrm{Mpc}^{-3}]$ & $\alpha$ & $\phi_{2}[10^{-3}\mathrm{Mpc}^{-3}]$ & $\alpha_{2}$\\
    
       \hline

       \textbf{EAGLE - All Galaxies; z = 0} & $11.03\tiny{\Vectorstack{+0.05 -0.04}}$ & $1.246\tiny{\Vectorstack{+0.121 -0.123}}$ & $-1.337\tiny{\Vectorstack{+0.018 -0.014}}$ & $\cdots$ & $\cdots$ \\
    
       \textbf{EAGLE - All Star Forming; z = 0} & $10.74\tiny{\Vectorstack{+0.04 -0.04}}$ & $1.589\tiny{\Vectorstack{+0.143 -0.135}}$ & $-1.236\tiny{\Vectorstack{+0.020 -0.022}}$ & $\cdots$ & $\cdots$ \\
    
       \textbf{EAGLE - All Quenched; z = 0} & $10.98\tiny{\Vectorstack{+0.16 -0.15}}$ & $0.592\tiny{\Vectorstack{+0.200 -0.175}}$ & $-0.904\tiny{\Vectorstack{+0.338 -0.238}}$ & $0.011\tiny{\Vectorstack{+0.041 -0.010}}$ & $-2.159\tiny{\Vectorstack{+0.319 -0.573}}$\\
    
       \textbf{EAGLE - Quenched Satellites; z = 0} & $10.60\tiny{\Vectorstack{+0.15 -0.53}}$ & $0.505\tiny{\Vectorstack{+0.68 -0.108}}$ & $-0.709\tiny{\Vectorstack{+1.38 -0.291}}$ & $0.0349\tiny{\Vectorstack{+0.197 -0.024}}$ & $-1.911\tiny{\Vectorstack{+0.384 -0.490}}$\\
    
       \textbf{EAGLE - Quenched Centrals; z = 0} & $11.24\tiny{\Vectorstack{+0.06 -0.05}}$ & $0.255\tiny{\Vectorstack{+0.020 -0.025}}$ & $-0.794\tiny{\Vectorstack{+0.031 -0.037}}$ & $\cdots$ & $\cdots$\\
    
       \textbf{EAGLE - All Galaxies; z = 1} & $11.12 \tiny{\Vectorstack{+0.03 -0.04}}$ & $0.761 \tiny{\Vectorstack{+0.085 -0.065}}$ & $-1.452 \tiny{\Vectorstack{+0.015 -0.012}}$ & $\cdots$ & $\cdots$ \\
    
       \textbf{EAGLE - All Star Forming; z = 1} & $10.84 \tiny{\Vectorstack{+0.02 -0.03}}$ & $1.029 \tiny{\Vectorstack{+0.082 -0.067}}$ & $-1.429 \tiny{\Vectorstack{+0.015 -0.012}}$ & $\cdots$ & $\cdots$ \\
    
       \textbf{EAGLE - All Quenched; z = 1} & $10.44 \tiny{\Vectorstack{+0.04 -0.04}}$ & $0.861 \tiny{\Vectorstack{+0.023 -0.020}}$ & $0.121 \tiny{\Vectorstack{+0.101 -0.092}}$ & $0.003 \tiny{\Vectorstack{+0.002 -0.001}}$ & $-2.705 \tiny{\Vectorstack{+0.152 -0.129}}$\\
    
       \textbf{EAGLE - Quenched Satellites; z = 1} &$ 10.24 \tiny{\Vectorstack{+0.06 -0.11}}$ & $0.364 \tiny{\Vectorstack{+0.026 -0.054}}$ & $0.375 \tiny{\Vectorstack{+0.322 -0.260}}$ & $0.023 \tiny{\Vectorstack{+0.021 -0.010}}$ & $-2.263 \tiny{\Vectorstack{+0.175 -0.179}}$\\
    
       \textbf{EAGLE - Quenched Centrals; z = 1} & $10.62 \tiny{\Vectorstack{+0.03 -0.04}}$ & $0.484 \tiny{\Vectorstack{+0.017 -0.018}}$ & $-0.015 \tiny{\Vectorstack{+0.092 -0.071}}$ & $\cdots$ & $\cdots$\\
    
       \textbf{EAGLE - All Galaxies; z = 2} & $11.09 \tiny{\Vectorstack{+0.07 -0.07}}$ & $0.318 \tiny{\Vectorstack{+0.049 -0.056}}$ & $-1.585 \tiny{\Vectorstack{+0.016 -0.021}}$ & $\cdots$ & $\cdots$ \\
    
       \textbf{EAGLE - All Star Forming; z = 2} & $10.80 \tiny{\Vectorstack{+0.10 -0.08}}$ & $0.445 \tiny{\Vectorstack{+0.106 -0.095}}$ & $-1.592 \tiny{\Vectorstack{+0.033 -0.030}}$ & $\cdots$ & $\cdots$ \\
    
       \textbf{EAGLE - All Quenched; z = 2} & $10.12 \tiny{\Vectorstack{+0.07 -0.08}}$ & $0.510 \tiny{\Vectorstack{+0.041 -0.052}}$ & $0.646 \tiny{\Vectorstack{+0.264 -0.204}}$ & $0.0006 \tiny{\Vectorstack{+0.0013 -0.0004}}$ & $-3.454 \tiny{\Vectorstack{+0.484 -0.533}}$\\
    
       \textbf{EAGLE - Quenched Satellites; z = 2} & $9.82 \tiny{\Vectorstack{+0.12 -0.07}}$ & $0.111 \tiny{\Vectorstack{+0.040 -0.025}}$ & $1.182 \tiny{\Vectorstack{+0.285 -0.558}}$ & $0.020 \tiny{\Vectorstack{+0.010 -0.012}}$ & $-2.232 \tiny{\Vectorstack{+0.193 -0.320}}$\\
    
       \textbf{EAGLE - Quenched Centrals; z = 2} &$ 10.25 \tiny{\Vectorstack{+0.07 -0.06}}$ & $0.390 \tiny{\Vectorstack{+0.019 -0.015}}$ & $0.429 \tiny{\Vectorstack{+0.188 -0.172}}$ & $\cdots$ & $\cdots$\\
      \hline

    \end{tblr}
    \label{tab:Schech_EAGLE_Param}
\end{table*}

\section{Machine Learning Reproducibility}\label{appendix:C}

For all Random Forest classification analyses we make use of the {\small SCIKIT-LEARN} Random Forest Classifier architecture~\citep{pedregosa11}. In order to achieve optimal performance, the Random Forest classifier allows for the fine tuning of multiple hyper-parameters. We include below a brief description for each of the relevant hyper-parameters (which are listed in Tables~\ref{tab:RF_Param_Cen_complete}, \ref{tab:RF_Param_Cen_2D}, \ref{tab:RF_Param_Cen_ML}, \& \ref{tab:RF_Param_Cen_ML_obs} for centrals and Tables~\ref{tab:RF_Param_Sat_complete}, \ref{tab:RF_Param_Sat_2D}, \ref{tab:RF_Param_Sat_ML}, \& \ref{tab:RF_Param_Sat_ML_obs} for satellites). \\

\textit{Method}: In all cases, we make use of the Random Forest Classifier from SCIKIT-LEARN, abbreviated to RFclass in the Tables.\\

\textit{Normalized}: Indicates wether the data is normalized prior to going through the Random Forest classifier. In all cases, we subtract the median and normalize by the inter-quartile range. This ensures the classifier will not be influenced by the different distributions or value ranges between features. \\

\textit{$\mathrm{N}_{int}$} : The number of times a random forest classification is run, altering the data randomly separated into the training and testing sample sets. In this case, we set all to, $\mathrm{N}_{int} = 100$. \\

\textit{Balanced}: Indicates whether the data is balanced to have an even amount of quiescent and star forming galaxies. While it is best to evenly sample the data prior to each Random Forest classification run, the dearth of quenched objects at higher redshift, especially quiescent satellites for z = 2, renders us unable to do so in some cases.\\

\textit{Train : Test}: The training to testing ratio prior to running the random forest classifier. For sufficiently large data sets, an even 50:50 split is ideal. However, for smaller amounts of data it may be preferable to allow for a larger training than testing data set~\citep[see][]{Bluck_2024}. As such, for certain parent sample sets, an uneven training:testing split is applied, prior to the Random Forest classification. In these cases, the training sample comprises up to 70\% of the total data set, and the testing data set is therefore the remaining 30\%.\\

\textit{Max Depth}: The maximum amount of decision splits which can occur within any given decision tree. This is set to \textit{Max Depth} = 200 for all of our Random Forest runs. This always exceeds the limit set by the \textit{min samples leaf} value (MSL, see below). \\

\textit{min-samples-leaf} - MSL: The minimum amount of data remaining in a node needed to attempt further splits. Although setting the MSL to its logical minimum of two seems intuitive, this would in general lead to overfitting. That is, while the training data would be optimally classified, the Random Forest algorithm would likely learn pathological attributes within the training data which do not scale to unseen testing data. To prevent this, we fine tune the MSL value to simultaneously maximize the AUC (further discussed below), which quantifies the training performance, and minimize the $\Delta$AUC (see below), which  quantifies the difference between the performance of the RF on the training and unseen data. \\

\textit{Random State}: The random state informs the random bootstrapping of data. Specifying the value of the random state assures the bootstrapping to be consistent through both iterations, and should enable other researches to reproduce the results precisely, even given their dependence on `random' variables. \\

\textit{AUC}: The area under true positive - false positive receiver operator curve (see, e.g., \citealt{Teimoorinia_2016}). The AUC reveals the performance of the Random Forest classification, with an AUC = 1 being a perfect classification and AUC = 0.5 indicating an entirely random performance. We include the mean training data AUC from the 100 iterations as the AUC column within each table. \\

\textit{$\Delta$AUC}: Quantifies the difference in the performances of the Random Forest classifier on the training and testing data sets. We attempt to minimize the $\Delta$AUC for all RF runs. While a threshold of $\Delta \mathrm{AUC} \leq 0.02$ is optimal as according to \citet{Bluck_2022}, we accept all tests with $|\Delta \mathrm{AUC}| \leq 0.07$ as a successful fit here. This change is due mainly to the sparsity of data available from certain parent sample sets. \\

To assure our RF classification results as reproducible as possible, we report information on said hyperparameters used for each RF test in Tables~\ref{tab:RF_Param_Cen_complete} - \ref{tab:RF_Param_Sat_ML_obs}. Additionally, we set \textit{max-features} = `None' for all classifications, so as to insure that all features are evaluated at every node of all decision trees of the Random Forest. As shown in \citet{Bluck_2022}, this is the optimal method available in RF analyses for determining the most causally connected feature for classification, enabling controlling for all other available features.

\begin{table*}
    \centering
        \caption{Random Forest hyper-parameters for the classifications tests performed for centrals of the complete data sets, presented in Figs. \ref{fig:TNG_RF} \& \ref{fig:EAGLE_RF}}
    \begin{tblr}{
             colspec = {l c c c c c c c c c c r},  
             row{1}  = {font=\bfseries, c},
             column{1} = {font = \bfseries, c},
             rowsep=0.5ex, colsep =0.5ex
             }
    
      \hline
      \textbf{Sample Set - Centrals} & Train:Test & Balanced & Normalized & Method & $\mathrm{N}_{int}$ & $\mathrm{N}_{trees}$ & Max Depth & MSL & Random State & AUC & $\Delta$ AUC\\
      \hline
       TNG; z = 0 & 50/50 & Yes & Yes & RFclass & 100 & 200 & 500 & 100 & 42 & 0.98 & 0.008\\

       TNG; z = 1 & 50/50 & Yes & Yes & RFclass & 100 & 200 & 500 & 35 & 42 & 0.98 & 0.014\\

       TNG; z = 2 & 50/50 & Yes & Yes & RFclass & 100 & 200 & 500 & 18 & 42 & 0.99 & 0.012\\

       EAGLE; z = 0 & 50/50 & Yes & Yes & RFclass & 100 & 200 & 500 & 95 & 42 & 0.84 & 0.022\\

       EAGLE; z = 1 & 50/50 & Yes & Yes & RFclass & 100 & 200 & 500 & 25 & 42 & 0.97 & 0.028\\

       EAGLE; z = 2 & 50/50 & Yes & Yes & RFclass & 100 & 200 & 500 & 35 & 42 & 0.96 & 0.027\\
      \hline
    \end{tblr}
    \label{tab:RF_Param_Cen_complete}
\end{table*}

\begin{table*}
    \centering
        \caption{Random Forest hyper-parameters for the classifications tests performed for satellites (`HM' \& `LM' are used for high and low-mass) of the complete data sets, presented in Figs. \ref{fig:TNG_RF} \& \ref{fig:EAGLE_RF}}
    \begin{tblr}{
             colspec = {l c c c c c c c c c c r},  
             row{1}  = {font=\bfseries, c},
             column{1} = {font = \bfseries, c},
             rowsep=0.5ex, colsep =0.5ex
             }
      \hline
      \textbf{Sample Set - Satellites} & Train:Test & Balanced & Normalized & Method & $\mathrm{N}_{int}$ & $\mathrm{N}_{trees}$ & Max Depth & MSL & Random State & AUC & $\Delta$ AUC\\
      \hline
       TNG HM; z = 0 & 50/50 & Yes & Yes & RFclass & 100 & 200 & 500 & 51 & 42 & 0.84 & 0.032\\

       TNG HM; z = 1 & 60/40 & Yes & Yes & RFclass & 100 & 200 & 500 & 64 & 42 & 0.88 & 0.030\\

       TNG HM; z = 2 & 70/30 & Yes & Yes & RFclass & 100 & 200 & 500 & 31 & 42 & 0.90 & 0.041\\

       TNG LM; z = 0 & 50/50 & Yes & Yes & RFclass & 100 & 200 & 500 & 300 & 42 & 0.87 & 0.007\\

       TNG LM; z = 1 & 50/50 & Yes & Yes & RFclass & 100 & 200 & 500 & 100 & 42 & 0.85 & 0.017\\

       TNG LM; z = 2 & 50/50 & Yes & Yes & RFclass & 100 & 200 & 500 & 47 & 42 & 0.89 & 0.026\\

       EAGLE HM; z = 0 & 70/30 & Yes & Yes & RFclass & 100 & 200 & 500 & 63 & 42 & 0.78 & 0.047\\

       EAGLE HM; z = 1 & 60/40 & Yes & Yes & RFclass & 100 & 200 & 500 & 52 & 42 & 0.85 & 0.038\\

       EAGLE HM; z = 2 & 50/50 & No & Yes & RFclass & 100 & 200 & 500 & 50 & 42 & 0.90 & 0.028\\

       EAGLE LM; z = 0 & 50/50 & Yes & Yes & RFclass & 100 & 200 & 500 & 200 & 42 & 0.88 & 0.009\\

       EAGLE LM; z = 1 & 70/30 & Yes & Yes & RFclass & 100 & 200 & 500 & 85 & 42 & 0.88 & 0.028\\

       EAGLE LM; z = 2 & 70/30 & No & Yes & RFclass & 100 & 200 & 500 & 81 & 42 & 0.83 & 0.057\\
      \hline
    \end{tblr}
    \label{tab:RF_Param_Sat_complete}
\end{table*}

\begin{table*}
    \centering
        \caption{Random Forest hyper-parameters for the classifications tests performed for centrals of the 2D-complete data sets, presented in Figs. \ref{fig:TNG_RF} \& \ref{fig:EAGLE_RF}}
    \begin{tblr}{
             colspec = {l c c c c c c c c c c r},  
             row{1}  = {font=\bfseries, c},
             column{1} = {font = \bfseries, c},
             rowsep=0.5ex, colsep =0.5ex
             }
    \hline
      \textbf{2D Sample Set - Centrals} & Train:Test & Balanced & Normalized & Method & $\mathrm{N}_{int}$ & $\mathrm{N}_{trees}$ & Max Depth & MSL & Random State & AUC & $\Delta$ AUC\\
      \hline
       TNG; z = 0 & 50/50 & Yes & Yes & RFclass & 100 & 200 & 500 & 100 & 42 & 0.97 & 0.006\\

       TNG; z = 1 & 50/50 & Yes & Yes & RFclass & 100 & 200 & 500 & 50 & 42 & 0.98 & 0.015\\

       TNG; z = 2 & 50/50 & Yes & Yes & RFclass & 100 & 200 & 500 & 15 & 42 & 0.99 & 0.011\\

       EAGLE; z = 0 & 50/50 & Yes & Yes & RFclass & 100 & 200 & 500 & 100 & 42 & 0.84 & 0.024\\

       EAGLE; z = 1 & 50/50 & Yes & Yes & RFclass & 100 & 200 & 500 & 50 & 42 & 0.96 & 0.027\\

       EAGLE; z = 2 & 50/50 & Yes & Yes & RFclass & 100 & 200 & 500 & 32 & 42 & 0.96 & 0.031\\
      \hline
    \end{tblr}
    \label{tab:RF_Param_Cen_2D}
\end{table*}

\begin{table*}
    \centering
        \caption{Random Forest hyper-parameters for the classifications tests performed for satellites (`HM' \& `LM' are used for high and low-mass) of the 2D-complete data sets, presented in Figs. \ref{fig:TNG_RF} \& \ref{fig:EAGLE_RF}}
    \begin{tblr}{
             colspec = {l c c c c c c c c c c r},  
             row{1}  = {font=\bfseries, c},
             column{1} = {font = \bfseries, c},
             rowsep=0.5ex, colsep =0.5ex
             }
      \hline
      \textbf{2D Sample Set - Satellites} & Train:Test & Balanced & Normalized & Method & $\mathrm{N}_{int}$ & $\mathrm{N}_{trees}$ & Max Depth & MSL & Random State & AUC & $\Delta$ AUC\\
      \hline
       TNG HM; z = 0 & 60/40 & Yes & Yes & RFclass & 100 & 200 & 500 & 54 & 42 & 0.85 & 0.024\\

       TNG HM; z = 1 & 50/50 & Yes & Yes & RFclass & 100 & 200 & 500 & 42 & 42 & 0.89 & 0.031\\

       TNG HM; z = 2 & 60/40 & No & Yes & RFclass & 100 & 200 & 500 & 40 & 42 & 0.89 & 0.031\\

       TNG LM; z = 0 & 50/50 & Yes & Yes & RFclass & 100 & 200 & 500 & 220 & 42 & 0.87 & 0.007\\

       TNG LM; z = 1 & 50/50 & Yes & Yes & RFclass & 100 & 200 & 500 & 170 & 42 & 0.84 & 0.010\\

       TNG LM; z = 2 & 50/50 & Yes & Yes & RFclass & 100 & 200 & 500 & 52 & 42 & 0.88 & 0.027\\

       EAGLE HM; z = 0 & 70/30 & Yes & Yes & RFclass & 100 & 200 & 500 & 64 & 42 & 0.79 & 0.05\\

       EAGLE HM; z = 1 & 60/40 & Yes & Yes & RFclass & 100 & 200 & 500 & 52 & 42 & 0.86 & 0.036\\

       EAGLE HM; z = 2 & 70/30 & No & Yes & RFclass & 100 & 200 & 500 & 50 & 42 & 0.89 & 0.039\\

       EAGLE LM; z = 0 & 50/50 & Yes & Yes & RFclass & 100 & 200 & 500 & 180 & 42 & 0.88 & 0.011\\

       EAGLE LM; z = 1 & 50/50 & Yes & Yes & RFclass & 100 & 200 & 500 & 65 & 42 & 0.87 & 0.021\\

       EAGLE LM; z = 2 & 70/30 & No & Yes & RFclass & 100 & 200 & 500 & 78 & 42 & 0.84 & 0.05\\
      \hline
    \end{tblr}
    \label{tab:RF_Param_Sat_2D}
\end{table*}

\begin{table*}
    \centering
        \caption{Random Forest hyper-parameters for the classifications tests performed for centrals of the MOONRISE-Like (ML) data sets, presented in Figs. \ref{fig:TNG_RF} \& \ref{fig:EAGLE_RF}}
    \begin{tblr}{
             colspec = {l c c c c c c c c c c r},  
             row{1}  = {font=\bfseries, c},
             column{1} = {font = \bfseries, c},
             rowsep=0.5ex, colsep =0.5ex
             }
      \hline
      \textbf{ML Sample Set - Centrals} & Train:Test & Balanced & Normalized & Method & $\mathrm{N}_{int}$ & $\mathrm{N}_{trees}$ & Max Depth & MSL & Random State & AUC & $\Delta$ AUC\\
      \hline
       TNG; z = 0 & 50/50 & Yes & Yes & RFclass & 100 & 200 & 500 & 70 & 42 & 0.96 & 0.013\\

       TNG; z = 1 & 50/50 & Yes & Yes & RFclass & 100 & 200 & 500 & 40 & 42 & 0.97 & 0.02\\

       TNG; z = 2 & 50/50 & Yes & Yes & RFclass & 100 & 200 & 500 & 15 & 42 & 0.99 & 0.015\\

       EAGLE; z = 0 & 50/50 & Yes & Yes & RFclass & 100 & 200 & 500 & 65 & 42 & 0.85 & 0.043\\

       EAGLE; z = 1 & 50/50 & Yes & Yes & RFclass & 100 & 200 & 500 & 25 & 42 & 0.96 & 0.031\\

       EAGLE; z = 2 & 50/50 & No & Yes & RFclass & 100 & 200 & 500 & 70 & 42 & 0.93 & 0.012\\
      \hline
    \end{tblr}
    \label{tab:RF_Param_Cen_ML}
\end{table*}

\begin{table*}
    \centering
        \caption{Random Forest hyper-parameters for the classifications tests performed for satellites of the MOONRISE-Like (ML) data sets, presented in Figs. \ref{fig:TNG_RF} \& \ref{fig:EAGLE_RF}}
    \begin{tblr}{
             colspec = {l c c c c c c c c c c r},  
             row{1}  = {font=\bfseries, c},
             column{1} = {font = \bfseries, c},
             rowsep=0.5ex, colsep =0.5ex
             }
      \hline
      \textbf{ML Sample Set - Satellites} & Train:Test & Balanced & Normalized & Method & $\mathrm{N}_{int}$ & $\mathrm{N}_{trees}$ & Max Depth & MSL & Random State & AUC & $\Delta$ AUC\\
      \hline
       TNG HM; z = 0 & 70/30 & Yes & Yes & RFclass & 100 & 200 & 500 & 44 & 42 & 0.83 & 0.024\\

       TNG HM; z = 1 & 50/50 & Yes & Yes & RFclass & 100 & 200 & 500 & 34 & 42 & 0.87 & 0.040\\

       TNG HM; z = 2 & 60/40 & No & Yes & RFclass & 100 & 200 & 500 & 20 & 42 & 0.90 & 0.052\\

       TNG LM; z = 0 & 50/50 & Yes & Yes & RFclass & 100 & 200 & 500 & 100 & 42 & 0.86 & 0.022\\

       TNG LM; z = 1 & 50/50 & Yes & Yes & RFclass & 100 & 200 & 500 & 48 & 42 & 0.83 & 0.032\\

       TNG LM; z = 2 & 70/30 & No & Yes & RFclass & 100 & 200 & 500 & 40 & 42 & 0.91 & 0.034\\      

       EAGLE HM; z = 0 & 70/30 & Yes & Yes & RFclass & 100 & 200 & 500 & 38 & 42 & 0.82 & 0.055\\

       EAGLE HM; z = 1 & 70/30 & Yes & Yes & RFclass & 100 & 200 & 500 & 40 & 42 & 0.86 & 0.053\\

       EAGLE HM; z = 2 & 60/40 & No & Yes & RFclass & 100 & 200 & 500 & 34 & 42 & 0.86 & 0.054\\

       EAGLE LM; z = 0 & 50/50 & Yes & Yes & RFclass & 100 & 200 & 500 & 90 & 42 & 0.86 & 0.021\\

       EAGLE ML - LM Sat; z = 1 & $\cdots$ & $\cdots$ & $\cdots$ & $\cdots$ & $\cdots$ & $\cdots$ & $\cdots$ & $\cdots$ & $\cdots$ & $\cdots$ & $\cdots$\\

       EAGLE ML - LM Sat; z = 2 & $\cdots$ & $\cdots$ & $\cdots$ & $\cdots$ & $\cdots$ & $\cdots$ & $\cdots$ & $\cdots$ & $\cdots$ & $\cdots$ & $\cdots$\\
      \hline
    \end{tblr}
    \label{tab:RF_Param_Sat_ML}
\end{table*}

\begin{table*}
    \centering
        \caption{Random Forest hyper-parameters for the classifications tests performed for centrals of the observation MOONRISE-Like ($\mathrm{ML_{obs}}$) data sets, presented in Figs. \ref{fig:TNG_RF_Moons} \& \ref{fig:EAGLE_RF_Moons}}
    \begin{tblr}{
             colspec = {l c c c c c c c c c c r},  
             row{1}  = {font=\bfseries, c},
             column{1} = {font = \bfseries, c},
             rowsep=0.5ex, colsep =0.5ex
             }
      \hline
      \textbf{$\mathrm{ML_{obs}}$ Sample Set - Centrals} & Train:Test & Balanced & Normalized & Method & $\mathrm{N}_{int}$ & $\mathrm{N}_{trees}$ & Max Depth & MSL & Random State & AUC & $\Delta$ AUC\\
      \hline
       TNG; z = 0 & 50/50 & Yes & Yes & RFclass & 100 & 200 & 500 & 120 & 42 & 0.91 & 0.010\\

       TNG; z = 1 & 50/50 & Yes & Yes & RFclass & 100 & 200 & 500 & 40 & 42 & 0.92 & 0.022\\

       TNG; z = 2 & 50/50 & Yes & Yes & RFclass & 100 & 200 & 500 & 15 & 42 & 0.98 & 0.025\\

       EAGLE; z = 0 & 50/50 & Yes & Yes & RFclass & 100 & 200 & 500 & 70 & 42 & 0.76 & 0.031\\

       EAGLE; z = 1 & 50/50 & Yes & Yes & RFclass & 100 & 200 & 500 & 60 & 42 & 0.84 & 0.032\\

       EAGLE; z = 2 & 50/50 & Yes & Yes & RFclass & 100 & 200 & 500 & 40 & 42 & 0.84 & 0.041\\
      \hline
    \end{tblr}
    \label{tab:RF_Param_Cen_ML_obs}
\end{table*}

\begin{table*}
    \centering
        \caption{Random Forest hyper-parameters for the classifications tests performed for satellites of the observation MOONRISE-Like ($\mathrm{ML_{obs}}$) data sets, presented in Figs. \ref{fig:TNG_RF_Moons} \& \ref{fig:EAGLE_RF_Moons}}
    \begin{tblr}{
             colspec = {l c c c c c c c c c c r},  
             row{1}  = {font=\bfseries, c},
             column{1} = {font = \bfseries, c},
             rowsep=0.5ex, colsep =0.5ex
             }
      \hline
      \textbf{$\mathrm{ML_{obs}}$ Sample Set - Satellites} & Train:Test & Balanced & Normalized & Method & $\mathrm{N}_{int}$ & $\mathrm{N}_{trees}$ & Max Depth & MSL & Random State & AUC & $\Delta$ AUC\\
      \hline
       TNG HM; z = 0 & 60/40 & No & Yes & RFclass & 100 & 200 & 500 & 66 & 42 & 0.74 & 0.054\\

       TNG HM; z = 1 & 70/30 & Yes & Yes & RFclass & 100 & 200 & 500 & 58 & 42 & 0.75 & 0.054\\

       TNG HM; z = 2 & 60/40 & No & Yes & RFclass & 100 & 200 & 500 & 36 & 42 & 0.75 & 0.058\\

       TNG LM; z = 0 & 50/50 & Yes & Yes & RFclass & 100 & 200 & 500 & 95 & 42 & 0.81 & 0.021\\

       TNG LM; z = 1 & 70/30 & Yes & Yes & RFclass & 100 & 200 & 500 & 49 & 42 & 0.80 & 0.048\\

       TNG LM; z = 2 & 70/30 & No & Yes & RFclass & 100 & 200 & 500 & 46 & 42 & 0.81 & 0.048\\      

       EAGLE HM; z = 0 & 70/30 & Yes & Yes & RFclass & 100 & 200 & 500 & 40 & 42 & 0.72 & 0.062\\

       EAGLE HM; z = 1 & 60/40 & No & Yes & RFclass & 100 & 200 & 500 & 49 & 42 & 0.74 & 0.07\\

       EAGLE HM; z = 2 & 70/30 & No & Yes & RFclass & 100 & 200 & 500 & 46 & 42 & 0.71 & 0.073\\

       EAGLE LM; z = 0 & 50/50 & Yes & Yes & RFclass & 100 & 200 & 500 & 70 & 42 & 0.81 & 0.027\\

       EAGLE LM; z = 1 & $\cdots$ & $\cdots$ & $\cdots$ & $\cdots$ & $\cdots$ & $\cdots$ & $\cdots$ & $\cdots$ & $\cdots$ & $\cdots$ & $\cdots$\\

       EAGLE LM; z = 2 & $\cdots$ & $\cdots$ & $\cdots$ & $\cdots$ & $\cdots$ & $\cdots$ & $\cdots$ & $\cdots$ & $\cdots$ & $\cdots$ & $\cdots$\\
      \hline
    \end{tblr}
    \label{tab:RF_Param_Sat_ML_obs}
\end{table*}

\newpage
\bsp	
\label{lastpage}
\end{document}